\documentclass[
11pt,
letterpaper,
abstract=true
]{scrartcl}

\usepackage[ngerman, english]{babel}
\usepackage[utf8]{inputenc}
\usepackage{graphicx}
\usepackage{url}
\usepackage{marvosym}
\usepackage{lmodern}
\usepackage{comment}
\usepackage{xcolor}
\usepackage[commandnameprefix=always]{changes}
\usepackage{changebar}
\usepackage[round]{natbib}

\usepackage{caption}
\usepackage{subcaption}
\usepackage[doublespacing]{setspace}

\usepackage{afterpage}
\usepackage{fancyref}
\usepackage{enumitem}	%?
\usepackage{tabularx}	%?
\usepackage{longtable}	% Tabellen über mehrere Seiten
\usepackage{booktabs}	% Tabellenlayout
\usepackage{dcolumn}	% In Regressionstabellen Zahlen am Dezimalpunkt zentrieren
\usepackage[para,online,flushleft]{threeparttable}
\usepackage{pdflscape} 	% Seiten im Querformat
\usepackage{amsmath, amssymb}

\usepackage{eurosym} % Official Euro-Symbol

\newcommand{\sym}[1]{{#1}} % for symbols in Table

\usepackage[left=2.5cm, right=2.5cm, bottom=2.5cm, top=2.5cm]{geometry} %Ränder % includehead, includefoot ??
\usepackage{graphicx} %Grafiken einbinden
\usepackage{mathtools} %Mathematik
\usepackage{mathpazo} %Mathematik
\usepackage{array}  %Column width

\newcolumntype{C}[1]{>{\centering\let\newline\\\arraybackslash\hspace{0pt}}m{#1}}
\newcolumntype{L}[1]{>{\raggedright\let\newline\\\arraybackslash\hspace{0pt}}m{#1}}

\usepackage[title]{appendix}
\usepackage{titletoc}
\usepackage[pdftex,bookmarks=true, colorlinks=false, pdfborder={0 0 0}]{hyperref} 	
\begin{document}

	%%%
	% Title page:
	
	\title{\LARGE  Firms' Risk Adjustments to Minimum Wage:
		\\ Financial Leverage and Labor Share Trade-off\vspace{0.5cm}}

	\author{Ying Liang % 
		\thanks{Graduate School of Economics, Finance and Management, Goethe University Frankfurt (GU Frankfurt) and Johannes Gutenberg-University Mainz (JGU Mainz) Address: JGU Mainz, Jakob-Welder-Weg 4, 55128 Mainz, e-mail: \texttt{liang.ying@uni-mainz.de}.
			This project is funded by the Chair of Applied Statistics and Econometrics at the University of Mainz, Freunde und Förderer of Goethe University Frankfurt, and the Frauenförderfonds FB 03 of the University of Mainz. I gratefully acknowledge helpful comments from Thorsten Schank, Tobias Berg, Andrej Gill, David Margolis, and participants at the seminar at the University of Mainz and the 19th Doctorissimes at the Paris School of Economics. Special thanks to Manfred Antoni and the Research Data Centre (FDZ) at the Institute for Employment Research (IAB) for their assistance with data provision.}\vspace{-0.3cm}
		\\
		\small{\emph {GSEFM, GU Frankfurt and JGU Mainz} }		
	}
	
	\vspace{-0.3cm}
	\date{August 2024}
	
	\maketitle
	
	\vspace{-0.9cm}
	\begin{abstract}
		This paper evaluates the impact of the German minimum wage policy on firms' financial leverage. By using a comprehensive firm-establishment-employee linked dataset and a difference-in-differences estimation with firm-level variation in treatment intensity, the analysis shows that the average minimum wage level reduces firms' financial leverage by about 0.5 to 0.9 percentage points, corresponding to 1 to 2 percent of the mean of financial leverage. Further investigation of the mechanism shows that the minimum wage does not lead to significant capital-labor substitution; therefore, the labor share increases. Firms react to the increased labor share by deleveraging. 
The results suggest that while the minimum wage benefits workers by allocating more earnings to the labor force, it also introduces greater operating risks and encourages conservative financial behavior among firms. [126 words]

		\vspace{0.3cm} 
	\end{abstract}
	
	\thispagestyle{empty}
	\clearpage
	\setcounter{page}{1}
	\doublespacing

	\startcontents[mainsections]

	\section{Introduction
		\label{sec:introduction}}

	In recent decades, scholars have paid attention to the relationship between corporate financial decision-making and labor market frictions. Compared to capital, labor, as a crucial production factor, possesses unique characteristics. For instance, labor differs from capital in terms of adjustment costs; it is influenced by labor market regulations such as employment protection laws and labor unions. Higher labor costs and associated adjustment costs can increase firms' risks during financial distress, thereby accelerating financial difficulties. Thus, in addition to the traditional determinants, labor market frictions have a significant impact on firms' capital structure decisions.

	The existing literature on labor and corporate finance predominantly concentrates on employment protection laws, while research on minimum wage and financial leverage is under investigation.\footnote{
Financial leverage is an indicator of a firm's capital structure. In this paper, financial leverage is defined as the ratio of total debts to total assets.} Moreover, studies on minimum wages tend to focus on labor market outcomes, often neglecting the analysis of capital structure and operating risks. This study investigates how the German statutory minimum wage influences firms' capital structure and explores the relationship between labor market regulations and the credit market. It sheds light on how the minimum wage affects firms' risk substitution, which directly influences their economic activities and the stability of the credit market.\footnote{Firms' capital structure is related to managers' incentive problems, with higher financial leverage mitigating agency costs \citep{grossman_corporate_1982} and thereby affecting firms' investment and performance \citep{harvey_effect_2004}. The financial leverage of non-financial firms is also important for credit market stability. For example, lower financial leverage among firms reduces the probability of default and enhances overall financial market stability.} Furthermore, the study contributes to understanding the deleveraging trend observed among German firms over the past decade.

	Firms exposed to the minimum wage may face increased total labor costs. If these firms cannot pass on all of the increased costs to consumers, they will suffer from decreased profits and a higher labor share, defined as the ratio of labor costs to value-added. An increased labor share can lead to greater business risks because labor costs cannot be fully adjusted during economic downturns, thereby increasing expected costs during financial distress. Meanwhile, higher financial leverage also escalates firms' business risks, making them more likely to default in adverse conditions due to fixed interest payments in each period. Labor share and financial leverage may, therefore, act as substitutes. To mitigate the risks exacerbated by the minimum wage, firms may choose to decrease their financial leverage ex-ante.

This paper examines the effect of the minimum wage on financial leverage by studying the implementation of the German statutory minimum wage policy in 2015. It utilizes a dataset that links firm financial information with administrative employment records in Germany. The identification method is a detrended difference-in-differences approach, with a continuous treatment measure at the firm level, known as the bite variable. This measure is calculated as the proportion of workers paid below the minimum wage before the policy implementation. I find that firms' financial leverage decreases by about 0.5 to 0.9 percentage points due to the average treatment level of the minimum wage, while the labor share increases by about 0.5 to 1.6 percentage points. The results are robust in relation to alternative measures of treatment or financial leverage and alternative sample restrictions.
	
	Moreover, further analysis of the mechanism reveals that the minimum wage boosts firms' cash holdings and lowers debt borrowing, leading to a reduction in financial leverage. Regarding the change in the labor share, the elasticity of substitution between labor and capital is estimated at 0.31, suggesting complementarity between labor and capital. This is consistent with my findings of increased labor share due to an exogenous wage increase. It is also found that total labor costs increase, while firms' profits decrease, suggesting that total value-added is distributed more towards labor. 
	
	In terms of heterogeneous effects, firstly, it is found that firms tend to reduce long-term debts instead of short-term ones, which is plausible since long-term debts require longer periods of interest payments and entail greater economic uncertainties with prolonged durations. Secondly, a firm's ability to adjust its labor flexibly is critical in determining its reaction to an increasing labor share. A flexible labor composition, such as occupations that are more easily outsourced, results in less reduction in financial leverage. Lastly, concerning firm size, small firms exhibit a higher increase in the labor share and demonstrate strong deleveraging behavior, reaffirming the relationship between the labor share and financial leverage. Overall, the results suggest that the minimum wage benefits employees by increasing their earnings share but introduces larger operating risks to firms and leads firms to adopt more conservative behaviors.
	
	\textbf{Related literature.}---My paper is situated within two strands of literature. The first pertains to the impact of labor frictions on firms' financial leverage, while the second examines the effect of the minimum wage on firms' responses.
	
	Previous studies have laid theoretical and empirical foundations regarding the impact of labor forces on firms' capital structure. \cite{favilukis_elephant_2020} use a dynamic stochastic general equilibrium (DSGE) model with heterogeneous firms to demonstrate that labor market frictions result in firms responding slowly to adjustments in labor costs. Therefore, high labor costs and sticky wages increase the likelihood of firms defaulting when facing negative shocks. Firms with a high labor share, therefore, tend to opt for a lower financial leverage ratio. \cite{berk_human_2010} also develops a model that describes the relationship between human capital costs and financial leverage, suggesting that labor-intensive firms will borrow less.
	
	On the empirical side, \cite{serfling_firing_2016} and \cite{simintzi_labor_2015} explore the effect of employment protection laws on firms' financial leverage and find that financial leverage has decreased following the implementation of these laws. \cite{kim_how_2020} shows that firms tend to increase their use of debt financing as the size of the local labor market grows. A larger labor market reduces the costs associated with workers' job loss, thereby lowering expenses related to financial distress.\footnote{For a comprehensive summary of papers on labor regulations and firms' capital structure, please see \cite{matsa_capital_2018}.}

	%subsection{German minimum wage studies about firms' response}
	Additionally, this paper links to the empirical literature evaluating the effects of minimum wage policies on firm-level outcomes. Internationally, studies find that minimum wage policies often negatively affect firms' profitability \citep{alexandre_minimum_2022, draca_minimum_2011} and capital investment \citep{gustafson_higher_2023}, while they tend to positively impact revenue and prices \citep{harasztosi_who_2019, leung_minimum_2021, renkin_pass-through_2022}. For Germany, \cite{bossler_german_2020} use the IAB Establishment Panel covering the years 2010 to 2015. They find no capital investment adjustment in the affected establishments and a very small reduction in human capital investment. Additionally, they show no effects of the minimum wage on productivity but observe a reduction in profitability. By utilizing the ifo Business Survey data, \cite{link_price_2019} finds that firms raised prices by 0.82\% in reaction to a 1\% increase in costs caused by the minimum wage.
	
	% \cite{kuhn_are_2021} uses aggregate data at the industry level and demonstrates no profit effects induced by the minimum wage.
	This paper makes several contributions to the existing literature. First, it expands the current body of research on labor costs and capital structure. While a significant amount of literature examines the impact of labor frictions in various forms on financial leverage, the impact of the minimum wage on financial leverage is a novel question that remains understudied.\footnote{To the best of my knowledge, there is only one existing paper that examines the effects of the minimum wage on financial leverage. In an early working paper \citep{gustafson_minimum_2017} version of \cite{gustafson_higher_2023}, the authors found that minimum wage increases in the US significantly reduce firms' net leverage. However, these findings are not included in the published version \citep{gustafson_higher_2023}.
My paper differs substantially from \cite{gustafson_minimum_2017}. First, while the minimum wage in the US is state-specific, it is nationwide with a uniform threshold in Germany. Second, I employ a different methodology. I use a continuous treatment measure, whereas they calculate the change in the minimum wage as the treatment measure. Lastly, I investigate the effects of the minimum wage on the labor share as the primary channel through which the policy impacts financial leverage. In contrast, \cite{gustafson_minimum_2017} does not explore this mechanism.} The minimum wage is a widely used policy tool that plays a crucial role in increasing wages. By focusing on the minimum wage, this paper demonstrates how a fundamental labor market regulation affects firms' capital structure. Second, the mechanism between the minimum wage and financial leverage is examined by quantifying the effect of the minimum wage on the labor share. Previous studies using policy as a quasi-experiment focus on employment protection laws and firing costs. Measuring firing costs and determining the extent to which the law increases these costs is challenging. This paper directly tests how the minimum wage affects the labor share and total labor costs, thus supporting theoretical predictions by considering the first-order adjustments within firms. Third, the study enriches the mechanism by providing the estimation of capital-labor substitution and examining the heterogeneous effects based on firms' flexibility in adjusting their labor and based on firm sizes. Lastly, the most comprehensive dataset available for Germany is used to study the firm-level response to the minimum wage. Other data sources, such as survey data, typically only collect information on a subset of the employees within a firm. The treatment intensity is often approximated using industry or regional-level variations based on the location of a firm's headquarters, leading to an imprecise measure. By calculating firm-level exposure with individual wage information for nearly all employees within a firm, the treatment variable is less prone to measurement error.

	The paper proceeds as follows. Section \ref{sec:theory} provides theoretical explanations of how the minimum wage affects firms' financial leverage. Section \ref{sec:back_data} introduces the institutional background of the minimum wage, followed by a description of the dataset. The empirical approach is described in Section \ref{sec:method}. Section \ref{sec: main_result} presents the main findings on the impact of the minimum wage on financial leverage and on the labor share. Further mechanisms investigation and heterogeneous effects are studied in Section \ref{sec:further}. Section \ref{sec: conclude} concludes.

\section{Theoretical background\label{sec:theory}}
	In the traditional corporate finance literature, firms' financial leverage ratio depends on the tax benefits and costs of financial distress. Specifically, commonly considered determinants of financial leverage include tax deductions, firm size,  tangibility, and profitability \citep{antoniou_determinants_2008}.\footnote{In general, the tax rate is positively related to financial leverage, as higher tax rates increase the benefits of debt borrowing due to tax deductibility of interest expenses. Firm size and tangibility also positively correlate with financial leverage because larger firms and firms with more tangible assets have a greater capacity to borrow. According to the pecking order theory, profitability negatively affects financial leverage, as firms increase debt borrowing when they lack sufficient internal resources. There are other factors that influence financial leverage; for a comprehensive summary, see \cite{antoniou_determinants_2008} and \cite{parsons_empirical_2009}.}
	In the past few decades, the labor force has increasingly been recognized as a significant factor that can impact a company's capital structure. From a theoretical background, labor frictions affect a firm's financial leverage in two ways. The positive effect is established by \cite{matsa_capital_2010}, where a firm's optimal capital structure is chosen as a strategic response to workers' bargaining power from the union. Higher liquidity will encourage workers to raise the demand for wage growth. Firms, therefore, tend to use more debt financing to reduce future cash flow and to strengthen firms' bargaining position against employees.

	\textbf{Leverage substitution.}---The negative effect is attributed to the substitution effect of operating and financial leverage \citep{mandelker_impact_1984, mauer_interactions_1994, chen_operating_2019, sarkar_relationship_2020}. Operating leverage is defined as the sensitivity of a firm's profits to changes in sales or the proportion of fixed costs to total costs \citep{hillier_corporate_2010}.\footnote{Two commonly used measures of operating leverage are (1) the change in EBIT/the change in sales or output and (2) fixed costs/(variable costs+fixed costs).} Regardless of the various methods used to measure operating leverage, the concept remains straightforward: firms with higher operating leverage are more sensitive to economic shocks. The leverage substitution theory suggests that both higher operating leverage and higher financial leverage increase the expected costs of financial distress. This is because higher operating leverage results in greater fixed costs, whereas higher financial leverage leads to higher coupon payments. Therefore, a trade-off exists between operating and financial leverage.

	Relative labor expenses and labor inflexibility give rise to a special form of operating leverage known as labor-induced leverage or the labor share \citep{donangelo_cross-section_2019, gourio_labor_2007}.\footnote{At the aggregate level, the labor share represents the ratio of returns to labor over the total output, such as GDP. In this paper, the labor share is a firm-level measure and denotes the proportion of labor costs to value-added. In the context of the theory part, the labor share primarily emphasizes the operating burden induced by labor expenses.} In response to macroeconomic changes, labor costs cannot be fully adjusted in a flexible manner, and revenue and labor costs do not move in lockstep. Consequently, profits decline when a shock hits the firm. A higher labor share and labor rigidity make profits more sensitive to shocks, which aligns with the concept of operating leverage.

	\textbf{The minimum wage, labor share, and financial leverage.}---The effect of the minimum wage on firms' financial leverage may work through the substitution theory. The legislation is an exogenous, compulsory policy that does not target union strength. The strategic channel proposed by \cite{matsa_capital_2010} does not play a role because firms must comply with the law regardless of their level of debt financing.
	
	Moreover, it is possible that the minimum wage policy is associated with higher labor-induced operating leverage (hereafter referred to as the labor share), leading firms to substitute these operating risks by deleveraging. The minimum wage policy may increase firms' total labor costs and, consequently, the labor share.\footnote{However, it is also possible that firms will adjust employment and use machines to replace labor, thereby offsetting the increased labor costs due to the minimum wage. Theoretically, it is unclear whether the minimum wage increases or decreases the labor share. According to a report by \cite{oecd_decoupling_2018}, in the short run, the minimum wage may elevate the labor share, but in the medium or long run, it may induce capital-labor substitution, thereby reducing the labor share. In Appendix \ref{app:mw_ls_theory}, a simple theoretical framework is provided, showing how an increase in wage affects the labor share. The direction of the effects depends on the elasticity of substitution between labor and capital. Empirically, \cite{petreski_minimum_2023} found that the minimum wage in North Macedonia increases the labor share in labor-intensive sectors but decreases it in capital-intensive sectors.} In addition, the minimum wage strengthens downward wage rigidity in Germany,\footnote{
Although wages are often considered rigid in Germany \citep{franz_reasons_2006}, \cite{jung_paying_2011} find that a wage premium exists for some plants under collective agreements in Western Germany. This wage premium, referred to as a "wage cushion," represents the difference between actual and contractual wages. The size of the wage cushion depends on factors such as labor demand and supply, as well as business cycles. Therefore, plants may adjust wages downward by reducing the wage cushion if labor supply exceeds demand or firms encounter adverse shocks.} preventing firms from adjusting wages below the minimum wage without violating the law. To compensate for the anticipated risks resulting from the rise in the labor share, firms tend to decrease their financial leverage.

	\textbf{Other channels through which the minimum wage affects financial leverage.}---Besides the leverage substitution theory, it is conceivable that the minimum wage impacts firms' financial leverage through alternative channels. Firstly, the rise in labor costs may prompt capital-labor substitution, leading firms to increase investments in assets, particularly fixed or tangible assets. This increase in tangible assets is typically associated with a positive correlation to financial leverage \citep{ozdagli_financial_2012}. This is because these assets can serve as collateral, enabling firms to secure higher levels of debt. Moreover, lenders may demand lower premiums when debt is backed by collateral \citep{antoniou_determinants_2008}, thereby making debt a favorable option for firms.

	Secondly, the minimum wage leads to reduced profitability, which is expected to correlate negatively with debt levels, as suggested by the pecking order theory. This theory posits that firms prioritize the use of retained earnings first, followed by issuing debt, with equity issuance being the last resort \citep{myers_capital_1984}. As the minimum wage increases labor costs, firms with limited retained earnings may have a greater incentive to borrow money to finance these elevated costs, potentially leading to an increase in financial leverage.

	In summary, the theoretical prediction of the minimum wage's effect on financial leverage is ambiguous. While the leverage trade-off theory predicts negative effects, other channels suggest a potential positive relationship between the minimum wage and financial leverage. This paper aims to empirically investigate the direct effect of the minimum wage on financial leverage and to explore the mechanism, focusing on whether the minimum wage increases firms' labor share.\footnote{It should be noted that this paper does not identify the causal mediation effects, namely the indirect effect of the minimum wage on financial leverage via labor share. While the treatment itself (minimum wage policy) is a quasi-natural experiment, it is crucial to note that the mediator (labor share) and the outcome variable (financial leverage) are endogenously correlated, given the presence of unobserved confounders and the reverse causality issue. The complex relationship between operating leverage and financial leverage is modeled in \cite{sarkar_relationship_2020}. A thorough review of causal mediation analysis in economics can be found in \cite{celli_causal_2022}.}

	\section{Institutional background and data\label{sec:back_data}}
	\subsection{Institutional background\label{sec:back}}
	
	\textbf{Timeline of the minimum wage.}---On January 1, 2015, Germany implemented its first statutory minimum wage, setting a minimum gross hourly wage of 8.5 euros. Prior to the implementation of the minimum wage, there may have been anticipation of the law in the preceding years. In late 2013, following the general election, the CDU (Christian Democratic Union of Germany)/CSU (Christian Social Union in Bavaria) and the SPD (Social Democratic Party) formed a grand coalition, with the SPD advocating for the minimum wage. In 2014, the statutory minimum wage was passed by the German Bundestag. Therefore, we may expect an anticipation effect in late 2013 or 2014. Since 2015, the minimum wage has been increased in the subsequent years. For instance, on January 1, 2017, it was adjusted to 8.84 \euro, and on January 1, 2019, it was increased to 9.19 \euro. Recently, it has been increased to 12 \euro, effective from October 1, 2022, and to 12.41 \euro\ from January 1, 2024.

	\textbf{Scope of application.}---Prior to 2015, Germany had only a few sector-specific minimum wage regulations in place. The new nationwide minimum wage is applicable to nearly all employees in the country, with a few exceptions. These exemptions include young workers under the age of 18, apprentices, interns, long-term unemployed individuals in the initial six months of starting work, and volunteers. Several industries, such as the meat industry, hairdressing, agricultural, and forestry industries, are allowed a transitional phase, with hourly wages required to be at least 8.5 \euro\ as of January 1, 2017. However, the number of employees who are exempted or affected by the transitional regulations was less than 1\% of all employees in 2015. Additionally, in April 2014, approximately 11.3\% of jobs were found to be paid below the minimum wage \citep{mindestlohnkommission_erster_2016}. According to the Federal Statistical Office of Germany (Destatis), by April 2015, this figure had decreased to about 2.7\%, and it has continued to decline over time. This trend suggests that the rate of non-compliance is relatively low.

	\subsection{Data\label{sec:data}}
	
	The data used in this analysis are linked data \citep{diegmann_mannheim_2024} combining the Amadeus data from Bureau van Dijk (BvD) with the employee history file (Beschäftigten-Historik, BeH) and the establishment history panel (Betriebs-Historik-Panel, BHP) from the Institute for Employment Research (IAB).\footnote{The Amadeus data is downloaded from \href{https://wrds-www.wharton.upenn.edu/}{WRDS platform}.} The linking procedure is conducted using a record linkage key that matches firms from the Amadeus with establishments from the BHP. The linkage key is provided by the IAB and generated based on the firm’s name and address \citep{diegmann_linking_2024}. This dataset includes three-dimensional information: employee-level records from the BeH, establishment-level characteristics from the BHP \citep{ganzer_establishment_2022}, and firm-level variables from the Amadeus.

	The employee-level data (BeH) are derived from records of the German social security system and contain information on the total workforce of regular workers, as well as on marginally employed workers in Germany. The recorded information includes employees’ gender, age, education, occupation, employment type, yearly working days, and daily wages. The establishment-level data (BHP) are generated based on the BeH and show the attributes of each establishment, such as the location, industry classification, and the number and age structure of employees in the establishment. The Amadeus dataset collects firms’ characteristics from their financial statements and annual reports, covering publicly listed and private firms in Europe, with only German firms retained. The firm-level data include rich variables such as firms’ ownership, address, industry, debt and asset amounts, and earnings before interest and taxes.
	
	This dataset is the first in Germany to merge firm information and administrative employment records on a large scale. Unlike other survey datasets that cover only a subset of workers within a firm, this dataset includes the entire workforce of each establishment. This feature allows for accurate measurement of the impact of the minimum wage policy at the firm level. Compared to regional-level measures of policy intensity, the firm-level impact variable offers two advantages. Firstly, within the same county, firms are affected differently by the minimum wage policy, which can only be captured by considering industry and firm heterogeneity when measuring the treatment variable. Secondly, regional-level treatment variables are typically headquarter-based \citep{gustafson_higher_2023}, making it challenging to capture how multi-establishment firms are affected when their establishments are located in several regions.
	\subsection{Sample restrictions and variables}

	The sample construction process involves several steps. Firstly, firms in the Amadeus with available data from any of the years between 2011 and 2018 are retained.\footnote{The year is determined based on the variable CLOSEDATE. If the closing date falls after June 1, the year component of the closing date is used; otherwise, the preceding year is applied.} For sample cleaning, the following criteria are implemented: firms in industries that are exempted from minimum wage rules are excluded.\footnote{These exempted industries have NACE codes 11, 12, 13, 14, 15, 16, 17, 21, 22, 23, 24, 101, 131, 132, 133, 139, 141, 142, 143, 782, 783, 813, and 960.} Financial firms are also excluded due to their different capital structure. Observations with unconsolidated financial information are retained, meaning financial reports that are not integrated with the company's subsidiaries, as the focus is on individual firms and labor costs are calculated for individual firms accordingly. Observations with financial reports following German generally accepted accounting principles (GAAP) are kept, whereas those following international financial reporting standards (IFRS) are excluded to ensure the consistency of financial information across firms. Firms not observed after 2014 are excluded,\footnote{This restriction is implemented for two reasons: firstly, because they do not contribute to the treatment effect, and secondly, because the IAB imposes a restriction on the maximum number of workers that can be provided in the data product, necessitating the exclusion of these firms.} and observations with missing information and nonsensical values, such as negative asset variables, debts, sales, cash, employment costs, or value-added, are removed.

	Subsequently, the remaining firms are matched with the BHP and the BeH datasets, aggregated at the firm-year level. The final sample under analysis constitutes an unbalanced panel, comprising 184,831 firm-year observations from 27,488 distinct firms.

	\textbf{Financial variables.}---Nominal financial variables, including assets, debts, income, and cash, are adjusted for inflation based on 2015 prices. All BvD financial variables are winsorized at the 1st and 99th percentile values for each year. In the main analysis, following \cite{erel_acquisitions_2015}, the total financial leverage (hereafter referred to as financial leverage) is defined as the sum of long-term debts and short-term liabilities over total assets. In the robustness check, alternative definitions of financial leverage are examined.
	
	\textbf{Labor-related variables.}---The BeH dataset includes information on workers' gross daily wages, as well as the start and end dates of their employment spells. However, the wage information is top-coded at the social security ceiling. Although censored wages do not directly impact the treatment measure, as only low-wage workers are affected by the minimum wage, they can influence the calculation of total labor costs. To address this issue, daily wages are imputed using the Stata package provided by \cite{dauth_preparing_2020}. Furthermore, the employment history data, organized as employment spells, are transformed into a yearly panel. During this transformation, only the main job with the highest wage is retained in the sample.

	Between 2011 and 2014, the BeH dataset includes a variable denoting total working hours, sourced from employers reporting to the German Social Accident Insurance. However, these reported working hours are subject to significant measurement error primarily due to differences in reporting schemes. In this study, I utilize a corrected version of daily working hours, as described by \cite{vom_berge_correction_2023}. Following the correction, the mean working hours in the IAB data closely align with those calculated using the German Structural Earnings Survey (SES), which provides higher-quality data on hours worked.
	The hourly wages are calculated as the daily wages divided by the daily working hours.

	The treatment intensity of the minimum wage on a firm is measured as the $Bite$, representing the share of workers whose gross hourly wages were below the minimum wage before the policy introduction. In order to rule out the potential anticipation effect, the $Bite$ variable is measured based on wages in 2013. This measure remains constant for a firm over time. When calculating the $Bite$, groups exempt from the minimum wage, such as workers under 18, interns, and apprentices, are excluded. To calculate overall labor costs, workers' annual total wages are aggregated at the firm level, including all employees. Subsequently, the firm's total annual labor costs are adjusted for inflation to the 2015 Euro value.

	The labor share is defined as the proportion of labor costs to firms' value-added \citep{donangelo_cross-section_2019, jager_labor_2021, favilukis_elephant_2020}, where value-added is the sum of labor costs and earnings before interest, taxes, depreciation, and amortization (EBITDA). Since the linked data includes two sources of annual labor cost variables, and IAB labor information is likely more accurate than the BvD information, I use the labor-related variables from the IAB and the financial data from the BvD. The formula for labor share is 
	\begin{equation}
		\text{Labor share}=\frac{\text{Labor costs (IAB)}}{\text{Labor costs (IAB)}+ \text{EBITDA(BvD)}}. 	\notag
	\end{equation}
	Moreover, as robustness evidence, I also add the results when using the labor costs and value-added variables directly from the BvD to calculate the alternative measure of labor share.
	\begin{equation}
		\text{Labor share (BvD)}=\frac{\text{Labor costs(BvD)}}{\text{Value-added(BvD)}}. 	\notag
	\end{equation}
	An overview of variables' definitions can be found in Appendix  \ref{app:definition}.
	
		\subsection{Summary statistics}
		\begin{table}[ht!]\centering			
			\caption{Cross-sectional summary statistics, year 2013}
			\label{tab:cross_summary}
			\begin{threeparttable}
				\begin{tabular}{l*{1}{c}}
					\hline\hline
					&        mean          \\
					\hline
					\hline
					\multicolumn{2}{l}{Treatment intensity}\\
					
					- Bite  &0.102\\
					- Minimum wage affected firm (Bite$>$0) & 0.706\\
					%Minimum wage workers (Hourly wage$<$8.5)  &0.083 \\
				
					Firm located in Eastern Germany & 0.191\\
					Single-establishment firm&0.779\\
					Firm size: $<50$ & 0.309 \\
					Firm size: $50-249$ &0.566\\
					Firm size: $>=250$ &0.125\\
					Corporation & 0.807\\
					Partnership & 0.163\\
					Other legal forms&0.030\\
					
					\hline
					Observations        & 27,488                  \\
					\hline\hline
				\end{tabular}

				\begin{tablenotes}		
					\textit{Notes}: Except for the variable $Bite$,  all variables are dummy variables. The standard deviation for $Bite$ is 0.174. \\					
					\textit{Data}: Linked data of BeH, BHP, and Amadeus, 2013.
				\end{tablenotes}
			\end{threeparttable}
			
		\end{table}
		
	\textbf{Treatment intensity.}---Table \ref{tab:cross_summary} lists cross-sectional descriptive statistics of firms in the year 2013. The average treatment intensity across all firms is 10.2\%. To simplify the interpretation of later regressions, the estimated effect sizes are interpreted as the magnitude resulting from a 10-percentage point increase in the bite variable. Additionally, 70.6\% of firms have at least one sub-minimum wage worker before the policy's introduction. At the worker level, 8.6\% of all workers were paid below the minimum wage, slightly lower than the approximately 10-percentage point reported in other papers \citep{mindestlohnkommission_erster_2016, bossler_wage_2023}.\footnote{The slightly lower proportion of sub-minimum wage workers in the sample reflects the underrepresentation of small firms, which tend to hire more minimum wage workers. According to \cite{bossler_devil_2024}, small establishments (with fewer than ten regular workers) have an average minimum wage worker share of 27.3\% in 2014. In contrast, larger establishments (with ten or more regular workers) have a share of 11.7\% minimum wage workers in the same year. The following paragraphs of this section will discuss the issue of underrepresentation in detail.}

		\textbf{Sample representativeness.}---Regarding the composition of firms, approximately 77.9\% of all firms have only one establishment, 14.2\% have two establishments, and the remaining 7.9\% have three or more establishments. Each year, the sample includes approximately 3.8 to 4 million employees. Regarding firm size, the final sample underrepresents small firms because approximately 97\% of all registered enterprises in Germany in 2013 were small enterprises with fewer than 50 employees.\footnote{According to the Structural Business Statistics Database (Eurostat), the number of enterprises in Germany ranged from 2.2 to 2.6 million from 2012 to 2018. The total number of employees in Germany ranged from 37.0 million to 40.6 million, according to the Federal Statistical Office of Germany. Thus, the sample covers approximately 10\% of all employees but only 1\% of all enterprises. This underrepresentation is due to the inclusion of only a small share of micro and small firms.}

		There are two reasons why the number of small firms is very limited in the sample. First, sole proprietorships are not included. This legal form represents firms founded by one person, making up over 65\% of all registered enterprise entities in 2013. However, excluding sole proprietorships is not an issue because they are not relevant to the research question per se; this category mostly includes freelancers, self-employed individuals, and sole traders. Second, small firms have more missing values in the BvD data. According to German business law (§ 267 Abs. 1 HGB and § 326 Abs. 1 HGB), small corporations (Kleine Kapitalgesellschaften) are only required to disclose balance sheet information and notes on the accounts.\footnote{Small corporations are defined as corporations with total assets up to EUR 6,000,000, sales revenue up to EUR 12,000,000 (in the 12 months before the reporting date), and a maximum of 50 employees on average per year.} The absence of small corporations is not a crucial problem either; as pointed out in section \ref{sec:further}, small firms react most to the minimum wage, and this underrepresentation only leads to an underestimation of the actual effects.

		Despite the unbalanced size distribution, my sample includes 68 publicly listed firms as well as private firms.\footnote{There were over 400 listed firms in Germany during the sample period. However, firms with consolidated financial information are excluded, resulting in the removal of the majority of these firms. Consequently, the final dataset contains only 68 listed firms.} Except for financial sectors, sectors B to S are all covered, and the sample is representative of the distribution of firms in the 16 federal states. The Appendix Figures \ref{fig:app_represent_sector} and \ref{fig:app_represent_state} illustrate the distribution of the sample across sector-size cells and state-size cells.

		\begin{figure}[ht!]
			\captionabove{Development of financial leverage, labor share, and labor costs over time}
			\label{fig:development}
			\begin{subfigure}{.5\textwidth}
				\centering
				% include the first image
				\includegraphics[width=.8\linewidth]{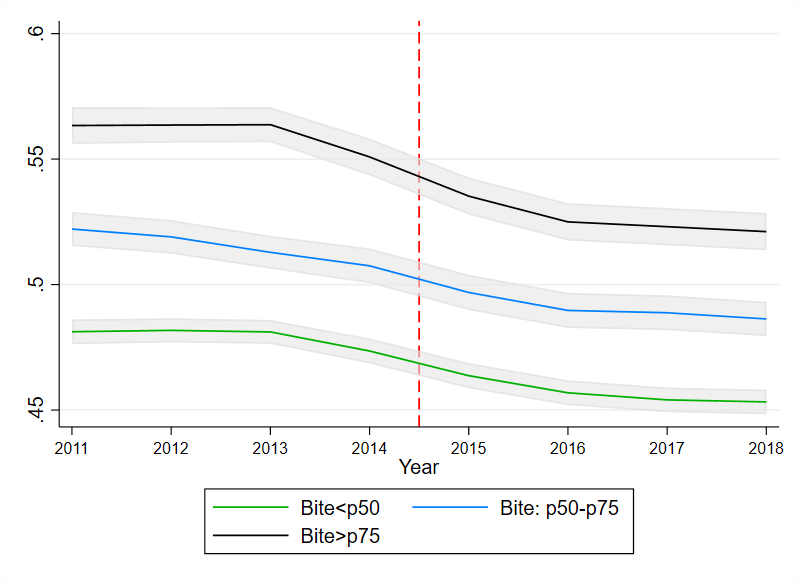}  
				\caption{Financial leverage}
				\label{fig:develop_lvg}
			\end{subfigure}
			\begin{subfigure}{.5\textwidth}
				\centering
				% includes the second image
				\includegraphics[width=.8\linewidth]{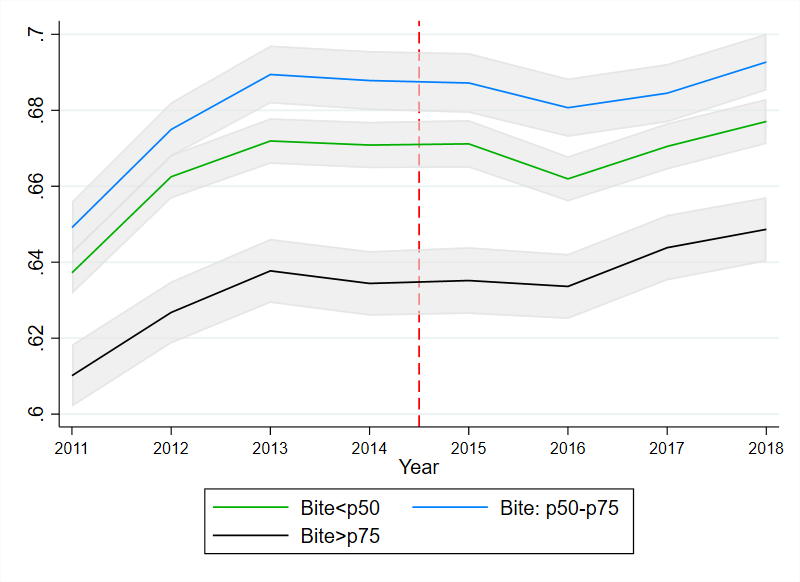}  
				\caption{Labor share}
				\label{fig:develop_ls}
			\end{subfigure}			
			\begin{subfigure}{.5\textwidth}
				\centering
				% include the third image
				\includegraphics[width=.8\linewidth]{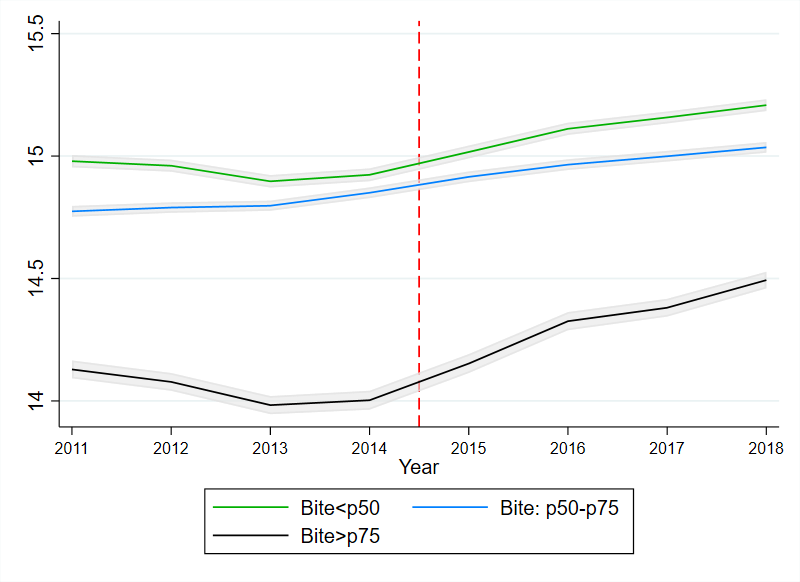}  
				\caption{Log of total labor costs}
				\label{fig:develop_log_tc}
			\end{subfigure}
			\begin{subfigure}{.5\textwidth}
				\centering
				% include the fourth image
				\includegraphics[width=.8\linewidth]{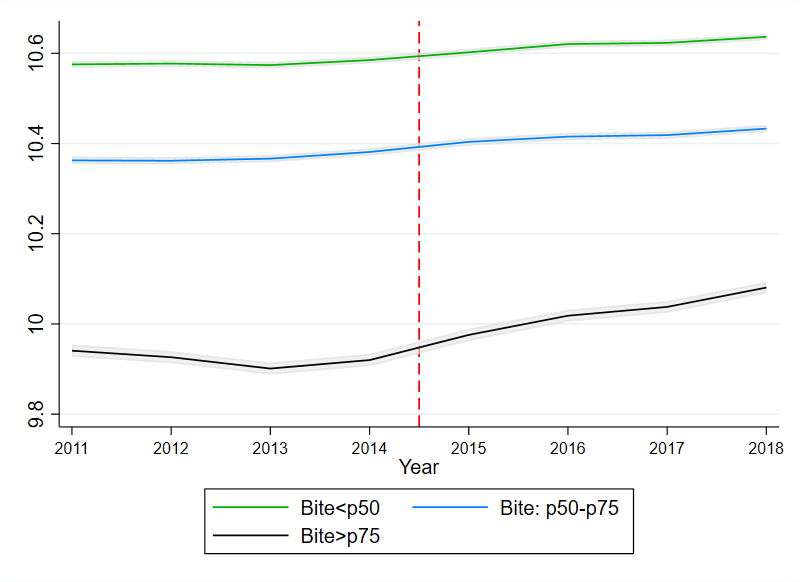}  
				\caption{Log of labor costs per worker}
				\label{fig:develop_tc_cp}
			\end{subfigure}			
			\caption*{\textit{Notes}:
				The figure depicts the trends in the mean of firms' financial leverage, labor share, log of total labor costs, and log of labor costs per worker for three subgroups with varying bite levels. p25, p50, and p75 denote the 25th, 50th, and 75th percentiles of the bite variable in 2013.  The bite values for the three groups are as follows: Bite $<$ p25 (0-0.03), Bite: p50-p75 (0.03-0.10), and Bite $>$ p75 (0.10-1). The gray shadow represents 95\% confidence intervals of the mean. The vertical red line indicates the year 2015, which marks the introduction of the minimum wage.				
				\\	\textit{Data}: Linked data from BeH, BHP, and Amadeus, 2011-2018.}			
		\end{figure}

		\textbf{Development of main variables.}---Figure \ref{fig:develop_lvg} illustrates the mean financial leverage from 2011 to 2018 for firms categorized into low, medium, and high bite levels. The plot highlights that the financial leverage increases with higher bite levels, indicating that firms with lower wages rely more on external financing. Notably, the financial leverage across all three groups decreases over the entire sample period.\footnote{Unlike market-oriented countries such as the US and UK, Germany operates as a bank-oriented economy, characterized by significantly higher leverage ratios for firms compared to their counterparts in market-oriented economies. Since the turn of the millennium, German firms, particularly large ones, have increased their equity capital proportion and decreased their reliance on bank debts \citep{deutsche_bank_trends_2018}. These changes in the German financial system and corporate finance could potentially be attributed to European integration and globalization \citep{schmidt_change_2019}. European financial markets are governed by market-oriented regulations, which in turn impact the German financial system. Additionally, globalization has influenced German firms, requiring them to meet international investors' expectations. Figure \ref{fig:develop_lvg} aligns with the broader trend of deleveraging among German firms.} In particular, in 2015, firms with a bite larger than 10\% experience a significant decline in financial leverage, suggesting the minimum wage may reduce firms' financial leverage.
		
		Figure \ref{fig:develop_ls} depicts the trend in average labor share over time.\footnote{The aggregate labor share in the sample ranges from 62.8\% to 66.1\%. According to the \href{https://www.rug.nl/ggdc/productivity/pwt/}{Penn World Table version 10.01}, the aggregate labor share for Germany from 2011 to 2018 ranges from 61.6\% to 64.1\%, measured as the ratio of labor compensation to GDP. The difference in labor share values between my sample and the Penn World Table may be due to different sample restrictions, as the Penn World Table covers all ranges of industries and employees. However, the trend in Figure \ref{fig:develop_ls} aligns with the trend in the Penn World Table shown in Appendix Figure \ref{fig:app_develop_ls}.} Overall, firms with a middle level of bite exhibit the highest labor share. A slight increase in labor share is observed for all three groups from 2015 to 2018, with a drop in labor share for low-bite firms in 2016. The treatment effect is not immediately apparent from Figure \ref{fig:develop_ls}. However, this descriptive graph only illustrates the raw distribution of treatment levels and does not account for factors such as industry-specific or regional-specific shocks. Turning to labor costs, Figures \ref{fig:develop_log_tc} and \ref{fig:develop_tc_cp} illustrate the mean of the log of annual labor costs and the log of labor costs per worker over the sample period. It is apparent that firms with higher bite levels exhibit lower total labor costs as well as labor costs per worker. From 2011 to 2014, firms with low and medium bite levels demonstrate relatively stable development in labor costs per worker, with high-bite firms even experiencing a decline. Plausibly, following the minimum wage intervention, high-bite firms exhibit a steeper growth in both total labor costs and labor costs per worker compared to the other two groups. These figures provide initial descriptive evidence suggesting that the minimum wage may influence firms' labor costs. Appendix \ref{app:summary} provides summary statistics for all variables, separated by pre-policy and post-policy periods.

		\section{Method\label{sec:method}}

		\subsection{Difference-in-differences estimation}
		This paper employs a difference-in-differences approach \citep{card_using_1992, caliendo_short-run_2018}:
		\begin{equation}\label{eq:did_base}
			y_{jt} = \delta_0 + \sum_{k\neq2013} \delta_{k} * Bite_{j} * Year_{k,t} + \sum_{k\neq2013} \gamma_{k} * Year_{k,t} + \phi * Bite_{j} + \alpha_{j}+ \theta_{c,t} + \lambda_{s,t} + \epsilon_{jt},
		\end{equation}
		
		where $y_{jt}$ represents the firm-level outcome variables for firm $j$ at year $t$, such as a firm's financial leverage. The $Bite_{j}$ is defined as the proportion of minimum wage workers in firm $j$ in 2013. This year was chosen to rule out potential anticipation effects, as the policy was already expected by the end of that year \citep{caliendo_short-run_2018}.
		The coefficients of interest are $\boldsymbol{\delta}_{k}$, where $\delta_{2011}$ and $\delta_{2012}$ indicate the placebo effects and whether the parallel trend assumption is satisfied. $\delta_{2014}$ displays the anticipation effect, and $\delta_{2015}$ to $\delta_{2018}$ represent the treatment effects in subsequent years.
		$\alpha_j$ denotes the firm-fixed effects, which control for firm-specific constant characteristics that would affect the leverage ratio, such as firm culture. The estimated effect is then identified from within-firm variations. To rule out the influence of the sector-year-specific change in financial leverage, I also control for fixed effects of two-digit industry-year dummies ($\lambda_{s,t}$). It is also possible that the local economic situations may affect firms' behaviors; therefore, county-year fixed effects $\theta_{c,t}$ are added.
		
		The outcome variables may develop differently for firms with different wage levels already before the minimum wage policy. After inspecting the bite-specific trend (see section \ref{sec: detect_trend}), a predetermined trend is subtracted from the outcome variables. This detrended method has been adopted in several minimum wage studies \citep{meer_effects_2016, monras_minimum_2019, bossler_wage_2023, dustmann_reallocation_2021}. It is achieved by using the data from the years 2011 to 2013 and running the regression
		\begin{equation}\label{eq:did_first_stage}
			y_{jt} = \beta_0 + \beta_{1} * Bite_{j} *T + \beta_{2} * Bite_{j} + \alpha_{j} + \theta_{c,t} + \lambda_{s,t} + u_{jt},
		\end{equation}
		where the estimated predetermined treatment-specific trend is $\hat{\beta_{1}} * Bite_{j} * T$, with $T=t-2010$. The estimated trend is subtracted from the $y_{jt}$ in Equation \ref{eq:did_base}. To remedy the serial correlation of observations from the same firm, standard errors are clustered at the firm level.

		\subsection{Detection of the pre-intervention trend} \label{sec: detect_trend}
		\begin{figure}[ht!]
			\captionabove{Coefficients of $Bite_{j} * Year_{k,t}$ in non-detrended DiD regressions}
			\label{fig:detect_trend}
			\begin{subfigure}{.5\textwidth}
				\centering
				% include the first image
				\includegraphics[width=.8\linewidth]{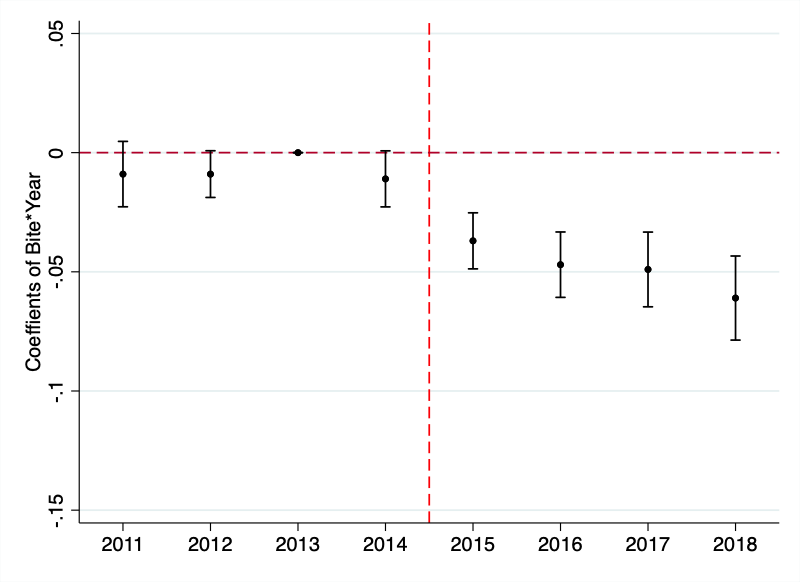}  
				\caption{Financial leverage}
				\label{fig:detect_lvg}
			\end{subfigure}
			\begin{subfigure}{.5\textwidth}
				\centering
				% include the second image
				\includegraphics[width=.8\linewidth]{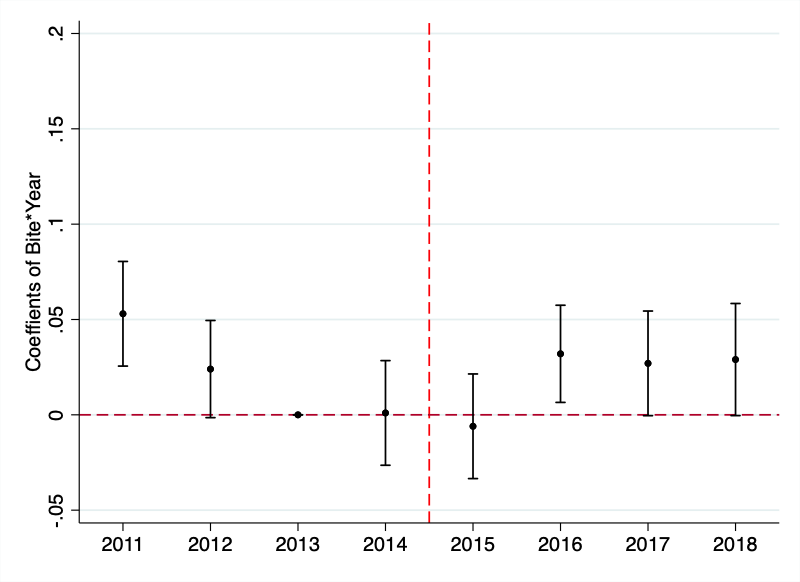}  
				\caption{Labor share}
				\label{fig:detect_ls}
			\end{subfigure}

			%			\begin{subfigure}{.5\textwidth}
				%				\centering
				%				% include third image
				%				\includegraphics[width=.8\linewidth]{fig/3_0_log_tc.png}  
				%				\caption{Log of total labor costs}
				%				\label{fig:detect_log_tc}
				%			\end{subfigure}
			%			\begin{subfigure}{.5\textwidth}
				%				\centering
				%				% include fourth image
				%				\includegraphics[width=.8\linewidth]{fig/3_0_lc_cp.png}  
				%				\caption{Log of labor costs per worker}
				%				\label{fig:detect_lc_cp}
				%			\end{subfigure}
			
			\caption*{\textit{Notes}: The figure displays the non-detrended difference-in-differences (DiD) regression coefficients of  $Bite_{j} * Year_{k,t}$ with 95\% confidence intervals. The dependent variables include (a) Financial leverage and (b) Labor share, with the year 2013 serving as the reference year. The regression results corresponding to this figure are presented in Appendix Table \ref{tab:full_non_detrend}.
						\\\textit{Data}: Linked data from BeH, BHP, and Amadeus, 2011-2018.}
			
		\end{figure}
		The difference-in-differences approach relies on the parallel trend assumption. This assumption implies that, without a minimum wage intervention, the financial leverage of firms with different treatment intensities would have developed in a parallel manner over the entire period. When controlling for a bite-specific trend in Equation \ref{eq:did_base}, the identification method is now based on the assumption that the predetermined bite-specific trend would have persisted had there been no minimum wage policy. Thus, the estimates of  $\delta_{2015}$ to $\delta_{2018}$, after subtracting the trend, represent the treatment effects.

		To demonstrate the importance of employing detrended regressions, I analyze the results from simple DiD regressions without controlling for trends. Figure \ref{fig:detect_trend} displays the coefficients and confidence intervals of $Bite_{j} * Year_{k,t}$ from the Equation \ref{eq:did_base}. The graph in Figure \ref{fig:detect_lvg} indicates a slightly increasing bite-specific trend in firms' financial leverage prior to the introduction of the minimum wage. This suggests that, even before the policy introduction, firms with a higher bite experienced a greater increase in financial leverage compared to those with a lower bite. Similarly, a decreasing trend is observed for the labor share, with the labor share of high-bite firms decreasing from 2011 to 2013. From both graphs, it is evident that the pre-policy trends reverse or halt after 2013, suggesting that in addition to the trend, the minimum wage has effects on these variables. The graphical presentation of non-detrended regressions emphasizes the importance of including a predetermined bite-specific trend in the DiD regression. Therefore, in the following analyses, this trend is subtracted from the outcome variable in all regressions.
		
		\section{Results \label{sec: main_result}}
  
   This section first examines the effect of the minimum wage on firms' financial leverage. Second, Section \ref{sec:main_labor_share} explores the mechanism of deleveraging by estimating the effect of the minimum wage on firms' labor share.
		
		\subsection{Capital structure \label{sec: main_result_fin}}

		%\textcolor{red}{compare the size to other studies}
		%is in line with the idea that 

		\begin{table}[ht!]\centering
			\caption{Minimum wage effect on financial leverage}
			\label{tab:baseline}
			\begin{threeparttable}
				\begin{tabular}{l*{5}{c}}
					\hline\hline
					&\multicolumn{5}{c}{Specifications}\\
					\cmidrule(lr){2-6}
					&\multicolumn{1}{c}{(1)}&\multicolumn{1}{c}{(2)}&\multicolumn{1}{c}{(3)}&\multicolumn{1}{c}{(4)}&\multicolumn{1}{c}{(5)}\\
					\hline
					$Bite*Year_{2011}$ &       0.001        &      -0.002         &       0.000         &       0.001         &       0.000         \\
					&     (0.006)         &     (0.006)         &     (0.006)         &     (0.006)         &     (0.006)         \\

					$Bite*Year_{2012}$ &      -0.004         &      -0.005         &      -0.005         &      -0.003         &      -0.004         \\
					&     (0.005)         &     (0.005)         &     (0.005)         &     (0.005)         &     (0.005)         \\[1ex] 
					
					$Bite*Year_{2013}$ &         \multicolumn{4}{c}{Reference} \\[1ex] 
					
					$Bite*Year_{2014}$  &      -0.015\sym{**} &      -0.017\sym{***}&      -0.016\sym{**} &      -0.015\sym{**} &      -0.016\sym{**} \\
					&     (0.005)         &     (0.005)         &     (0.005)         &     (0.005)         &     (0.005 )         \\
					
					$Bite*Year_{2015}$  &      -0.046\sym{***}&      -0.051\sym{***}&      -0.042\sym{***}&      -0.046\sym{***}&      -0.043\sym{***}\\
					&     (0.006)         &     (0.006)         &     (0.006)         &     (0.006)         &     (0.006)         \\

					$Bite*Year_{2016}$ &            -0.061\sym{***}&      -0.072\sym{***}&      -0.063\sym{***}&      -0.059\sym{***}&      -0.063\sym{***}\\
					&     (0.007)         &     (0.007)         &     (0.007)         &     (0.007)         &     (0.007)         \\
					
					$Bite*Year_{2017}$ &      -0.070\sym{***}&      -0.086\sym{***}&      -0.072\sym{***}&      -0.067\sym{***}&      -0.072\sym{***}\\
					&     (0.007)         &     (0.007)         &     (0.007)         &     (0.007)         &     (0.007)         \\
					
					$Bite*Year_{2018}$ &      -0.088\sym{***}&      -0.106\sym{***}&      -0.090\sym{***}&      -0.086\sym{***}&      -0.090\sym{***}\\
					&     (0.008)         &     (0.008)         &     (0.008)         &     (0.008)         &     (0.008)         \\

					$Constant$          &         0.496\sym{***}&      -0.401\sym{***}&       0.496\sym{***}&       0.496\sym{***}&       0.504\sym{***}\\
					&     (0.000)         &     (0.048)         &     (0.000)         &     (0.000)         &     (0.001)         \\

					\hline
					Firm FE             &         Yes         &         Yes         &         Yes         &         Yes         &         Yes         \\
					Firm controls       &          No         &         Yes         &          No         &          No         &          No         \\
					County-year FE      &         Yes         &         Yes         &         Yes         &          No         &          No         \\
					Industry-year FE    &         Yes         &         Yes         &          No         &         Yes         &          No         \\
					Observations        &     184,702         &     184,702         &     184,703        &     184,702         &     184,703        \\
					
					\hline\hline
				\end{tabular}
				
				\begin{tablenotes}
					\item \textit{Notes:} Difference-in-differences regressions. The dependent variable in all five columns is financial leverage (total debts/total assets). A predetermined bite-specific trend is subtracted in all regressions. Firm controls in the second column include the logarithm of total assets, the ratio of tangible assets to total assets (tangibility), the cash assets ratio (cash ratio), and ROA. Firms are assigned to the county where their largest establishment is located. Industries are categorized with a two-digit industry code. The number of observations deviates from the summary statistics since singletons are dropped. Firm-level clustered standard errors are in parentheses.  \sym{*}, \sym{**}, and \sym{***} denote statistical significance at 5\%, 1\% and 0.1\%, respectively.
					\\\textit{Data}: Linked data of BeH, BHP, and Amadeus, 2011-2018.
				\end{tablenotes}
			\end{threeparttable}
		\end{table}

		Table \ref{tab:baseline} presents the effects of the minimum wage on firms' financial leverage. The first column represents the preferred specification (same as Equation \ref{eq:did_base} with detrended $y_{jt}$). The second column additionally controls for a set of covariates, including ROA, cash ratio, tangibility, and total assets, commonly seen in the traditional corporate finance literature. However, since the minimum wage may also impact these covariates, controlling for post-treatment covariates may bias the treatment effects of interest. Hence, they are excluded from the main specification. Furthermore, column 3 controls for firm fixed effects and county-year fixed effects. Column 4 controls for firm fixed effects and industry-year fixed effects. Column 5 controls only for firm fixed effects. Despite the similar effect sizes across all specifications, column 1 is preferred because it identifies the treatment effect while ruling out the influence from firm constant components and industry or regional-specific shocks in each year.
		
		The main result shows that the reduction in financial leverage has been evident since 2015, attributable to the impact of the minimum wage. The anticipation effects of the minimum wage are found to be smaller in magnitude compared to the treatment effects observed in other years. With regards to a 10-percentage point increase in the bite variable, firms' financial leverage decreases by 0.5 to 0.9 percentage points,\footnote{The effect is calculated as 0.1*0.046 and 0.1*0.088.} which corresponds to 1 to 2 percent of the average financial leverage. The empirical results confirm that firms reduce their external financing rate in response to the minimum wage. Additionally, firm financial leverage displays a continuous decline from 2015 to 2018. This observation may be attributed to the ongoing increase in the labor share or the fact that high-bite firms did not complete the deleveraging process in the short term, but rather over a medium to long-term period.
				
		\subsection{Mechanism: labor share \label{sec:main_labor_share}}

		\begin{figure}[ht!]
			\captionabove{Minimum wage effects on labor-related outcomes }
			\label{fig:mech_ls}
			\begin{subfigure}{.5\textwidth}
				\centering
				% include the first image
				\includegraphics[width=.8\linewidth]{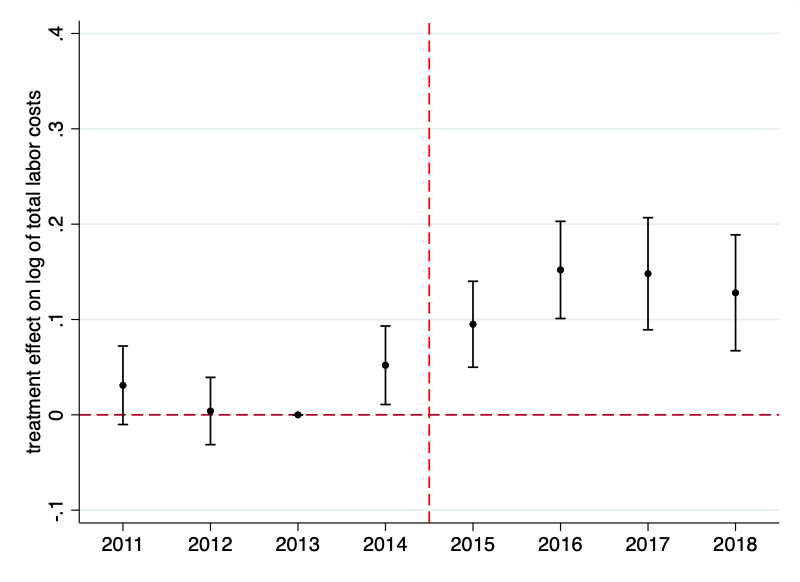}  
				\caption{Log of total labor costs}
				\label{fig:mech_log_tc}
			\end{subfigure}
			\begin{subfigure}{.5\textwidth}
				\centering
				% includes the second image
				\includegraphics[width=.8\linewidth]{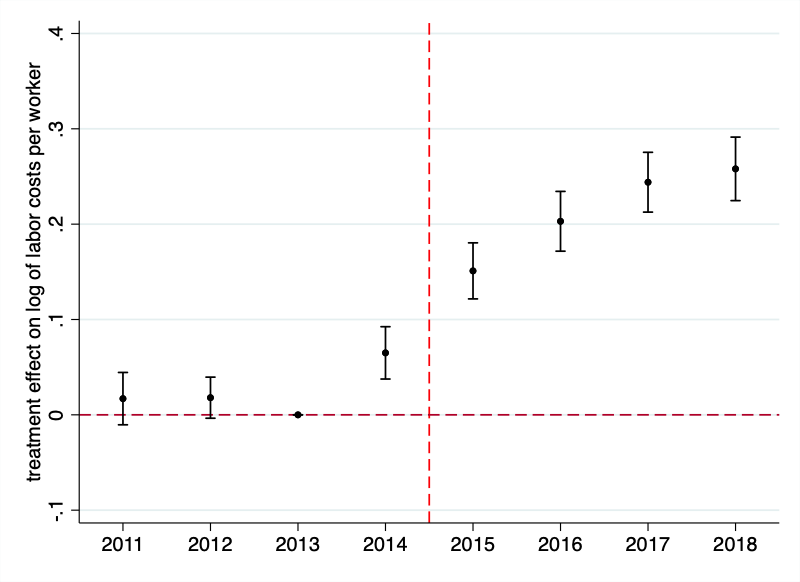}  
				\caption{Log of labor costs per worker}
				\label{fig:mech_lc_cp}
			\end{subfigure}
						\begin{subfigure}{.5\textwidth}
				\centering
				
				% include the third image
				\includegraphics[width=.8\linewidth]{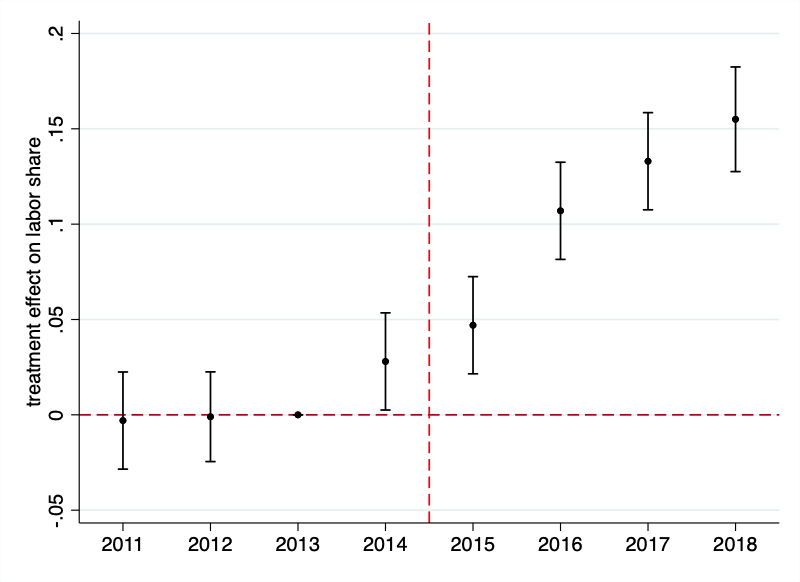}  
				\caption{Labor share}
				\label{fig:mech_labor_lvg_a}
			\end{subfigure}
			\begin{subfigure}{.5\textwidth}
				\centering
				% include the fourth image
				\includegraphics[width=.8\linewidth]{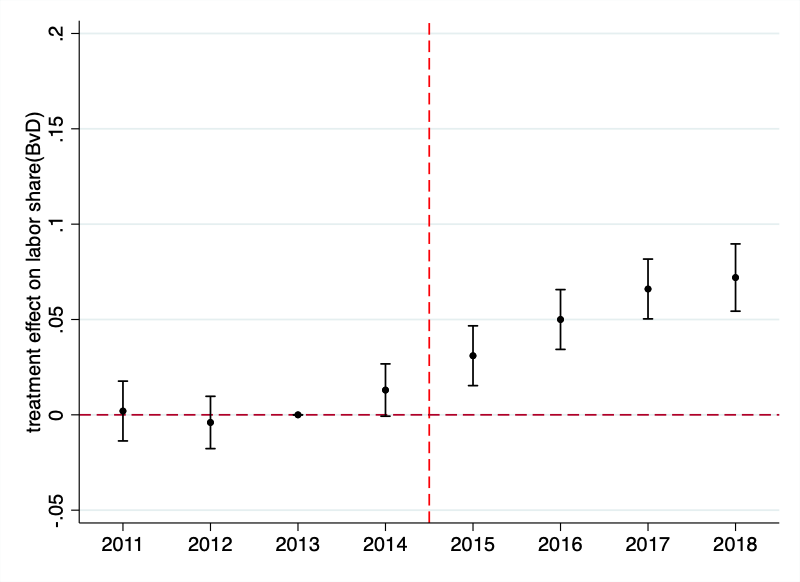}  
				\caption{Labor share (BvD)}
				\label{fig:mech_labor_lvg_b}
			\end{subfigure}
			
			\caption*{\textit{Notes}: The figure displays the detrended difference-in-differences (DiD) regression coefficients of $Bite_{j} * Year_{k,t}$ with 95\% confidence intervals. The dependent variables include:
				(a) Log of total labor costs,
				(b) Log of labor costs per worker,
				(c) Labor share calculated as total labor costs divided by the sum of total labor costs and EBITDA,
				(d) Labor share (BvD) calculated as total labor costs divided by value-added.
				The year 2013 serves as the reference year. The regression results corresponding to this figure are presented in Appendix Table \ref{tab:full_mech_ls}.
				\\\textit{Data}: Linked data from BeH, BHP, and Amadeus, 2011-2018.}
		\end{figure}
				In this section, the mechanism of deleveraging is studied by focusing on the labor share as the outcome variable of the minimum wage effect. Other possible channels will be investigated in section \ref{sec:further}. 
		
	 Figure \ref{fig:mech_log_tc} presents the treatment effects on the log of total labor costs. The effect size can be quantified as follows: for a 10-percentage point increase in the bite, the total labor costs increase by approximately 1\% to 1.5\%. Similarly, Figure \ref{fig:mech_lc_cp} displays that labor costs per employee increase by about 1.5 to 2.6\%.\footnote{Comparing Figure \ref{fig:mech_log_tc} with Figure \ref{fig:mech_lc_cp}, the higher treatment effects on the log of labor costs per worker suggest that the affected firms may reduce employment. The employment effect is examined and discussed in Section \ref{sec:bd_labor_share}.}

		Regarding the labor share measured by IAB data, it increases by 0.5 to 1.6 percentage points for a 10-percentage point increase in the bite, while there are smaller increases in the labor share (BvD), potentially due to the inclusion of wages of workers in foreign countries in the BvD data. An anticipation effect is observed for all four variables in the year 2014.\footnote{Previous studies have yielded inconclusive results regarding the anticipation effects of the German minimum wage on wages. \cite{caliendo_short-run_2018} find no anticipation effects on hourly wages, whereas \cite{bossler_wage_2023} identify small anticipation effects on monthly wages.} Furthermore, the effect sizes on labor share become larger from 2016 to 2018, suggesting that the minimum wage continuously impacts firms' labor share rather than having only a short-term effect. This finding supports the theory that the minimum wage increases firms' labor share, leading to a decrease in their financial leverage. It also aligns with the results in Section \ref{sec: main_result_fin}, which show a continuous reduction in firms' leverage over time.
		\subsection{Robustness checks\label{sec:robust}}
		%\textcolor{red}{Here should add the results when wages are imputed.}
	I conduct several robustness checks to validate the main findings on the effects of the minimum wage on financial leverage and labor share. These checks include using alternative measures of treatment intensity and financial leverage, applying different sample restrictions, and analyzing the impact of other concurrent policies.
		\subsubsection{Alternative treatment measures}
		 One of the treatment measures is a working hours weighted gap variable, calculated as 
		
		\begin{equation}\label{eq:gap}
			gap_{j,2013}=\frac{\sum_{i\in j}^{}h_{i,2013}Max\{0,8.5-wage_{i,2013}\}}{\sum_{i\in j}h_{i,2013}wage_{i,2013}} , 	
		\end{equation} 
		where $h_{i,2013}$ denotes worker $i$'s weekly working hours in the year 2013, $wage_{i,2013}$ is the hourly wage.
		Moreover, the bite variable averaged over the years 2011 to 2014, as well as the averaged gap variable \citep{dustmann_reallocation_2021}, are also used to check the robustness of the main results. The gap measure captures not only the share of affected individuals but also the wage increase necessary for firms to comply with the minimum wage. The estimated effects are shown in Appendix \ref{app: robust} and are in line with the main findings.

		\subsubsection{Alternative measures for financial leverage}

		In addition to defining financial leverage as total debts divided by total assets in the main results, the effect of the minimum wage on three alternative measures of financial leverage is also tested. It is arguable what kind of debts should be included when calculating the leverage ratio. First, a broader concept of liabilities is used. The financial leverage is then defined as
		\begin{equation}
			\text{Broader financial leverage}=	\frac{\text{Total liabilities}}{\text{Total assets}}. 	\notag
		\end{equation}   
		Total liabilities are the sum of long-term liabilities and short-term liabilities, where, in addition to the long-term debts, the provisions are included as part of the long-term liabilities. Even though provisions are counted as liabilities, they pose less risk than debts to firms and are firms' internal estimates. Therefore, they are not included in the main analysis. However, it is still worth examining whether financial leverage decreases after the minimum wage introduction when considering a broader definition of liabilities. Second, I define the liabilities in a narrower sense, namely only including long-term debts; correspondingly, the financial leverage is 
		\begin{equation}
			\text{Long-term leverage}=	\frac{\text{Long-term\ debts}}{\text{Total assets}}. 	\notag
		\end{equation} 
		Third, I also use the net leverage ratio as a dependent variable, which measures the leverage ratio net of firms' cash holding. This measure thus also accounts for the change in cash reserves. It is defined as:
		\begin{equation}
			\text{Net leverage}=	\frac{(\text{Total debts-cash})}{\text{Total assets}}. 	\notag
		\end{equation}

		Appendix Table \ref{tab: robust_other_fin_lvg} reports the treatment effects on other measures of financial leverage. Concerning the broader financial leverage, the treatment effects are almost the same as in the main results, showing that including the provision does not alter the main conclusion. The effects on long-term leverage are smaller in size but still significant. The decrease in net leverage also suggests that firms continue to deleverage when we take into account the level of cash holdings.
		\subsubsection{Alternative sample restrictions}
		\textbf{Restriction on unsuccessful matching.}---The linkage between IAB employment data and BvD financial data provides an opportunity to analyze all workers within a firm. However, unsuccessful matches could occur, such as when not all establishments of a firm are successfully matched \citep{antoni_orbis-adiab_2018}. If the final sample includes only a small proportion of employees from a firm, the bite variable would be imprecisely measured.
		
		The BvD data does not provide information about the number of establishments a firm has, making it difficult to validate the matching quality. An alternative approach is to compare the employee count information in the two datasets. The IAB employment variable measures the number of employees registered in the German social security system, whereas the BvD variable includes employees both domestically and abroad \citep{jager_labor_2021}.\footnote{However, when comparing the two variables, there are observations where employment in IAB exceeds employment in BvD. Due to the accuracy of the IAB data and the definition of the two employment variables, this suggests significant measurement errors in the employment information provided by the BvD. Additionally, only workers in Germany are relevant to the minimum wage policy. Therefore, employment data from IAB is used for the main analysis.} Since firms that do not adhere to German local accounting standards are excluded, the sample mostly comprises firms operating primarily in Germany. If the IAB employment variable is substantially lower than it is in the BvD dataset, it may indicate a significant loss of workers in the matched sample. To address this concern, a robustness check is conducted by excluding firms where the share of IAB employment is less than 30 percent of the BvD employment. The results are presented in Appendix Table \ref{tab:app_link_rate}, showing findings closely aligned with the main results.

		\textbf{Non-imputed wages.}---Furthermore, the wage imputation procedure may alter the minimum wage effects on labor-related outcomes. Because it directly changes the top-censored daily wages and consequently affects the value of total labor costs, labor costs per worker, and labor share. I also conducted robustness checks using non-imputed wages. The results are similar to the main results and are presented in Appendix Table \ref{tab:app_non_impute}.

		\textbf{Firms' exit.}---According to the leverage substitution theory, higher labor share and higher financial leverage both increase the likelihood of firms defaulting. Consequently, we may expect a higher exit rate among firms with higher financial leverage. Additionally, if an increase in the minimum wage raises the labor share, we may also expect a higher probability of exiting for firms with a higher bite. Furthermore, a selection bias may occur if firms are selected based on financial leverage and if those that exit the market respond differently to the minimum wage compared to those that remain in the market.
  
Therefore, I conduct cross-sectional regressions in Appendix Table \ref{tab:app_exit} to investigate whether the bite variable and financial leverage are positively related to firms' exit probability and to examine whether firms' exit biases the estimated deleveraging results.
 The dependent variable indicates whether a firm exited the market between 2016 and 2020. This variable is based on the year of the firm's last appearance in BHP data.\footnote{Since the sample consists of firms that operated at least until 2015, it is impossible to conduct a placebo test to assess whether the minimum wage has an impact on firms' exit probability before 2015. The information on a firm's last appearance is derived from the variable $lzt\_jahr$. I use the year of the last remaining establishment's appearance as the firm's exit year. The reason for exit could be either the closure of all the firm's establishments or a change in its legal form. However, the dataset does not provide information on the specific reasons for the disappearance of these establishments. Therefore, the analysis is to provide suggestive evidence of the correlation between financial leverage, minimum wage, and firms' exit probability.} Control variables are measured as average values during the pre-policy period.

  Coefficients of the $Bite$ and $Avg.leverage$ exhibit that firms with higher bite or higher financial leverage are more likely to exit the market, which aligns with theoretical predictions. Specifically, a 10-percentage point increase in bite corresponds to a 0.46-percentage point increase in the probability of exiting. 
Moreover, the estimated extent of deleveraging would be biased if firms that exit the market do not reduce their financial leverage or even increase it, suggesting that minimum wage worsens their financial condition. The risk of default is particularly severe for firms already in financial distress. Thus, the larger the bite, the more financially constrained firms should be more likely to default, which would be reflected by a positive coefficient for the interaction term $Avg.\ fin. \ leverage * Bite$. However, the results show that firms with higher pre-policy financial leverage are less likely to exit the market as the bite increases. This suggests that firms may deleverage in response to the minimum wage and manage their overall risks effectively. Thus, firm exits do not bias the main findings of this study.

		\subsubsection{Concurrent policies}
		
		One potential threat to identification is the presence of other policies that affect financial leverage and may also be correlated with the firms' bite variable. In September 2013, the election led to a new coalition government in Germany, resulting in a series of policy changes beyond the stationary minimum wage policy. However, the most significant reforms, such as the Energy Transition (Energiewende) promoting the shift to renewable energies, are unrelated to the treatment measure of the minimum wage. Moreover, controlling for industry-year fixed effects rules out potential effects from industry-specific policies. Additionally, changes in taxation rates can influence firms' financial leverage. Yet, between 2013 and 2015, there were no major reforms concerning the corporation tax rate, particularly no specific policies targeting low or high-wage firms. Moreover, municipalities in Germany have the authority to set their own local business tax rates, but the adjustments could occur in any year and are not specific to the year 2015 or to firms based on their wage structure. Therefore, the observed effects are unlikely to stem from spurious correlations arising from concurrent policies.
		\section{Further analysis on mechanism \label{sec:further}}

		In Section \ref{sec: main_result}, the analysis shows that the minimum wage reduces financial leverage and discusses the mechanism through an increased labor share. Additional evidence is provided in this section to further explore this mechanism and illustrate the leverage substitution theory and other potential channels between minimum wage and financial leverage.
		Firstly, the section separately dissects the factors contributing to changes in financial leverage and labor share. 	
		Secondly, heterogeneous effects are examined. The analysis tests mechanisms from the perspectives of the time length of debts, the firm's ability to adjust its labor, and firms of different sizes.
		
		The identification approach is simplified in this section. The treatment period is no longer divided into separate years but combined into a post-period (2015-2018). The pre-period (2011-2013) serves as a reference, and a $Bite*Year_{2014}$ term is still included to account for potential anticipation effects. The other control variables remain unchanged, and a bite-specific trend is subtracted from the outcome variables, as detailed in Equation \ref{eq:did_first_stage}. Results from the approach with interactions of the bite and individual years can be found in Appendix \ref{app:event_graphs}.
		
		\subsection{Examinations on financial leverage changes \label{sec:bd_fin_lvg}}
		
		\begin{table}[ht!]\centering
			\caption{Minimum wage effects on log total debts and log assets}
			\label{tab:debts_assets}
			\begin{threeparttable}
				\begin{tabular}{l*{5}{c}}
					
					\hline\hline
					&\multicolumn{1}{p{2.5cm}}{\centering Log total\\debts}&\multicolumn{1}{p{2.5cm}}{\centering Log total\\assets}&\multicolumn{1}{p{2.5cm}}{\centering Log fixed\\assets}&\multicolumn{1}{p{2.5cm}}{\centering Log current\\assets}&\multicolumn{1}{p{2.5cm}}{\centering Log \\cash}\\
                   & (1)&(2)&(3)&(4)&(5)\\
					
					\hline
					$Bite*Post$   &       -0.090\sym{**} &       0.067\sym{***}&       0.003         &       0.071\sym{***}&       0.284\sym{***}\\
						&     (0.024)         &     (0.015)         &     (0.028)         &     (0.018)         &     (0.054 )         \\
	
					%	$Bite*Year_{2014}$      &      -0.037         &      -0.004         &       0.009         &      -0.011         &       0.196\sym{***}\\
					%	&     (0.023)         &     (0.013)         &     (0.023)         &     (0.017)         &     (0.055 )         \\	[1ex] 
					%	$Bite*Pre$&         \multicolumn{5}{c}{Reference} \\[1ex] 
					%	$Constant$           &      15.453\sym{***}&      16.389\sym{***}&      14.793\sym{***}&      15.791\sym{***}&      13.174\sym{***}\\
					%	&     (0.001)         &     (0.001)         &     (0.001)         &     (0.001)         &     (0.003)         \\
					\hline
					Observations        &      184,702         &      184,702         &      184,702         &      184,702         &      184,702         \\
					
					\hline\hline
				\end{tabular}
				
				\begin{tablenotes}
					\item \textit{Notes:} Difference-in-differences regressions. The dependent variables are displayed above each column. A predetermined bite-specific trend is subtracted in all regressions. Firm fixed effects, county-year, and industry-year fixed effects are controlled. Firms are assigned to the county where their largest establishment is located. Industries are categorized with a two-digit industry code. Firm-level clustered standard errors are in parentheses. \sym{*}, \sym{**}, and \sym{***} denote statistical significance at 5\%, 1\% and 0.1\%, respectively. The full regression results are presented in Appendix Table \ref{tab:full_debts_assets}.
					\\\textit{Data}: Linked data of BeH, BHP, and Amadeus, 2011-2018.
				\end{tablenotes}
			\end{threeparttable}
			
		\end{table}
		According to the definition of financial leverage, changes in the leverage ratio stem from alterations in both debts and total assets. Therefore, I investigate the impact of the minimum wage on total debts and total assets separately. Columns 1 and 2 of Table \ref{tab:debts_assets} outline the regression results, demonstrating a decrease in debt borrowing and an increase in total assets. 
		
		The increase in total assets may be due to an increase in fixed or current assets. However, there is no observed impact of the minimum wage on fixed assets, as displayed in column 3 of Table \ref{tab:debts_assets}. The surge in total assets primarily stems from the expansion of current assets, which includes, for instance, cash in hand, bank balances, trade receivables, and other liquid assets. Notably, a substantial increase in cash is found, as depicted in column 5. Thus, the reduction in debts is partly credited to the decline in total debts and also to the expanding current assets, especially cash reserves.
		\subsection{Examinations on labor share changes \label{sec:bd_labor_share}}
		\begin{table}[ht!]\centering
			
			\caption{Minimum wage effects on labor-related outcomes and log EBITDA}
			\label{tab:bd_labor_share}
			\begin{threeparttable}
				\begin{tabular}{l*{6}{c}}
					\hline\hline
					&\multicolumn{1}{p{2cm}}{\centering Log \\employment} &\multicolumn{1}{p{2.2cm}}{\centering Log (fixed\\assets/empl.)} & \multicolumn{1}{p{2cm}}{\centering Log (labor\\costs/empl.)}
					&	\multicolumn{1}{p{2cm}}{\centering Log value\\added} &\multicolumn{1}{p{2cm}}{\centering Log\\EBITDA} &\multicolumn{1}{p{2cm}}{\centering Log total\\labor costs}\\
                        & (1)&(2)&(3)&(4)&(5)&(6)\\

					\hline
					
						$Bite*Post$    &    -0.081\sym{***}&       0.063\sym{*} & 0.201\sym{***}  &       0.070\sym{***}&      -0.180\sym{***}&       0.120\sym{***}\\
						&     (0.018)         &     (0.023) & (0.013)			&     (0.017)         &     (0.029)         &     (0.023)         \\
					%	$Bite*Year_{2014}$       &    -0.014         &      0.013         &       0.057\sym{***}  & -0.006  &-0.036  & 0.043\sym{*} \\
				%		&  (0.014)          &     (0.022)         &   (0.013)      &(0.016) &(0.029)& (0.019)     \\[1ex] 
				%		$Bite*Pre$&       \multicolumn{6}{c}{Reference} \\[1ex] 
				%		$Constant$            &     4.391\sym{***}&    10.387\sym{***}&      10.381\sym{***} &15.346\sym{***} &14.187\sym{***}&14.77\sym{***}   \\
				%		&     (0.001)         &     (0.001)         &     (0.001)    & (0.001)  & (0.002)  & (0.001)      \\
					\hline
						Observations        &          184,702   &          184,702     &      184,702    &      189,045         &      169,645         &      184,702         \\
						
						\hline\hline
					
				\end{tabular}
				\begin{tablenotes}
					\item \textit{Notes:} Difference-in-differences regressions. The dependent variables are displayed above each column. A predetermined bite-specific trend is subtracted in all regressions. Firm fixed effects, county-year, and industry-year fixed effects are controlled. Firms are assigned to the county where their largest establishment is located. Industries are categorized with a two-digit industry code. Firm-level clustered standard errors are in parentheses. \sym{*}, \sym{**}, and \sym{***} denote statistical significance at 5\%, 1\% and 0.1\%, respectively. The full regression results are presented in Appendix Table \ref{tab:full_value_added}.
					\\\textit{Data}: Linked data of BeH, BHP, and Amadeus, 2011-2018.
				\end{tablenotes}
			\end{threeparttable}
			
		\end{table}
		
		\textbf{Capital-labor substitution.}---To thoroughly examine the change in labor share, I integrate the analysis with the theoretical model described in Appendix \ref{app:mw_ls_theory}. It mentions that, under the scenario where the marginal product equals wages, the minimum wage increases labor share if the elasticity of substitution between labor and capital ($\sigma$) is smaller than one, indicating that they are complements. In such cases, firms do not substitute labor with capital. Second, if there is bargaining over employment, namely, if firms face difficulties in reducing employment to equate wages with the marginal product of labor, then the labor share will be even higher than it is in the first scenario. In both scenarios, the effect direction of wage increase on labor share depends on whether $\sigma$ is larger than one.

		Therefore, it is crucial to investigate whether capital-labor substitution occurs and estimate how large $\sigma$ is. Although the data does not directly provide indicators for firms' capital investments, we can approximate it using the logarithm of fixed assets. Fixed assets include capital stock, such as properties and equipment, which result from the investment. Estimates from column 3 of Table \ref{tab:debts_assets} suggest that firms are not significantly increasing their investment in fixed capital in response to the minimum wage. Moreover, column 1 in Table \ref{tab:bd_labor_share} demonstrates that employment decreases by 0.81 percentage points for a 10-percentage point increase in the bite. Existing literature finds a negligible employment effect \citep{dustmann_reallocation_2021, bossler_devil_2024} at the regional level.\footnote{Appendix Table \ref{tab:app_empl_region} reports the employment regression at the regional level. No significant reduction in employment is found, which is consistent with other studies.}  However, the insignificant effect at the regional level could result from worker reallocation \citep{dustmann_reallocation_2021}, implying that employment reduction at affected firms is possible. Column 2 in Table  \ref{tab:bd_labor_share} directly estimates the minimum wage effects on the capital-labor ratio, measured as the log of fixed assets scaling by employment. Together with the coefficient from the wage regression of column 3, we can calculate the elasticity of substitution between labor and capital, assuming the price of capital remains constant, and first-order conditions hold:
		%and the marginal product of labor equals to wage
		\begin{equation}
			\sigma = \frac{d \ln(\frac{K}{L})}{d \ln(\frac{w}{r})}. \notag
		\end{equation}   			
		The change in $\ln(\frac{K}{L})$ is measured as 0.063 from Table \ref{tab:bd_labor_share} and the change in $\ln(\frac{w}{r})$ is measured as 0.201 from column 3. Thus, $\sigma$ is equal to 0.31. This value, being less than one,\footnote{The results are consistent with recent studies that obtain that $\sigma$ is less than unity in most developed countries. Both \cite{muck_elasticity_2017,bellocchi_can_2023} estimate the value of $\sigma$ in Germany to be around 0.5.} indicates that labor and capital are complements. This empirical finding is consistent with the theoretical framework and points out that, at least in the sample period, we do not observe that capital substitutes labor. Therefore, an exogenous increase in wages leads to an increased labor share.

		\textbf{Break down labor share changes.}---Changes in labor share can be attributed to shifts in value-added or adjustments in labor costs or EBITDA, as value-added equals total labor costs plus EBITDA. Columns 4 to 6 of Table \ref{tab:bd_labor_share} show that the minimum wage implementation significantly increases firms' value-added, mainly through higher total labor costs. However, it also leads to a decrease in EBITDA. These results suggest that while the minimum wage boosts returns to labor (total labor costs), it reduces returns to capital (EBITDA). This points to a larger pie and more distribution of the pie towards labor. 
		
		\subsection{Other channels: tangible assets and profitability \label{sec:other_channels}}
		
		\begin{table}[ht!]\centering
			
			\caption{Minimum wage effects on log tangible assets and profitability}
			\label{tab:other}
			\begin{threeparttable}
				\begin{tabular}{l*{4}{c}}
					\hline\hline

					&	\multicolumn{1}{p{3cm}}{\centering Log tangible\\assets} &	\multicolumn{1}{p{3cm}}{\centering EBITDA\\/ Total A.}  &	\multicolumn{1}{p{3cm}}{\centering EBIT\\/Total A.}	&\multicolumn{1}{p{3cm}}{\centering Net Income\\/Total A.} \\
                        & (1)&(2)&(3)&(4)\\

					\hline
					
					$Bite*Post$                       &     -0.036        &      -0.036\sym{***}&      -0.034\sym{***}&      -0.027\sym{***}\\
					&     (0.030)         &     (0.004)         &     (0.004)         &     (0.003)         \\
					
					\hline
					Observations        &      184,702         &      184,702         &      184,702         &      173,208         \\
					\hline\hline
					
				\end{tabular}
				\begin{tablenotes}
					\item \textit{Notes:} Difference-in-differences regressions. The dependent variables are displayed above each column. A predetermined bite-specific trend is subtracted in all regressions. Firm fixed effects, county-year, and industry-year fixed effects are controlled. Firms are assigned to the county where their largest establishment is located. Industries are categorized with a two-digit industry code. Firm-level clustered standard errors are in parentheses. \sym{*}, \sym{**}, and \sym{***} denote statistical significance at 5\%, 1\% and 0.1\%, respectively. The full regression results are presented in Appendix Table \ref{tab:full_other}.
					\\\textit{Data}: Linked data of BeH, BHP, and Amadeus, 2011-2018.
				\end{tablenotes}
			\end{threeparttable}
			
		\end{table}
		
		Apart from the leverage substitution theory, the minimum wage could also influence firms' capital structure through increased tangible assets and decreased profitability. This section investigates whether the minimum wage also affects these two variables.

		Despite the theory suggesting that the minimum wage might elevate tangible assets, enabling firms to borrow more and increase financial leverage, the estimated results in column 1 of Table \ref{tab:other} do not provide supporting evidence for this. Columns 2 to 4 of Table \ref{tab:other} demonstrate reduced profitability across various measures.\footnote{Since all profitability measures have total assets as a denominator in Table \ref{tab:other}, Appendix Table \ref{tab:app_profit} demonstrates robust results when revenue is used as the denominator.} According to the pecking order theory, firms prefer to use retained profits first before resorting to debts. Consequently, declining profits may lead to increased debt borrowing, potentially raising financial leverage. However, this effect does not offset the negative impact of the leverage substitution channel. Therefore, increased labor share is the main channel through which the minimum wage affects financial leverage.

		\subsection{Heterogeneities: long-term debts
			\label{sec:long-short}}
		\begin{table}[ht!]\centering
			
			\caption{Minimum wage effects on long/short term liabilities}
			\label{tab:long}
			\begin{threeparttable}
				\begin{tabular}{l*{2}{c}}
					\hline\hline
					&	\multicolumn{1}{p{4.5cm}}{\centering Log long-term debts} &\multicolumn{1}{p{4.5cm}}{\centering Log short-term liabilities}\\
                        & (1)&(2)\\

					\hline
					$Bite*Post$          &      -0.727\sym{**} &      0.113         \\
					&     (0.180)         &     (0.109)         \\
					
					%	$Bite*Year_{2014}$     &      -0.283         &       0.155         \\
					%	&     (0.170)         &     (0.127)         \\[1ex] 
					%	$Bite*Pre$&     \multicolumn{2}{c}{Reference} \\[1ex] 
					%	$Constant$            &     9.263\sym{***}&      14.533\sym{***}\\
					%	&     (0.010)         &     (0.006)         \\
					\hline
					Observations        &      184,702         &      184,702         \\

					\hline\hline
					
				\end{tabular}
				\begin{tablenotes}
					\item \textit{Notes:} Difference-in-differences regressions. The dependent variables are displayed above each column. A predetermined bite-specific trend is subtracted in all regressions. Firm fixed effects, county-year, and industry-year fixed effects are controlled. Firms are assigned to the county where their largest establishment is located. Industries are categorized with a two-digit industry code. Firm-level clustered standard errors are in parentheses. \sym{*}, \sym{**}, and \sym{***} denote statistical significance at 5\%, 1\% and 0.1\%, respectively. The full regression results are presented in Appendix Table \ref{tab:full_long}.
					\\\textit{Data}: Linked data of BeH, BHP, and Amadeus, 2011-2018.
				\end{tablenotes}
			\end{threeparttable}
			
		\end{table}

		Firms can borrow long-term debts (maturing in more than one year) and short-term debts (maturing within one year). The duration of debts is relevant when assessing the risks associated with borrowing. As pointed out by \cite{yazdanfar_debt_2015}, long-term debt is particularly risky because it is more likely that firms will encounter negative shocks over the long term, such as obtaining lower profits but still having to pay the interest on their long-term debt \citep{favilukis_elephant_2020}. In contrast, firms face less uncertainty in the short run. Therefore, in response to increasing labor share, firms may tend to reduce their long-term debt first to mitigate risks in the longer term.
		
		Table \ref{tab:long} examines the minimum wage effects on log long-term debts and short-term liabilities separately. The coefficient for long-term debts is not only significant but also much larger than that for short-term liabilities, confirming the hypothesis that firms tend to primarily reduce risky long-term debts.

		\subsection{Heterogeneities: firms flexibility in adjusting labor \label{sec:outsource}}
		\begin{table}[ht!]\centering
			
			\caption{Minimum wage effects on financial leverage and labor share}
			\label{tab:os}
			\begin{threeparttable}
				\begin{tabular}{l*{4}{c}}
					\hline\hline
					
					&\multicolumn{2}{c}{Higher share of OS jobs}&\multicolumn{2}{c}{Lower share of OS jobs}\\
					\cmidrule(lr){2-3} \cmidrule(lr){4-5}
					
					&\multicolumn{1}{c}{Financial leverage}&\multicolumn{1}{c}{Labor share}&\multicolumn{1}{c}{Financial leverage}&\multicolumn{1}{c}{Labor share}\\
					                   & (1)&(2)&(3)&(4)\\

								\hline
						$Bite*Post$          &      -0.027\sym{***} &       0.124\sym{***}&      -0.133\sym{***}&       0.109\sym{***}\\
						&     (0.008)         &     (0.012)         &     (0.011)         &     (0.016)       \\
					%	$Bite*Year_{2014}$              &      -0.007         &       0.036\sym{*}  &      -0.035\sym{***}&       0.040         \\
				%		&     (0.007)         &     (0.016)         &     (0.009)         &     (0.021)         \\[1ex] 
				%		$Bite*Pre$&      \multicolumn{4}{c}{Reference} \\[1ex] 
				%		$Constant$           &       0.504\sym{***}&       0.660\sym{***}&       0.481\sym{***}&       0.665\sym{***}\\
				%		&     (0.000)         &     (0.001)         &     (0.000)         &     (0.001)         \\
						\hline
						Observations        &       91,586         &       91,586        &       81,608       &       81,608          \\
						\hline\hline
					
				\end{tabular}
				\begin{tablenotes}
					\item \textit{Notes:} Difference-in-differences regressions. The dependent variables are displayed above each column. The sample is split based on firms' share of outsourceable (OS) jobs. A predetermined bite-specific trend is subtracted in all regressions. Firm fixed effects, county-year, and industry-year fixed effects are controlled. Firms are assigned to the county where their largest establishment is located. Industries are categorized with a two-digit industry code. Firm-level clustered standard errors are in parentheses. \sym{*}, \sym{**}, and \sym{***} denote statistical significance at 5\%, 1\% and 0.1\%, respectively. The full regression results are presented in Appendix Table \ref{tab:full_os}.
					\\\textit{Data}: Linked data of BeH, BHP, and Amadeus, 2011-2018.
				\end{tablenotes}
			\end{threeparttable}
			
		\end{table}

		High labor share amplifies business risks due to the inflexibility of adjusting employment and the rigidity of adjusting wages. However, due to the different compositions of labor or occupations, labor costs for some firms may be less inflexible than others. For example, under negative shocks, firms may find it advantageous to outsource certain tasks. Outsourcing allows firms to avoid maintaining a large in-house labor force and provides them with greater flexibility in adjusting their production scale to respond to changes in the economic landscape rapidly. Consequently, outsourcing might serve to counteract the operating risks brought about by the minimum wage.
		
		In this section, I distinguish between firms with a large share of outsourceable occupations and those with only a small share. The hypothesis is that the former possess greater flexibility in labor adjustments, and for the same level of impact, they engage in less deleveraging.\footnote{\label{footnote_os}It is also possible that the effects of the minimum wage vary between firms with a higher or lower share of outsourceable occupations due to (1) nonlinear impacts of the minimum wage bite, or (2) different impacts at various quantiles of the financial leverage distribution. Appendix Table \ref{tab:summary_subsample} columns 1 to 2 presents the descriptive statistics for firms with a higher or lower share of outsourceable occupations. While the average pre-policy financial leverage is similar among them, the average level of the minimum wage bite differs significantly. Firms with a lower share of outsourceable occupations have a much lower level of the bite. Therefore, I conduct regressions using both the linear and quadratic terms of the bite to examine its nonlinear effects on financial leverage. The results indicate that the larger effects observed in firms with a lower share of outsourceable occupations cannot be attributed to the pre-policy levels of the bite in these firms. For a detailed comparison, please see Appendix \ref{app_non_linear}.} 
		
		Following \cite{goldschmidt_rise_2017}, my focus primarily rests on low-wage outsourcing occupations, namely food, cleaning, security, and logistics occupations,\footnote{The occupational code, initially based on the KldB 1988 standard, is converted to KldB 2010. Outsourceable occupations fall within categories 514, 541, 631, 632, 633, 831, 832, 942, 946, 223, 273, 292, 293, 341, 513, 516, 521, 525, 531, and 913.} as the minimum wage has the most substantial impact on these low-skilled jobs. Moreover, I exclude potential business service firms that provide outsourcing services.\footnote{The industry code for business service firms providing food, cleaning, security, and logistics services is initially based on the wz2003 standard and is converted to wz2008. These business service firms fall within categories 562, 812, 801, 802, 803, 749, 493, 494, 522, 521, 781, 782, 783, 799, and 853.} The occupations eligible for outsourcing are classified using a 3-digit occupation code, while business service firms are categorized based on the 3-digit industry code. All classification codes are provided by \cite{goldschmidt_rise_2017}. Firms are split based on whether their share of outsourceable occupations is larger than the sample median value (13.6\%) of the year 2013.

		Table \ref{tab:os} displays the impact of the minimum wage on financial leverage, differentiating between firms with higher and lower shares of outsourceable occupations. Firms with greater outsourcing potential and labor flexibility exhibit a significantly lower response in their financial leverage ratio compared to those with less flexibility. Quantitatively, a 10-percentage point increase in the minimum wage bite results in a mere 0.27 percentage points decrease in financial leverage for highly outsourceable firms, whereas firms with fewer outsourceable occupations experience a more substantial 1.33 percentage points decrease. Additional findings on labor share demonstrate a similar increase in the operational burden for these two types of firms. Therefore, the difference in coefficients observed in financial leverage regressions cannot be attributed to disparities in the effects of the minimum wage on labor share. This indicates that the reduction in financial leverage is significantly correlated with the inflexible nature of the labor force. The correlation is weakened when labor can be easily adjusted, such as through outsourcing.
		
		\subsection{Heterogeneities: firm sizes\label{sec:size}}
		
		\begin{table}[ht!]\centering
			\caption{Minimum wage effect on financial leverage and labor share}
			\label{tab:size_fi_lvg}
			\begin{threeparttable}
				
				\begin{tabular}{l*{3}{c}}
					\hline\hline
					&\multicolumn{1}{c}{Firm size: $<$50}&\multicolumn{1}{c}{Firm size: 50-249}&\multicolumn{1}{c}{Firm size: $>=$250}\\
                        & (1)&(2)&(3)\\

					\hline
					Financial leverage & &&\\
					\hline
					$Bite*Post$           &      -0.148\sym{***}&      -0.019\sym{*}         &       -0.017         \\
					&     (0.011)         &     (0.009)         &     (0.017)         \\
					%	$Bite*Year_{2014}$            &      -0.046\sym{***}&       -0.001         &      -0.026         \\
					%	&     (0.010)         &     (0.007)         &     (0.014)         \\[1ex] 
					%	$Bite*Pre$&       \multicolumn{3}{c}{Reference} \\[1ex] 
					%	$Constant$           &       0.519\sym{***}&       0.497\sym{***}&       0.426\sym{***}\\
					%	&     (0.001)         &     (0.000)         &     (0.001)         \\
					\hline
					Labor share & &&\\
					\hline
					
					\hline
					$Bite*Post$           &       0.184\sym{***}&       0.085\sym{***}&      -0.019         \\
					&     (0.019)         &     (0.012)         &     (0.013)         \\
					%	$Bite*Year_{2014}$          &       0.068\sym{**} &       0.015         &      -0.036\sym{*}         \\
					%	&     (0.026)         &     (0.013)         &     (0.014)         \\[1ex] 
					%	$Bite*Pre$&         \multicolumn{3}{c}{Reference} \\[1ex] 
					%	$Constant$           &       0.496\sym{***}&       0.711\sym{***}&       0.786\sym{***}\\
					%	&     (0.001)         &     (0.001)         &     (0.001)         \\
					\hline
					Observations        &       39,280         &       93,670         &       19,274         \\
					
					\hline
					Log EBITDA & &&\\
					\hline
					$Bite*Post$            &      -0.230\sym{***}&      -0.136\sym{**} &       0.118         \\
					&     (0.058)         &     (0.043)         &     (0.092)         \\
					
					%	$Bite*Year_{2014}$         &      -0.089         &      -0.014         &       0.158         \\
					%	&     (0.058)         &     (0.042)         &     (0.098)         \\[1ex]
					%	$Bite*Pre$&           \multicolumn{3}{c}{Reference} \\[1ex] 
					%	$Constant$            &      13.531\sym{***}&      14.195\sym{***}&      15.259\sym{***}\\
					%	&     (0.004)         &     (0.002)         &     (0.005)         \\
					\hline
					Observations        &       35,593        &       87,303         &       17,389        \\

					\hline\hline
				\end{tabular}
				\begin{tablenotes}
					\item \textit{Notes:} Difference-in-differences regressions. The dependent variables are displayed above each panel. The sample is split based on firm size categories. A predetermined bite-specific trend is subtracted in all regressions. Firm fixed effects, county-year, and industry-year fixed effects are controlled. Firms are assigned to the county where their largest establishment is located. Industries are categorized with a two-digit industry code. Firm-level clustered standard errors are in parentheses. \sym{*}, \sym{**}, and \sym{***} denote statistical significance at 5\%, 1\% and 0.1\%, respectively. The full regression results are presented in Appendix Table \ref{tab:full_size_fi_lvg}.
					\\\textit{Data}: Linked data of BeH, BHP, and Amadeus, 2011-2018.
				\end{tablenotes}
			\end{threeparttable}
			
		\end{table}		
		
		Despite the declining trend in financial leverage over the past two decades, bank financing remains crucial for small and medium-sized firms in Germany. Unlike listed firms, relatively small firms encounter challenges when attempting to transition to alternative financing methods or accessing the capital market. Therefore, at the same level of minimum wage influence, small firms may face difficulties in deleveraging and may respond less to the minimum wage.
		
		However, on the other side, given that small firms have less market power and limited ability to transfer rising labor costs to prices and maintain previous profit levels, this suggests that small firms face a more substantial increase in labor leverage. Consequently, they may have a higher incentive to decrease their financial leverage.\footnote{As Note \ref{footnote_os} discusses, the heterogeneous effects of the minimum wage on firms of different sizes could also be due to variations in the pre-policy levels of the bite or financial leverage. Appendix Table \ref{tab:summary_subsample} columns 3 to 5 display the means of these two variables. While the bite does not differ substantially, larger firms have significantly lower financial leverage. To examine the minimum wage effect on the distribution of financial leverage, I conduct unconditional quantile regressions \citep{firpo_unconditional_2009}. As illustrated in Appendix Figure \ref{app_ucq}, the heterogeneous effects are not attributable to the distributional effect of the minimum wage.}
		
		In this section, I assess whether firms of different sizes adjust their leverage differently in response to the same level of the bite variable. The analysis is conducted by splitting the sample in terms of firm sizes in 2013. The upper panel of Table \ref{tab:size_fi_lvg} demonstrates how the minimum wage affects financial leverage depending on firm sizes. Significant effects are observed for firms with fewer than 50 employees and for firms with 50-249 employees. Effect sizes decrease as firm sizes increase. Since the sample is underrepresentative of small firms, the estimates for the entire sample underestimate the true deleveraging effect. Appendix Table \ref{app: weighted_reg} presents the weighted regression using size-sector weights and estimating the effect of the minimum wage on firms' financial leverage. The effect size is significantly larger than those reported in the main results.

		When examining the labor share results, it becomes evident that small firms are more profoundly affected by the minimum wage, whereas the minimum wage has no effect on firms with more than 250 employees. These findings reveal that risks induced by the minimum wage are more pronounced in small firms, consequently incentivizing them to engage in greater deleveraging efforts.

		Additionally, I also examine whether firms' EBITDA is differentially affected by the minimum wage. If small firms have less market power, their earnings should be reduced more. The lower panel in Table \ref{tab:size_fi_lvg} displays that, for the same level of impact, the EBITDA of small firms decreases much more significantly than those of medium-sized firms, while the largest firms are unaffected.
	
		\section{Conclusion\label{sec: conclude}}
 I investigate the impact of the minimum wage on firms' financial leverage by using the firm-establishment-employee linked data and the difference-in-differences estimation method with firm-level variations of the minimum wage exposure. Firms face a trade-off between financial leverage and the increasing labor share resulting from the minimum wage, both of which amplify the expected costs of financial distress. To mitigate the risks caused by the rising labor share, firms reduce their financial leverage.

I find that the average minimum wage treatment level leads to a decrease in the financial leverage by 0.5 to 0.9 percentage points and to an increase in the labor share by 0.5 to 1.6 percentage points. Descriptively, in my sample period, the mean of financial leverage experiences a reduction of 2 percentage points between pre- and post-intervention, while the labor share increases by 1 percentage point.\footnote{See Appendix Table \ref{tab:summar}.} Comparing the developments in financial leverage and labor share at the aggregate level with the average minimum wage effects, the minimum wage contributes a non-negligible part to the deleveraging trend and the increasing labor share in Germany in recent years.

		Moreover, I explore the mechanism further. I find that the minimum wage reduces firms' debt borrowing and increases cash holdings. Regarding the labor share, the elasticity of substitution between labor and capital is estimated at 0.31, suggesting labor and capital are complements. Additionally, changes in the labor share result in increased total labor costs but decreased profits, indicating a shift toward labor in firms' total value-added. Other channels through which the minimum wage affects financial leverage are found to play a minor role compared to the labor share.
		
		Furthermore, heterogeneous effects indicate that firms tend to decrease long-term debts instead of short-term debts due to the higher risks associated with long-term debts. The flexibility of adjusting labor is crucial for firms responding to rising labor share. A more flexible labor composition, including outsourced occupations, reduces leverage substitution behavior. Small firms show a notable increase in the labor share and a larger deleveraging behavior, reinforcing the link between the labor share and financial leverage.
		
		In summary, these results establish that firms' corporate decisions are responsive to labor market policies. The minimum wage benefits employees overall, with increased total value-added and more earnings shifted to the labor force. However, for firms, the minimum wage makes them less flexible in adjusting costs and imposes larger operating burdens. Consequently, they exhibit more conservative behavior to offset the associated risks. These findings are derived from an examination of the impact of the German minimum wage; they may also be applicable to countries with characteristics similar to Germany's, such as those with strong employment protection laws and high compliance rates with such policies. However, caution is warranted when extrapolating these findings to countries with weak employment protection, where the minimum wage may prompt significant capital-labor substitution, potentially leading to different conclusions. Additionally, this study focuses on the effects observed within four years following the introduction of the policy; further research is needed to understand the long-term implications.
		
		\stopcontents[mainsections]

		%%%%%%%%%%%%%%%%%%%%%%%%%%%%%%%%%%%%%%%%%%%%%%%%%
		\clearpage
		\singlespacing
		\bibliographystyle{apalike}
		\bibliography{mw_lvg}
		
		%%%%%%%%%%%%%%%%%%%%%%%%%%%%%%%%%%%%%%%%%%%%%%%%%
		
		\clearpage

		%\begin{refsection} % define new bibliography for appendix

		\begin{center}
			
			\Large
			Online Appendix for \\
			\vspace{0.6cm}
			
			Firms' Risk Adjustments to Minimum Wage:
			\\ Financial Leverage and Labor Share Trade-off\\
			
			\vspace{0.6cm}
			\normalsize
			by \\
			Ying Liang
			
			\vspace{1.0cm}
			\Large 
			\textbf{Content}
			\normalsize
			
			\singlespacing
			
			\startcontents[appendices]
			\printcontents[appendices]{l}{1}{\setcounter{tocdepth}{2}}

		\end{center}
		
		\clearpage

		%%%%%%%%%%%%%%%%%%%%%%%%%%%%%%%%%%%%%%%%%%%%%%%%%

		%%%%%%%%%%%%%%%%%%%%%%%%%%%%%%%%%%%%%%%%%%%%%%%%%
		%%%%% These commands start the appendix and change the Table & Figure numbering

		%%%%%%%%%%%%%%%%%%%%%%%%%%%%%%%%%%%%%%%%%%%%%%%%%
		\doublespacing
		
		\newpage
		\appendix
		\section{Theoretical framework: minimum wage and labor share}\label{app:mw_ls_theory}
		\renewcommand*{\thefigure}{\thesection\arabic{figure}}
		\renewcommand*{\thetable}{\thesection\arabic{table}}
  		\renewcommand*{\theequation}{\thesection\arabic{equation}}
		\setcounter{figure}{0}
		\setcounter{table}{0}
  \setcounter{equation}{0}

  In this section, I use the framework developed by \cite{bentolila_explaining_2003}, which is also employed by \cite{petreski_minimum_2023} to analyze the impact of the minimum wage on the labor share.

Starting with the simple scenario, firms face an increasing wage level due to the minimum wage, and they have the option to adjust employment so that the equilibrium condition still holds that the real wage is equal to the marginal product of labor. The CES (constant elasticity of substitution) production function is defined as 
\begin{equation} \label{eq:ces_production}
	Y_j=\left[\alpha (A_{j}K_j)^{\epsilon}+(1-\alpha)(B_{j} L_j)^{\epsilon}\right]^{\frac{1}{\epsilon}}  \notag
\end{equation}
where $Y_j$ denotes output,  $K_j$ is quantity of capital input, and $L_j$ is quantity of labor. $A_j$ represents capital-augmenting technical progress. $B_j$ is labor-augmenting technical progress. $\alpha$ is the share parameter and $\epsilon$ the substitution parameter.
Labor share is defined as the share of total labor income to output:
\begin{equation}
	LS_j=\frac{w_{j}L_j}{Y_{j}p_{j}} \notag
\end{equation}
Firms solve the profit maximization problem:
%Then normalize equation \ref{eq:ces_production} by the amount of labor $ L_t$:
%	\begin{equation} \label{eq:ces_production_norm}
	%	\frac{Y_j}{L_j}=A_t\left[\alpha	\frac{K_j}{L_j}^{\epsilon}+(1-\alpha)^{\epsilon}\right]^{\frac{1}{\epsilon}}
	%\end{equation}
	
	\begin{equation} \label{eq:profit_max}
		\max \pi_j=Y_{j}p_{j}-w_j L_j-r_j K_{j}			\notag
	\end{equation}
	$w_j$ and $r_j$ are the prices of labor and capital, namely wage and interest. $p_j$ is the price of output.
	Take the partial derivative with respect to $L_j$:
	
	\begin{equation} \label{eq:w_j}
		\frac{w_j}{p_{j}}=B_j^{\epsilon}(1-\alpha){(\frac{L_j}{Y_j})}^{\epsilon-1} 				\notag
	\end{equation}
	
	%		\begin{equation}
		%			\frac{r_j}{p_{j}}=A_j^{\epsilon}\alpha{(\frac{K_j}{Y_j})}^{\epsilon-1} 
		%		\end{equation}
	
	\begin{equation} \label{eq:Li_Yi}
		\frac{L_j}{Y_j}={(\frac{w_j}{p_j})}^{\frac{1}{\epsilon-1}}{B_j}^{\frac{\epsilon}{(1-\epsilon)}}(1-\alpha)^{\frac{1}{1-\epsilon}}			
	\end{equation}
	Insert Equation \ref{eq:Li_Yi} to the definition of labor share:
	
	\begin{equation}\label{eq:ls_old}
		LS_j=\frac{w_{j}L_j}{Y_{j}p_{j}}={(\frac{w_j}{p_j})}^{\frac{\epsilon}{\epsilon-1}}{B_j}^{\frac{\epsilon}{(1-\epsilon)}}(1-\alpha)^{\frac{1}{1-\epsilon}} 
	\end{equation}
	Take the partial derivative of labor share with respect to real wage:
	\begin{equation} \label{eq:relation}
		\frac{\partial LS_j}{\partial \frac{w_j}{p_j}}=\frac{\epsilon}{\epsilon-1}{(\frac{w_j}{p_j})}^{\frac{1}{\epsilon-1}}{B_j}^{\frac{\epsilon}{(1-\epsilon)}}(1-\alpha)^{\frac{1}{1-\epsilon}} 	
	\end{equation}
	The direction of the partial derivative depends on $\epsilon/(\epsilon-1)$. The elasticity of substitution between labor and capital is  $\sigma=1/(1-\epsilon)$. If $\epsilon \in (0,1)$,  $\sigma>1$ and labor and capital are substitutes. In this case, the increase in wage leads to decreased labor share. If $\epsilon \in (-\infty,0)$,  $\sigma<1$ and labor and capital are complements. In this case, the increased wage results in increased labor share.
	
	However, if firms are not able to adjust employment fully, then there would be a wedge between the marginal product of labor and real wages. \cite{bentolila_explaining_2003} also discussed the situation where firms and workers bargain over both wages and employment. In this case, employment is set such that the marginal product of labor is equal to its opportunity cost ($\frac{\bar{w_j}}{p_j}$):
	\begin{equation} \label{eq:barw_i}
		\frac{\bar{w_j}}{p_{j}}=B_j^{\epsilon}(1-\alpha){(\frac{L_j}{Y_j})}^{\epsilon-1} 				\notag
	\end{equation}
	Assume workers' bargaining power is $\theta$, with a Nash-bargaining model, the real wage is then the weighted average of opportunity costs and the average labor product:
	
	\begin{equation} 
		\frac{w_j}{p_{j}}=\theta\frac{Y_j}{L_j} 	+(1-\theta)(B_j^{\epsilon}(1-\alpha){(\frac{L_j}{Y_j})}^{\epsilon-1}) 	\notag
	\end{equation}
	Denote the labor share in Equation \ref{eq:ls_old} as $LS_{j}$ and the new labor share in bargaining set-up as $LS_{new}$, 
	
	\begin{equation}\label{eq:new_ls}
		LS_{new}=\theta+(1-\theta) LS_{j}=LS_j+(1-LS_j)\theta 	\notag
	\end{equation}
	The labor share becomes larger if workers bargain employment with firms. Moreover, introducing the bargaining power of workers over employment does not affect the conclusion from Equation \ref{eq:relation} that the effect's direction of increase in real wages on labor share depends on $\sigma$.
			\clearpage
			\section{Summary statistics}\label{app:summary_statistics}
		\renewcommand*{\thefigure}{\thesection\arabic{figure}}
			\renewcommand*{\thetable}{\thesection\arabic{table}}
			\setcounter{figure}{0}
			\setcounter{table}{0}
			
			\subsection{Sample representativeness}\label{app:represent}
			\begin{figure}[ht!]
				\captionabove{Development of aggregate labor share in Germany}
				%1-------------------------------------
				
				\centering
				% include the fourth image
				\includegraphics[width=.7\linewidth]{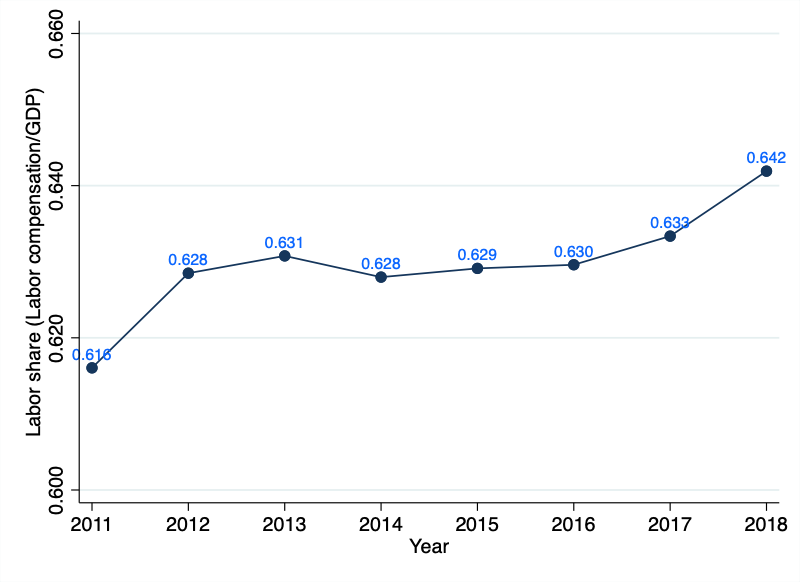}  
				\label{fig:app_develop_ls}				
				\caption*{
					\textit{Notes}: The figure displays Germany's aggregate labor share (labor compensation/GDP). All industries are included. 
					
					\textit{Data}: \href{https://www.rug.nl/ggdc/productivity/pwt/}{Penn World Table version 10.01}. }
			\end{figure}		
			
			\begin{figure}[ht!]
				\captionabove{Sample proportion to population, in sector-size cells}
    \label{app:fig_size_cell}
				%1-------------------------------------
				\centering					
				\includegraphics[width=1.1\linewidth]{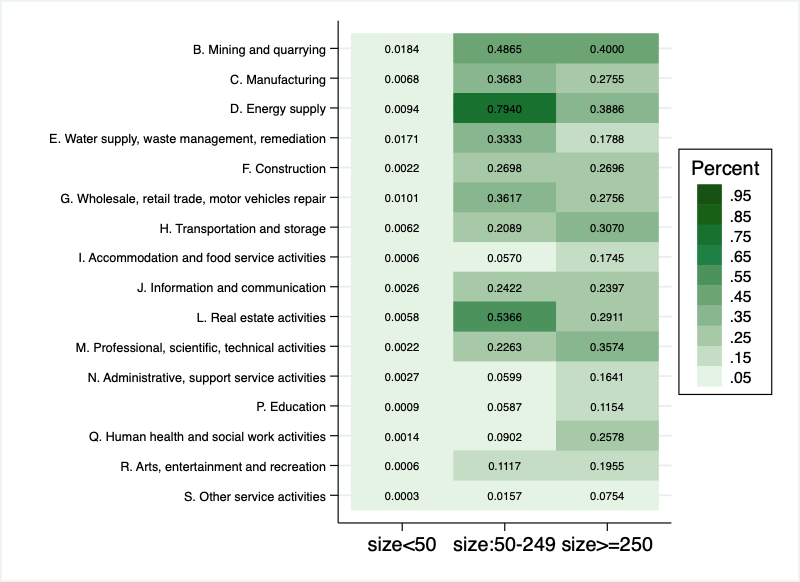}  
				\label{fig:app_represent_sector}					
				\caption*{\textit{Notes}: The figure displays the ratio of the number of firms in the sample (2018) over the number of firms in the population (2019). Sector K financial firms are dropped from the sample as well as from the population data. Cells for sector O (Public Sector) are not shown in the figure, as the number of public firms in some cells is below the threshold required for publication due to IAB data protection rules. Population data are from the German Business Register, extracted form \href{https://www.destatis.de/EN/Themes/Economic-Sectors-Enterprises/Enterprises/Business-Register/_node.html265826}{GENESIS-Online database}.  The year 2018 is not available in the GENESIS database. Thus, the closest year, 2019, is chosen. 
			\\	\textit{Data}: Linked data from BeH, BHP, Amadeus(2018), and GENESIS-Online database. }
			\end{figure}		
			\begin{figure}[ht!]
				\captionabove{Sample proportion to population, in state-size cells}
				%1-------------------------------------
				\centering
				% include the fourth image
				\includegraphics[width=1.1\linewidth]{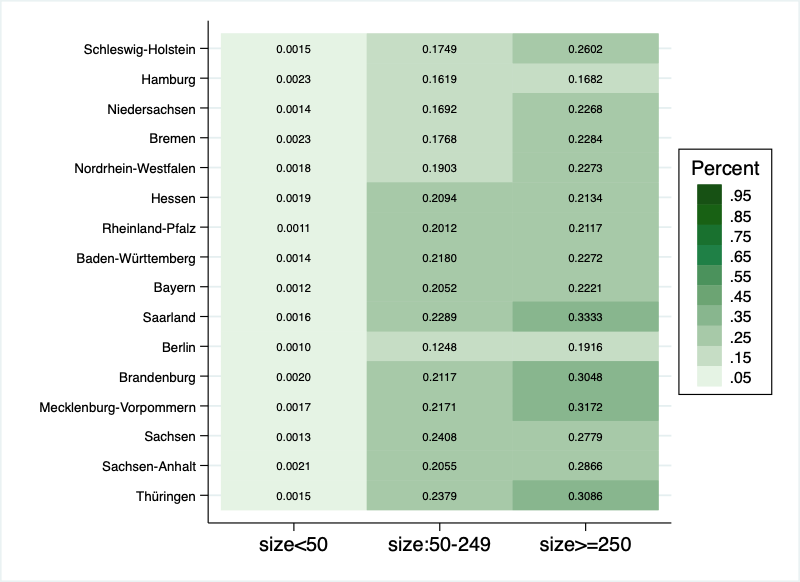}  
				
				\label{fig:app_represent_state}			
				
				\caption*{\textit{Notes}: The figure displays the ratio of the number of firms in the sample (2018) over the number of firms in the population (2019). Sector K financial firms are dropped from the sample as well as from the population data. Population data are from the German Business Register, extracted form \href{https://www.destatis.de/EN/Themes/Economic-Sectors-Enterprises/Enterprises/Business-Register/_node.html265826}{GENESIS-Online database}.  The year 2018 is not available in the GENESIS database. Thus, the closest year, 2019, is chosen. 
				
					\textit{Data}: Linked data from BeH, BHP, Amadeus(2018), and GENESIS-Online database.}
			\end{figure}		
			
			\clearpage
			%=================================================================
   \begin{landscape}

			\subsection{Variables definitions}\label{app:definition}
			\begin{table}[ht!]\centering
   \caption{Vairables definitions and sources}
				\resizebox{1.3\textheight}{!}{
					\begin{threeparttable}
						\begin{tabular}{|c|c|c|c|}
							\hline
							Variable & Definition & Original variables used& Source \\[1ex]
							\hline
							
							\hline
							\multicolumn{4}{|c|}{Financial leverages} \\[1ex] 
							\hline
							Financial leverage & $\frac{\text{(long-term debts+short-term liabilities)}}{\text{total assets}}$ &  $\frac{\text{(LTDB+CULI)}}{\text{TOAS}}$ & BvD\\[1ex]
							\hline
							Broader fin. leverage & $\frac{\text{(long-term liabilities+short-term liabilities)}}{\text{total assets}}$ &$\frac{\text{(NCLI+CULI)}}{\text{TOAS}}$ &BvD\\[1ex]
							\hline
							Long-term leverage &$\frac{\text{(long-term debts)}}{\text{total assets}}$&$\frac{\text{(LTDB)}}{\text{TOAS}}$&BvD\\[1ex]
							\hline
							Net leverage& $\frac{\text{(long-term debts+short-term liabilities-cash)}}{\text{total assets}}$&$\frac{\text{(LTDB+CULI-CASH)}}{\text{TOAS}}$ &BvD \\[1ex]
							\hline
							\multicolumn{4}{|c|}{Labor related variables} \\[1ex] 
							\hline
							Labor share& total labor costs/value-added&$\frac{\text{(TLC)}}{\text{TLC+EBTA}}$ &BvD,IAB\\
							\hline
							Total  labor costs per worker&total labor costs/employment&&IAB\\
       \hline
							Total  labor costs(TLC)& total annual gross wages&&IAB\\
							
							\hline
							Employment &number of workers &&IAB\\
       \hline
							value-added& total labor costs + EBITDA&TLC+EBTA&BvD, IAB\\
       \hline
							Capital-labor ratio & fixed assets/employment &$\frac{\text{(FIAS)}}{\text{number of workers}}$ &BvD, IAB\\
       \hline
       Outsourceable occupations &\multicolumn{1}{p{7cm}|}{\centering food, cleaning, security, and logistics occupations \citep{goldschmidt_rise_2017}} &&IAB\\
							\hline
							Bite &\multicolumn{1}{p{7cm}|}{\centering share of sub-minimum wage workers, based on gross hourly wages}&& IAB\\
							\hline
							Gap &		$\frac{\sum_{i\in j}^{}h_{i,2013}Max\{0,8.5-wage_{i,2013}\}}{\sum_{i\in j}h_{i,2013}wage_{i,2013}}$&&IAB\\[1ex]
							\hline

							\hline
						\end{tabular}
						
						\begin{tablenotes}
							\item \textit{Notes:} Total labor costs, total labor costs per worker, employment, bite, outsourceable occupations, and gap variables are self-calculated from the IAB data.
						\end{tablenotes}
					\end{threeparttable}
					
 				}
			\end{table}

			\clearpage

   		%=================================================================	
			\begin{table}[ht!]\centering
      \caption{Variable's definitions and sources, cont'd}

				\resizebox{1.3\textheight}{!}{
					\begin{threeparttable}
						\begin{tabular}{|c|c|c|c|}
							\hline
							Variable & Definition & Original variables used& Source \\[1ex]
							\hline

							\hline
							\multicolumn{4}{|c|}{Other variables} \\[1ex] 
       \hline
       Total debts&Long-term debts and short-term liabilities&LTDB+CULI&BvD\\
							\hline
							Long-term liabilities & long-term debts and provisions &NCLI &BvD\\
							\hline
							Short-term liabilities &\multicolumn{1}{p{7cm}|}{\centering  short-term debts, trade payables,\\ and other current liabilities}
							&CULI&BvD\\
       \hline
							Total assets	& fixed assets+current assets&TOAS&BvD\\
       \hline
							Fixed assets	& tangible assets+intangible assets&FIAS&BvD\\
							\hline
							Tangible assets&\multicolumn{1}{p{7cm}|}{\centering physical assets, such as property, equipment, and inventory}&IFAS&BvD\\
       	\hline
							Intangible assets&\multicolumn{1}{p{7cm}|}{\centering non-physical assets, such as patents and trademarks}&TFAS&BvD\\
       \hline
							Current assets &\multicolumn{1}{p{7cm}|}{\centering cash, trade receivables and other current assets}&CUAS&BvD\\
       \hline
							Cash	&\multicolumn{1}{p{7cm}|}{\centering  cash-in-hand, central Bank balances, \\bank balances and cheques}&CASH&BvD\\
							\hline
							EBITDA &\multicolumn{1}{p{7cm}|}{\centering earnings before interest, taxes,\\ depreciation and amortization} & EBTA &BvD\\

							\hline
							Net income&profit/loss - minority interest&PL&BvD\\
							\hline
							EBIT	& earnings before interest and taxes&EBIT&BvD\\
							\hline
							ROA	&EBIT/total assets&$\frac{\text{(EBIT)}}{\text{TOAS}}$&BvD\\
							\hline
						\end{tabular}
						
						\begin{tablenotes}
							\item \textit{Notes:} Total labor costs, total labor costs per worker, employment, bite, and gap variables are self-calculated from the IAB data.
						\end{tablenotes}
					\end{threeparttable}
					
 				}
			\end{table}
    \end{landscape}

			\clearpage
			\begin{landscape}
				
				\subsection{Summary statistics table}\label{app:summary}
				\renewcommand*{\thefigure}{\thesection\arabic{figure}}
				\renewcommand*{\thetable}{\thesection\arabic{table}}
				\setcounter{figure}{0}
				\setcounter{table}{0}
				%=================================================================			
				
				\begin{table}[ht!]\centering			
					\caption{Summary Statistics}
					\label{tab:summar}
					\resizebox{1.25\textheight}{!}{
						\begin{threeparttable}
							\begin{tabular}{l*{2}{ccccc}}
								\hline\hline
								&  \multicolumn{5}{c}{  Pre-period: 2011-2014}              &          \multicolumn{5}{c}{ Post-period: 2015-2018}          \\
								\cmidrule(lr){2-6}\cmidrule(lr){7-11}
								&        mean&          sd&         p25&         p50&         p75&        mean&          sd&          p25&         p50&         p75 \\
								\hline
								\hline
								Financial Leverage variables &                     &            &            &            &           &           &&&&            \\
								\hline 
								Financial leverage  &       0.508&       0.274&       0.286&       0.507&       0.720&       0.482&       0.272&       0.258&       0.473&       0.690\\
								Financial  leverage (broad def.)&       0.629&       0.264&       0.434&       0.650&       0.830&       0.604&       0.268&       0.399&       0.616&       0.808\\
								Long-term leverage  &       0.154&       0.211&       0.000&       0.056&       0.238&       0.132&       0.188&       0.000&       0.038&       0.202\\
								Net leverage        &       0.376&       0.354&       0.126&       0.403&       0.648&       0.346&       0.352&       0.097&       0.368&       0.614\\
								\hline
								Labor-related variables &           &                        &            &            &           &     &&&&               \\
								\hline
								Labor share         &       0.657&       0.320&       0.469&       0.669&       0.826&       0.667&       0.320&       0.480&       0.679&       0.833\\
								Log average annual labor costs per employee worker&      10.365&       0.498&      10.124&      10.396&      10.656&      10.427&       0.467&      10.189&      10.447&      10.704\\
								Log total annual labor costs&      14.686&       1.363&      14.123&      14.800&      15.440&      14.898&       1.294&      14.335&      14.982&      15.613\\
								Log employment (IAB)&       4.321&       1.222&       3.761&       4.431&       5.030&       4.471&       1.179&       3.951&       4.544&       5.147\\
								Log value-added (IAB)\textsuperscript{†}&      15.264&       1.164&      14.712&      15.297&      15.905&      15.454&       1.103&      14.897&      15.462&      16.076\\
								Log capital-labor ratio&10.357&1.820&9.246&10.330&11.319&10.432&1.793&9.363&10.427&11.358\\
								Share of outsourceable workers&       0.227&       0.247&       0.051&       0.135&       0.308&       0.241&       0.256&   0.058&       0.146&       0.333\\									
								\hline					
								Other variables           &            &            &           &         &            &              &&&&                     \\ 
								\hline					
								Log total debts  &      15.392&       1.497&      14.638&      15.435&      16.215&      15.532&       1.457&      14 .759&      15.560&      16.354\\
								Log long-term debts & 9.407 & 6.860&0.000&13.032&14.822&9.153&6.966&0.000&12.855&14.832\\
								Log short-term liabilities&14.302&3.450&14.028&14.992&15.823&14.861&2.474&14.265&15.158&15.980\\
								Log total assets    &      16.292&       1.314&      15.657&      16.190&      16.944&      16.503&       1.232&      15.839&      16.380&      17.114\\
								Log fixed assets&14.678&2.026&13.531&14.821&15.903&14.901&1.961&13.824&15.034&16.075\\
								Log tangible assets&14.275&2.166&13.020&14.489&15.638&14.480&2.118&13.284&14.702&15.812\\
         Log intangible assets&		8.915&4.345&7.690&10.032&11.712&9.235&4.205&8.042&10.297&11.938\\
								Log current assets&15.715&1.297&15.103&15.725&16.413&15.925&1.228&15.300&15.904&16.579\\
								Log cash         &      13.072&       2.316&      11.865&      13.476&     14.719&      13.309&       2.316&      12.182&      13.758&      14.928\\
								Log EBITDA\textsuperscript{†}&14.117&1.423&13.398&14.155&14.954&14.290&1.382&13.560&14.314&15.110\\		
                                   
                                    Log EBIT\textsuperscript{†}&13.702&1.520&12.903&13.779&14.642&13.871&1.497&13.059&13.940&14.803\\
                                    ROA\textsuperscript{†} &0.087&0.123&0.025&0.067&0.136&0.084&0.119&0.023&0.064&0.120\\
                                    Log net income&13.126&1.660&12.299&13.303&14.181&13.342&1.617&12.534&13.509&14.362\\
								\hline
								Observations        & 95,239        &&&                 &            &       89,592                &            &        &&         \\
								\hline\hline
							\end{tabular}					
							\begin{tablenotes}						
								\textit{Notes}: Appendix \ref{app:definition} contains variables' definitions. "Observations" indicates the number of observations in the main sample. \textsuperscript{†} denotes variables that have missing values and the number of observations is smaller than it is in the main sample.	\\
								\textit{Data}: Linked data of BeH, BHP, and Amadeus, 2011-2018.
							\end{tablenotes}
						\end{threeparttable}
					}
				\end{table}
			\end{landscape}
			%=================================================================	

				\section{Robustness checks}\label{app: robust}
				\renewcommand*{\thefigure}{\thesection\arabic{figure}}
				\renewcommand*{\thetable}{\thesection\arabic{table}}
				\setcounter{figure}{0}
				\setcounter{table}{0}
				
				\subsection{Alternative treatment measures}\label{app: robust_other_bite}			
The gap variable is also a firm-level measure, and it is the mean of workers' wage gap, which is defined as in Equation \ref{eq:gap}. The worker's gap measures how much a worker's hourly wage should increase so that it reaches the minimum wage threshold of 8.5 \euro. Additionally, I use the average gap \citep{dustmann_reallocation_2021} and average bite variables as further measures of treatment intensity.
\begin{equation}
    Average\ gap=\frac{1}{4}\sum_{2011}^{2014}gap_{j,t} \notag
\end{equation}

\begin{equation}
    Average\ bite=\frac{1}{4}\sum_{2011}^{2014}bite_{j,t} \notag
\end{equation}
    The gap variable is sensitive to very low hourly wages, as such low wages can lead to outliers with extremely large gap values. Moreover, the gap variable aggregated at the firm level involves more measurement error than at the regional level, especially in small firms with only a few employees. Due to the two reasons mentioned above, the gap and the average gap are winsorized at the 99th percentile for each year. 
Appendix Table \ref{app:tab_gap} columns (1) and (2) report the effects of the minimum wage policy on financial leverage and labor outcomes when using the gap as the measure of treatment. The coefficient of the interaction terms ranges from -0.11 to -0.22. The mean of the gap variable measured in 2013 is 0.02. The minimum wage reduces firms' financial leverage by about 0.22 to 0.44 percentage points due to the mean level of the treatment. Similarly, the labor share is reduced by 0.29 to 0.96 percentage points for a mean level of the gap. The mean of the average gap is about 0.26. Therefore, the average minimum wage effect on financial leverage is about 3.5 to 7 percentage points, as shown in column (5). Overall, the effect size is smaller when using the gap measure, but when using the average gap, it is close to the main findings.

In addition, columns (3) and (4) report the effects when using average bite as the treatment measure. The mean level of the averaged bite is the same as the bite, and the coefficient sizes are also very close to the main results. Therefore, the effect sizes are almost the same as in the main results.
\begin{landscape}

				\begin{table}[ht!]\centering
					\caption{Minimum wage effect on financial leverage and labor costs, using gap variable}
     \label{app:tab_gap}
					\resizebox{1.5\textheight}{!}{
						\begin{threeparttable}
							\begin{tabular}{l*{6}{c}}
								\hline\hline
        		&\multicolumn{2}{c}{Gap}&\multicolumn{2}{c}{Average Bite}&\multicolumn{2}{c}{Average Gap}\\
						\cmidrule(lr){2-3} \cmidrule(lr){4-5} \cmidrule(lr){6-7}
								&\multicolumn{1}{p{3.5cm}}{\centering Financial\\leverage}&\multicolumn{1}{p{3.5cm}}{\centering Labor \\share}&\multicolumn{1}{p{3.5cm}}{\centering Financial\\leverage}&\multicolumn{1}{p{3.5cm}}{\centering Labor \\share}&\multicolumn{1}{p{3.5cm}}{\centering Financial\\leverage}&\multicolumn{1}{p{3.5cm}}{\centering Labor \\share}\\
                           & (1)&(2)&(3)&(4)&(5)&(6)\\

									\hline
								$Treatment*Year_{2011}$&       -0.014        &       0.022         &       -0.001         &       0.002  &-0.019 & 0.067      \\
								                        &      (0.026)        &     (0.022)        &    (0.007)         &     (0.014)    &(0.027) &(0.053)   \\
								$Treatment*Year_{2012}$&       -0.031         &      -0.001         &       -0.003         &      -0.006   &-0.036 & 0.016        \\
							                        	&      (0.019)         &    (0.022)         &       (0.005)         &     (0.012)     &(0.021) &(0.044)           \\[1ex] 
								$Treatment*Year_{2013}$&         \multicolumn{6}{c}{Reference} \\[1ex] 
								
								$Treatment*Year_{2014}$&       -0.029 &               0.089\sym{*}         &       -0.017\sym{**} &       0.040\sym{*}  & -0.045\sym{**} &       0.066\\
								                       &      (0.015)   &            (0.042)         &     (0.006)         &     (0.013)     &(0.020) &(0.045)          \\
								$Treatment*Year_{2015}$&       -0.105\sym{***}&       0.143\sym{***} &      -0.049\sym{***}&       0.048\sym{***}&  -0.135\sym{***}&       0.085\sym{*}\\
								                         &     (0.018)         &     (0.041)         &     (0.006)         &     (0.013)     &(0.021) &(0.041)        \\
								
								$Treatment*Year_{2016}$&       -0.156\sym{***}&       0.325\sym{***}       &       -0.067\sym{***}&       0.105\sym{***} &  -0.198\sym{***}&       0.235\sym{***}\\
								                          &     (0.023)         &     (0.041)         &    (0.008)         &     (0.013)   &(0.025) &(0.046)            \\
								$Treatment*Year_{2017}$&   -0.170\sym{***}&       0.398\sym{***}      &     -0.074\sym{***}&       0.127\sym{***} & -0.211\sym{***}&       0.291\sym{***}\\
							                           	&     (0.026)         &     (0.045)       &  (0.008)         &     (0.014)     &(0.027) &(0.049)         \\
								$Treatment*Year_{2018}$&       -0.216\sym{***}&       0.478\sym{***}          &     -0.094\sym{***}&       0.142\sym{***} &  -0.273\sym{***}&       0.348\sym{***} \\
								                          &     (0.031)         &     (0.056)         &   (0.009)         &     (0.014)   &(0.034) &(0.059)             \\
								$Constant$            &       0.494\sym{***}&       0.656\sym{***}&        0.496\sym{***}&       0.668\sym{***}&  0.492\sym{***}&       0.642\sym{***}\\
								                           &     (0.002)         &     (0.004)         &     (0.001)         &     (0.001) &  (0.003)         &     (0.006)         \\
								\hline
								Observations        &      184,702         &      184,702         &       179,059         &     179,059       &       179,059         &     179,059   \\
								
								\hline\hline
								
							\end{tabular}

							\begin{tablenotes}
								\item \textit{Notes:} Difference-in-differences regressions. The dependent variables are displayed above each column. Treatment intensity is the gap for columns (1) and (2), the average bite for columns (3) and (4), and the average gap for columns (5) and (6). A predetermined gap-specific trend is subtracted in all regressions. Firm fixed effects, county-year, and industry-year fixed effects are controlled. Firms are assigned to the county where their largest establishment is located. Industries are categorized with a two-digit industry code. Firm-level clustered standard errors are in parentheses. \sym{*}, \sym{**}, and \sym{***} denote statistical significance at 5\%, 1\% and 0.1\%, respectively.
								\\\textit{Data}: Linked data of BeH, BHP, and Amadeus, 2011-2018.
							\end{tablenotes}
						\end{threeparttable}
						
					}
				\end{table}

				\clearpage
				
				%===========================================
    				\subsection{Alternative measures of financial leverage}\label{app: robust_other_fin_lvg}			

	\begin{table}[ht!]\centering
				
				\caption{Minimum wage effect on different financial leverages.}
				\label{tab: robust_other_fin_lvg}
    					\resizebox{0.65\textheight}{!}{

				\begin{threeparttable}
					\begin{tabular}{l*{3}{c}}
						\hline\hline
						
						&\multicolumn{1}{p{2.5cm}}{\centering Total liabilities/\\total A.}&\multicolumn{1}{p{2.5cm}}{\centering Total debts/\\total A.}&\multicolumn{1}{p{2.5cm}}{\centering (Long-term debts-cash)/\\total A.}\\
                         & (1)&(2)&(3)\\

						\hline
						$Bite*Year_{2011}$ &        0.002         &      -0.006         &      -0.000         \\
						&     (0.006)         &     (0.008)         &     (0.008)         \\
						$Bite*Year_{2012}$ &     -0.004         &      -0.005         &      -0.003         \\
						&     (0.005)         &     (0.007)         &     (0.007)         \\[1ex] 
						$Bite*Year_{2013}$ &         \multicolumn{3}{c}{Reference} \\[1ex] 
						$Bite*Year_{2014}$ &     -0.012\sym{*}  &      -0.012         &      -0.016\sym{*}  \\
						&     (0.005)         &     (0.007)         &     (0.007)         \\
						$Bite*Year_{2015}$ &     -0.042\sym{***}&      -0.025\sym{***}  &      -0.034\sym{***}\\
						&     (0.006)         &     (0.007)         &     (0.008)         \\
						$Bite*Year_{2016}$ &   -0.056\sym{***}&      -0.027\sym{**} &      -0.038\sym{***}\\
						&     (0.007)         &     (0.008)         &     (0.009)         \\
						$Bite*Year_{2017}$ &     -0.066\sym{***}&      -0.040\sym{***}&      -0.043\sym{***}\\
						&     (0.007)         &     (0.008)         &     (0.009)         \\
						$Bite*Year_{2018}$ &      -0.086\sym{***}&      -0.033\sym{*}  &      -0.053\sym{***}\\
						&     (0.008)         &     (0.009)         &     (0.011)         \\
						$Constant$            &       0.618\sym{***}&       0.145\sym{***}&       0.365\sym{***}\\
						&     (0.000)         &     (0.000)         &     (0.000)         \\
						\hline
						Observations        &      184,702         &      184,702         &      184,702         \\
						
						\hline\hline
					\end{tabular}
					
					\begin{tablenotes}
						\item \textit{Notes:} Difference-in-differences regressions. The dependent variables are displayed above each column. A predetermined bite-specific trend is subtracted in all regressions. Firm fixed effects, county-year, and industry-year fixed effects are controlled. Firms are assigned to the county where their largest establishment is located. Industries are categorized with a two-digit industry code. Firm-level clustered standard errors are in parentheses. \sym{*}, \sym{**}, and \sym{***} denote statistical significance at 5\%, 1\% and 0.1\%, respectively.
						\\\textit{Data}: Linked data of BeH, BHP, and Amadeus, 2011-2018.
					\end{tablenotes}
				\end{threeparttable}
				}
			\end{table}
			\clearpage			%================================================================================================================

				%===========================================

			\subsection{Alternative sample restrictions}\label{app: robust_other_restriction}
				\begin{table}[ht!]\centering
					
					\caption{Minimum wage effect on financial leverage and labor costs,\\additional sample restriction: drop if the linking rate$<$0.3}
					\label{tab:app_link_rate}
					\begin{threeparttable}
						\begin{tabular}{l*{4}{c}}
							\hline\hline
							&\multicolumn{1}{p{3.5cm}}{\centering Financial\\leverage}&\multicolumn{1}{p{3.5cm}}{\centering Labor \\share}&\multicolumn{1}{p{3.5cm}}{\centering Log total\\labor costs}&\multicolumn{1}{p{3.5cm}}{\centering Log labor\\costs/worker}\\
       & (1)&(2)&(3)&(4)\\
							\hline
							$Bite*Year_{2011}$ &       0.002         &      -0.001         &       0.031         &       0.016         \\
							&     (0.007)         &     (0.013)         &     (0.021)         &     (0.014)         \\
							$Bite*Year_{2012}$&      -0.003         &      0.002         &       0.010         &       0.019         \\
							&     (0.005)         &     (0.012)         &     (0.017)         &     (0.011)         \\[1ex] 
							$Bite*Year_{2013}$&         \multicolumn{4}{c}{Reference} \\[1ex] 
							$Bite*Year_{2014}$&      -0.015\sym{**} &       0.031\sym{*}  &       0.067\sym{**}&       0.066\sym{***}\\
							&     (0.005)         &     (0.013)         &     (0.021)         &     (0.014)         \\
							$Bite*Year_{2015}$&      -0.049\sym{***}&       0.054\sym{***}&       0.184\sym{***}&       0.198\sym{***}\\
							&     (0.006)         &     (0.013)         &     (0.023)         &     (0.016)         \\
							
							$Bite*Year_{2016}$ &      -0.070\sym{***}&       0.115\sym{***}&       0.184\sym{***}&       0.198\sym{***}\\
							&     (0.007)         &     (0.013)         &     (0.026)         &     (0.016)         \\
							
							$Bite*Year_{2017}$&      -0.077\sym{***}&       0.145\sym{***}&       0.170\sym{***}&       0.224\sym{***}\\
							&     (0.008)         &     (0.014)         &     (0.029)         &     (0.016)         \\
							$Bite*Year_{2018}$&      -0.097\sym{***}&       0.174\sym{***}&       0.160\sym{***}&       0.247\sym{***}\\
							&     (0.009)         &     (0.014)         &     (0.030)         &     (0.017)         \\
							$Constant$            &       0.497\sym{***}&       0.674\sym{***}&      14.802\sym{***}&      10.387\sym{***}\\
							&     (0.000)         &     (0.001)         &     (0.001)         &     (0.001)         \\
							\hline
							Observations        &      178,745         &      178,745         &      178,745         &      178,745         \\
							
							\hline\hline
							
						\end{tabular}

						\begin{tablenotes}
							\item \textit{Notes:} Difference-in-differences regressions. The dependent variables are displayed above each column. A predetermined gap-specific trend is subtracted in all regressions. Firm fixed effects, county-year, and industry-year fixed effects are controlled. Firms are assigned to the county where their largest establishment is located. Industries are categorized with a two-digit industry code. Firm-level clustered standard errors are in parentheses. \sym{*}, \sym{**}, and \sym{***} denote statistical significance at 5\%, 1\% and 0.1\%, respectively.
							\\\textit{Data}: Linked data of BeH, BHP, and Amadeus, 2011-2018.
						\end{tablenotes}
					\end{threeparttable}

				\end{table}

   %================================================================================================================

				\begin{table}[ht!]\centering
					\caption{Minimum wage effect on financial leverage and labor costs,\\no wage imputation}
					\label{tab:app_non_impute}
					
					\begin{threeparttable}
						\begin{tabular}{l*{3}{c}}
							\hline\hline
							&\multicolumn{1}{p{2.5cm}}{\centering Labor \\share}&\multicolumn{1}{p{2.5cm}}{\centering Log total\\labor costs}&\multicolumn{1}{p{2.5cm}}{\centering Log labor\\costs/worker}\\
       & (1)&(2)&(3)\\
							
							%&\multicolumn{1}{c}{Labor share}&\multicolumn{1}{c}{Labor share(BvD)}&\multicolumn{1}{c}{Log total labor costs}&\multicolumn{1}{c}{Log labor costs per worker}\\
							\hline
							$Bite*Year_{2011}$ &      -0.003         &      0.031         &       0.017            \\
							&     (0.014)         &     (0.021)         &     (0.013)                \\
							
							$Bite*Year_{2012}$ &   0.005         &      0.003         &       0.016               \\
							&     (0.013)         &     (0.018)         &     (0.011)                 \\[1ex] 
							
							$Bite*Year_{2013}$ &         \multicolumn{3}{c}{Reference} \\[1ex] 
							
							$Bite*Year_{2014}$ &       0.026  &       0.055\sym{**}  &       0.062\sym{***}      \\
							&     (0.014)         &     (0.020)         &     (0.014)           \\
							
							$Bite*Year_{2015}$ &        0.055\sym{***}&       0.105\sym{***}&       0.157\sym{***}\\
							&      (0.014)         &     (0.023)         &     (0.015)             \\
							
							$Bite*Year_{2016}$ &      0.113\sym{***}&       0.159\sym{***}&             0.204\sym{***}\\
							&     (0.013)         &     (0.026)         &     (0.016)               \\
							
							$Bite*Year_{2017}$ &       0.138\sym{***}&       0.161\sym{***}&           0.249\sym{***}\\
							&     (0.014)         &     (0.030)         &     (0.016)               \\
							
							$Bite*Year_{2018}$ &       0.161\sym{***}&       0.142\sym{***}&          0.257\sym{***}\\
							&     (0.015)         &     (0.031)         &     (0.016)                \\
							
							$Constant$            &       0.653\sym{***}&       14.705\sym{***}&         10.313\sym{***}\\
							&     (0.001)         &     (0.001)         &     (0.001)             \\

							\hline
							Observations        &      184,702         &        184,702          &      184,702            \\
							\hline\hline
						\end{tabular}
						\begin{tablenotes}
							\item \textit{Notes:} Difference-in-differences regressions. The dependent variables are displayed above each column. A predetermined bite-specific trend is subtracted in all regressions. Firm fixed effects, county-year, and industry-year fixed effects are controlled. Firms are assigned to the county where their largest establishment is located. Industries are categorized with a two-digit industry code. Firm-level clustered standard errors are in parentheses. \sym{*}, \sym{**}, and \sym{***} denote statistical significance at 5\%, 1\% and 0.1\%, respectively.
							\\\textit{Data}: Linked data of BeH, BHP, and Amadeus, 2011-2018.
						\end{tablenotes}
					\end{threeparttable}
				\end{table}
				
			\end{landscape}

\clearpage

			\begin{table}[ht!]\centering
				
				\caption{Minimum wage bite and firm exit}
				\label{tab:app_exit}
				
				\begin{threeparttable}
					\begin{tabular}{l*{2}{c}}
						\hline\hline
						&\multicolumn{1}{c}{Exit (=1)}&\multicolumn{1}{c}{Exit (=1)(without covariates)}\\
      & (1)&(2)\\
						\hline
						$Bite$              &       0.046\sym{**} &       0.078\sym{***}\\
						&     (0.015)         &     (0.015)         \\
						
						$Avg. \ fin.\ leverage$        &       0.020\sym{***}  &       0.042\sym{***}\\
						&     (0.006)         &     (0.005)         \\
						$Avg. \ fin.\ leverage*Bite$   &      -0.0  -0.048\sym{*}  &      -0.070\sym{**}         \\
						&     (0.023)         &     (0.023)         \\
						$Avg.\ cash$            &      -0.034\sym{***}&                     \\
						&     (0.009)         &                     \\
						$Avg. \ ROA $           &      -0.083\sym{***}&                     \\
						&     (0.011)         &                     \\
						$Avg. \ total \ assets$   &      -0.003\sym{**} &                     \\
						&     (0.001)         &                     \\
						$Avg.\  log \ empl.$       &      -0.016\sym{***}&                     \\
						&     (0.001)         &                     \\
						$Constant$            &       0.154\sym{***}&       0.013\sym{***}\\
						&     (0.018)         &     (0.003)         \\
						\hline
						Observations        &   26,033         &   26,033       \\
						\hline\hline
						
					\end{tabular}				
					\begin{tablenotes}
						\item \textit{Notes:}  The dependent variable for both columns is a dummy variable indicating whether all of a firm's establishments exit the market between 2016 and 2020. County and industry fixed effects are controlled. Firms are assigned to the county where their largest establishment is located. Industries are categorized with a two-digit industry code.  \sym{*}, \sym{**}, and \sym{***} denote statistical significance at 5\%, 1\% and 0.1\%, respectively.
						\\\textit{Data}: Linked data of BeH, BHP, and Amadeus, 2011-2018.
					\end{tablenotes}
				\end{threeparttable}
				
			\end{table}
			\clearpage
			%===========================================
			\begin{landscape}
				\clearpage
				\section{Further analysis on mechanism}\label{app:further}
				\renewcommand*{\thefigure}{\thesection\arabic{figure}}
				\renewcommand*{\thetable}{\thesection\arabic{table}}
				\setcounter{figure}{0}
				\setcounter{table}{0}

				\subsection{Other measures of profitability}\label{app:further_profit}
				\begin{table}[ht!]\centering
					
					\caption{Minimum wage effect on profitability}
					\label{tab:app_profit}
					
					\begin{threeparttable}
						\begin{tabular}{l*{3}{c}}
							\hline\hline
							&\multicolumn{1}{c}{EBITDA/Revenue}&\multicolumn{1}{c}{EBIT/Revenue}&\multicolumn{1}{c}{Net Income/Revenue}\\
             & (1)&(2)&(3)\\

							\hline
							$Bite*Post$           &      -0.032\sym{***}&      -0.026\sym{***}&      -0.023\sym{***}\\
							&     (0.004)         &     (0.004)         &     (0.004)               \\
							$Bite*Year_{2014}$      &      -0.008         &      -0.004         &      -0.0046        \\
							&     (0.004)         &     (0.004)         &     (0.004)             \\[1ex] 
							$Bite*Pre$&       \multicolumn{3}{c}{Reference} \\[1ex] 
							$Constant$            &       0.093\sym{***}&       0.052\sym{***}&       0.031\sym{***}\\
							&     (0.000)         &     (0.000)         &     (0.000)              \\
							\hline
							Observations        &       91,252         &       91,252         &       79,265               \\
							\hline\hline
						\end{tabular}
						\begin{tablenotes}
							\item \textit{Notes:} Difference-in-differences regressions.  The dependent variables are displayed above each column. A predetermined average gap-specific trend is subtracted in all regressions. Firm fixed effects, county-year, and industry-year fixed effects are controlled. Firms are assigned to the county where their largest establishment is located. Industries are categorized with a two-digit industry code. Firm-level clustered standard errors are in parentheses.  \sym{*}, \sym{**}, and \sym{***} denote statistical significance at 5\%, 1\% and 0.1\%, respectively.
							\\\textit{Data}: Linked data of BeH, BHP, and Amadeus, 2011-2018.
						\end{tablenotes}
					\end{threeparttable}
					
				\end{table}
			\end{landscape}		
			
			\clearpage

   \subsection{Employment regression at the regional level}\label{app:empl_region}

  The table below presents the regional-level employment regression results. The regression equation is:

   	\begin{equation}\label{eq:empl_region}
			y_{rt} = \delta_0 + \delta_1 * Bite_{r} * Post_{r,t} + \delta_2 * Bite_{r} * Year_{r,2014} + \phi * Bite_{r}  + \epsilon_{rt}, \notag
		\end{equation}
   
   where $y_{rt}$ denotes the log employment aggregated at the county level, the $Bite$ variable is defined as the share of workers earning less than 8.5\ \euro \ per hour in 2013 within a county $r$. Additionally, A predetermined bite-specific trend is subtracted from the dependent variable.
The results, consistent with other studies on the German minimum wage, indicate that the minimum wage policy did not reduce employment at the regional level.
   \begin{table}[ht!]\centering
					
					\caption{Minimum wage effect on employment}
					\label{tab:app_empl_region}
					
					\begin{threeparttable}
						\begin{tabular}{l*{1}{c}}
							\hline\hline
							&\multicolumn{1}{c}{Log employment}\\
                    & (1)\\

							\hline
							$Bite*Post$           &      0.477\\
							&     (0.245)            \\
							$Bite*Year_{2014}$      &      0.095          \\
							&     (0.371)                  \\[1ex] 
							$Bite*Pre$&       \multicolumn{1}{c}{Reference} \\[1ex] 
                                $Bite$ & -3.086\sym{***} \\
                                &(0.364)\\
							$Constant$            &       9.096\sym{***}\\
							&     (0.043)                 \\
							\hline
							Observations        &       3,200                \\
							\hline\hline
						\end{tabular}
						\begin{tablenotes}
							\item \textit{Notes:} Difference-in-differences regressions.  The dependent variable is log employment at the regional level. \sym{*}, \sym{**}, and \sym{***} denote statistical significance at 5\%, 1\% and 0.1\%, respectively.
							\\\textit{Data}: Linked data of BeH, BHP, and Amadeus, 2011-2018.
						\end{tablenotes}
					\end{threeparttable}
					
				\end{table}
    \clearpage
		 \subsection{Minimum wage effect on financial leverage, weighted regression}\label{app:weighted}

The following regression is a weighted regression. The weighting factor is derived from the stratification of three size groups and 16 sectors, consistent with the cells in Appendix Figure  \ref{app:fig_size_cell}. It is calculated by dividing the number of firms in the population within each cell by the number of firms in the sample for the corresponding cell. The population data, showing the firm distribution in 2019, are sourced from the German Business Register and extracted from the \href{https://www.destatis.de/EN/Themes/Economic-Sectors-Enterprises/Enterprises/Business-Register/_node.html265826}{GENESIS-Online database}. Since data for 2018 are not available in the GENESIS database, the closest available year, 2019, is used.

   \begin{table}[ht!]\centering
					
					\caption{Minimum wage effect on financial leverage, weighted regression}
					\label{app: weighted_reg}
										\begin{threeparttable}
						\begin{tabular}{l*{1}{c}}
							\hline\hline
							&\multicolumn{1}{c}{Financial leverage}\\
                    & (1)\\

									\hline
							$Bite*Year_{2011}$ &      0.032     \\
							&     (0.023)                        \\							
							$Bite*Year_{2012}$ &   -0.023                     \\
							&     (0.0242)                        \\[1ex] 							
							$Bite*Year_{2013}$ &         \multicolumn{1}{c}{Reference} \\[1ex] 
							
							$Bite*Year_{2014}$ &       -0.057\sym{**}            \\
							&     (0.015)                  \\
							$Bite*Year_{2015}$ &        -0.132\sym{***}    \\
							&     (0.019)                        \\					
							$Bite*Year_{2016}$ &      -0.190\sym{***}\\
							&     (0.024)                 \\							
							$Bite*Year_{2017}$ &       -0.252\sym{***}   \\
							&     (0.025)                    \\							
							$Bite*Year_{2018}$ &       -0.333\sym{***}    \\
							&     (0.031)                      \\
							$Constant$            &       0.482\sym{***}  \\
							&     (0.002)                 \\

							\hline
							Observations        &      184,702            \\
							\hline\hline
						\end{tabular}
						\begin{tablenotes}
							\item \textit{Notes:} Weighted difference-in-differences regressions. The dependent variable is the firm's financial leverage. Firm fixed effects, county-year, and industry-year fixed effects are controlled. Firms are assigned to the county where their largest establishment is located. Industries are categorized with a two-digit industry code. Firm-level clustered standard errors are in parentheses. \sym{*}, \sym{**}, and \sym{***} denote statistical significance at 5\%, 1\% and 0.1\%, respectively.
						\\\textit{Data}: Linked data of BeH, BHP, and Amadeus, 2011-2018.
						\end{tablenotes}
					\end{threeparttable}
					
				\end{table}
    \clearpage	
%=================================================================
\subsection{Descriptive statistics for subsamples}

\begin{table}[ht!]\centering			
					\caption{Descriptive statistics for subsamples}
					\label{tab:summary_subsample}
						\begin{threeparttable}
							\begin{tabular}{l*{5}{c}}
								\hline\hline
				&\multicolumn{1}{c}{High OS firms} 	&\multicolumn{1}{c}{Low OS firms}
	&\multicolumn{1}{c}{Firm size$<$50}   	&\multicolumn{1}{c}{Firm size: 50-249} &\multicolumn{1}{c}{Firm size$>=$250}\\
  & (1)&(2)&(3)&(4)&(5)
         \\
								
								\hline
								\hline
								Financial leverage &      0.511              &   0.490         & 0.540            &   0.503         &     0.439    \\  &  (0.270)       & (0.271)&(0.287) &(0.265)&        (0.252)    \\
								
								Bite  &    0.115   &  0.063    & 0.114    &   0.085    &   0.091   \\&   (0.178)    &    (0.140)   &(0.206) &   (0.144)   &     (0.180) \\
							
								\hline
								Observations        &91,586        &81,608&39,280&93,670                 &19,274                  \\
								\hline\hline
							\end{tabular}					
							\begin{tablenotes}						
								\textit{Notes}: The table above presents the mean and standard deviation (in parentheses) of pre-policy financial leverage and bite. Column (1) represents the subsample of firms with a higher share of outsourceable occupations. Column (2) represents the subsample of firms with a lower share of outsourceable occupations. Columns (3) and (4) represent subsamples of firms based on different firm sizes. \\
								\textit{Data}: Linked data of BeH, BHP, and Amadeus, 2011-2018.
							\end{tablenotes}
						\end{threeparttable}
					
				\end{table}

    \clearpage
%=================================================================

\subsection{Nonlinear effects of the minimum wage on firms' financial leverage \label{app_non_linear}}
Appendix Table \ref{tab:summary_subsample} indicates that the mean of the bite variable is significantly higher for firms with a larger share of outsourceable occupations. To rule out the possibility that different bite levels cause heterogeneous effects of the minimum wage on firms with varying abilities to adjust their labor force, I examine the nonlinear effects of the minimum wage on firms' financial leverage. The following regression analyzes the impact of the minimum wage on firms' financial leverage, adding a quadratic term of bite and interaction terms for the period and the quadratic bite.

A positive coefficient is found for $Bite^2*Post$, which suggests that the higher the bite, the smaller the absolute marginal effect size of the bite on financial leverage. The marginal effect of bite on financial leverage is $-0.101+0.106*Bite$. When inserting the average bite firms with higher or lower outsourceable occupations, the marginal effects are -0.094 and -0.089, respectively. While from Table \ref{tab:os}, the corresponding marginal effects are -0.027 and -0.133. Therefore, the differences in bite level cannot explain the heterogeneous effects of the minimum wage based on the share of outsourceable occupations in firms.
   \begin{table}[ht!]\centering
					
					\caption{Minimum wage effect on financial leverage}
					\label{app:tab_nonlinear}
					\begin{threeparttable}
						\begin{tabular}{l*{1}{c}}
							\hline\hline
							&\multicolumn{1}{c}{Financial leverage}\\\
                    & (1)\\

									\hline
							$Bite*Post$ &      -0.101\sym{***}   \\
							&     (0.015)                        \\							
							$Bite*Year_{2014}$ &   -0.045\sym{**}                     \\
							&     (0.013)                        \\[1ex] 							
							$Bite*Pre$ &         \multicolumn{1}{c}{Reference} \\[1ex] 
							
							$Bite^2*Post$ &       0.053\sym{**}            \\
							&     (0.020)                  \\
							$Bite^2*Year_{2014}$ &        0.042\sym{*}    \\
							&     (0.018)                        \\					
							$Bite^2*Pre$ &      \multicolumn{1}{c}{Reference} \\[1ex] 					
							
							$Constant$            &       0.497\sym{***}  \\
							&     (0.000)                 \\

							\hline
							Observations        &      184,702            \\
							\hline\hline
						\end{tabular}
						\begin{tablenotes}
							\item \textit{Notes:} Difference-in-differences regressions. The dependent variable is the firm's financial leverage. A predetermined bite-specific trend is subtracted. Firm fixed effects, county-year, and industry-year fixed effects are controlled. Firms are assigned to the county where their largest establishment is located. Industries are categorized with a two-digit industry code. Firm-level clustered standard errors are in parentheses. \sym{*}, \sym{**}, and \sym{***} denote statistical significance at 5\%, 1\% and 0.1\%, respectively.
						\\\textit{Data}: Linked data of BeH, BHP, and Amadeus, 2011-2018.
						\end{tablenotes}
					\end{threeparttable}
					
				\end{table}

\clearpage
%=================================================================%====================================================
\subsection{Unconditional quantile regressions on financial leverage}

As presented in Appendix Table \ref{tab:summary_subsample}, the mean of pre-policy financial leverage varies among firms of different sizes. For small firms with fewer than 50 employees, the mean pre-policy financial leverage is 0.54 and is around the 55th percentile of the distribution of financial leverage. For the largest firms with more than 250 employees, it is around the 45th percentile. I conduct unconditional quantile \citep{firpo_unconditional_2009} regressions to investigate different effects along the financial leverage distribution. The regression equation is
	\begin{equation}\label{eq:ucq}
			RIF(y_{jt}, \tau) =\delta_0 + \delta_1 * Bite_{j} * Post_{j,t} + \delta_2 * Bite_{j} * Year_{j,2014} + \phi * Bite_{j}+ \alpha_{j}+ \theta_{c,t} + \lambda_{s,t} + \epsilon_{jt}, \notag
		\end{equation}
   
 where the dependent variable is a recentered influence function (RIF) of financial leverage for different percentiles, and a predetermined bite-specific trend is subtracted from the dependent variable. The following figure illustrates the treatment effects of the minimum wage on the 40th to 60th percentiles of financial leverage using unconditional quantile regressions. The mean of the coefficients for $Bite*Post$ is -0.060 for the 40th to 50th percentiles and -0.042 for the 51st to 60th percentiles. Therefore, the largest treatment effect among small firms is not driven by distributional effects on financial leverage.

		\begin{figure}[ht!]
				\captionabove{Unconditional quantile regressions on financial leverage, the 40th to 60th pecentiles}
            \label{app_ucq}
					\centering
					\includegraphics[width=.8\linewidth]{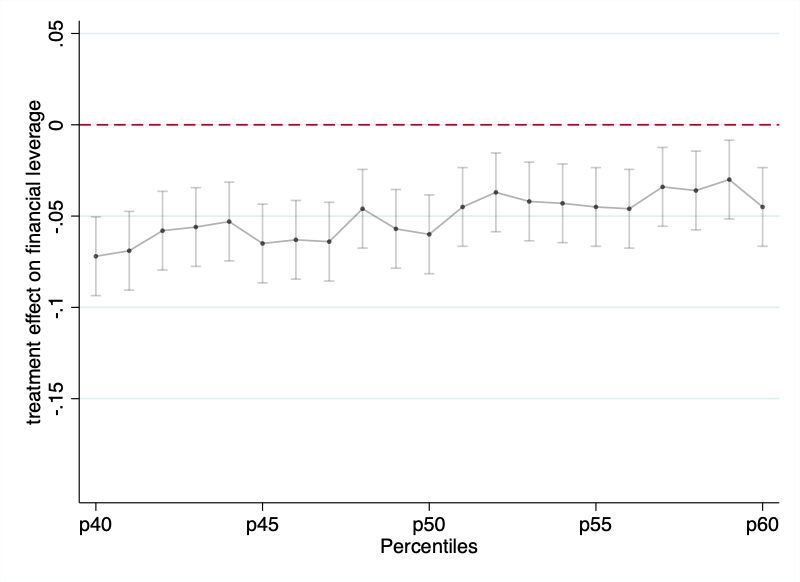}  
				\caption*{\textit{Notes}: The figure displays the detrended RIF difference-in-differences (DiD) regression coefficients of $Bite*Post$ with 95\% confidence intervals.
 			\\	\textit{Data}: Linked data of  BeH, BHP, and Amadeus, 2011-2018.}
			\end{figure}

%=================================================================%====================================================
			\clearpage
			\section{Full regression tables}\label{app:mech_ls}
			\renewcommand*{\thefigure}{\thesection\arabic{figure}}
			\renewcommand*{\thetable}{\thesection\arabic{table}}
			\setcounter{figure}{0}
			\setcounter{table}{0}
			%=================================================================	
			\begin{table}[ht!]\centering
				
				\caption{
					Minimum wage effects on financial leverage and labor share, non-detrended.
					\\Regression table for figure \ref{fig:detect_trend}.
				}
    \label{tab:full_non_detrend}
				\begin{threeparttable}
					\begin{tabular}{l*{2}{c}}
						\hline\hline
						&\multicolumn{1}{c}{Financial leverage}&\multicolumn{1}{c}{Labor share}\\
                   & (1)&(2)\\

						\hline
						$Bite*Year_{2011}$ &           -0.011         &       0.048\sym{***}\\
						&     (0.006)         &     (0.013)         \\
						$Bite*Year_{2012}$ &    -0.010         &       0.024\sym{*}         \\
						&     (0.005)         &     (0.012)         \\[1ex] 
						$Bite*Year_{2013}$ &         \multicolumn{2}{c}{Reference} \\[1ex] 
						$Bite*Year_{2014}$ &   -0.009 &       0.003         \\
						&     (0.005)         &     (0.013)         \\
						$Bite*Year_{2015}$ &      -0.034\sym{***}&      -0.004         \\
						&     (0.006)         &     (0.013)         \\
						$Bite*Year_{2016}$ &    -0.044\sym{***}&       0.030\sym{*}  \\
						&     (0.007)         &     (0.013)         \\
						$Bite*Year_{2017}$ &       -0.046\sym{***}&       0.030\sym{*}           \\
						&     (0.007)         &     (0.013)         \\
						$Bite*Year_{2018}$ &       -0.059\sym{***}&       0.027         \\
						&     (0.009)         &     (0.014)         \\
						$Constant$             &       0.498\sym{***}&       0.660\sym{***}\\
						&     (0.000)         &     (0.001)         \\
						\hline
						Observations         &      184,702         &      184,702         \\
						\hline\hline
						
					\end{tabular}
					
					\begin{tablenotes}
						\item \textit{Notes:} Difference-in-differences regressions. The dependent variables are displayed above each column. A predetermined bite-specific trend is subtracted in all regressions. Firm fixed effects, county-year, and industry-year fixed effects are controlled. Firms are assigned to the county where their largest establishment is located. Industries are categorized with a two-digit industry code. Firm-level clustered standard errors are in parentheses. \sym{*}, \sym{**}, and \sym{***} denote statistical significance at 5\%, 1\% and 0.1\%, respectively.
						\\\textit{Data}: Linked data of BeH, BHP, and Amadeus, 2011-2018.
					\end{tablenotes}
				\end{threeparttable}
				
			\end{table}

			\begin{table}[ht!]\centering
				
				\caption{Minimum wage effects on labor-related outcomes.
					\\Regression table for figure \ref{fig:mech_ls}. }
\label{tab:full_mech_ls}
    \begin{threeparttable}
					\begin{tabular}{l*{4}{c}}
						\hline\hline
						&\multicolumn{1}{p{2.5cm}}{\centering Labor \\share}&\multicolumn{1}{p{2.5cm}}{\centering Labor \\share(BvD)}&\multicolumn{1}{p{2.5cm}}{\centering Log total\\labor costs}&\multicolumn{1}{p{2.5cm}}{\centering Log labor\\costs/worker}\\
						             & (1)&(2)&(3)&(4)\\

						%&\multicolumn{1}{c}{Labor share}&\multicolumn{1}{c}{Labor share(BvD)}&\multicolumn{1}{c}{Log total labor costs}&\multicolumn{1}{c}{Log labor costs per worker}\\
						\hline
						$Bite*Year_{2011}$ &      -0.003         &      0.002         &       0.031         &       0.017         \\
						&     (0.013)         &     (0.008)         &     (0.021)         &     (0.014)         \\
						
						$Bite*Year_{2012}$ &   -0.001         &      -0.004         &       0.004         &       0.018         \\
						&     (0.012)         &     (0.007)         &     (0.018)         &     (0.011)         \\[1ex] 
						
						$Bite*Year_{2013}$ &         \multicolumn{4}{c}{Reference} \\[1ex] 
						
						$Bite*Year_{2014}$ &       0.028\sym{*}  &       0.013\sym{*}  &       0.052\sym{*}  &       0.065\sym{***}\\
						&     (0.013)         &     (0.007)         &     (0.021)         &     (0.014)         \\
						
						$Bite*Year_{2015}$ &        0.047\sym{***}&       0.031\sym{***}&       0.095\sym{***}&       0.151\sym{***}\\
						&     (0.013)         &     (0.008)         &     (0.023)         &     (0.015)         \\
						
						$Bite*Year_{2016}$ &      0.107\sym{***}&       0.050\sym{***}&       0.152\sym{***}&       0.203\sym{***}\\
						&     (0.013)         &     (0.008)         &     (0.026)         &     (0.016)         \\
						
						$Bite*Year_{2017}$ &       0.133\sym{***}&       0.066\sym{***}&       0.148\sym{***}&       0.244\sym{***}\\
						&     (0.013)         &     (0.008)         &     (0.030)         &     (0.016)         \\
						
						$Bite*Year_{2018}$ &       0.155\sym{***}&       0.072\sym{***}&       0.128\sym{***}&       0.258\sym{***}\\
						&     (0.014)         &     (0.009)         &     (0.031)         &     (0.017)         \\
						
						$Constant$            &       0.667\sym{***}&       0.719\sym{***}&      14.776\sym{***}&      10.386\sym{***}\\
						&     (0.001)         &     (0.000)         &     (0.002)         &     (0.001)         \\

						\hline
						Observations        &      184,702         &      160,140         &      184,702         &      184,702         \\
						\hline\hline
					\end{tabular}
					\begin{tablenotes}
						\item \textit{Notes:} Difference-in-differences regressions. The dependent variables are displayed above each column. A predetermined bite-specific trend is subtracted in all regressions. Firm fixed effects, county-year, and industry-year fixed effects are controlled. Firms are assigned to the county where their largest establishment is located. Industries are categorized with a two-digit industry code. Firm-level clustered standard errors are in parentheses. \sym{*}, \sym{**}, and \sym{***} denote statistical significance at 5\%, 1\% and 0.1\%, respectively.
						\\\textit{Data}: Linked data of BeH, BHP, and Amadeus, 2011-2018.
					\end{tablenotes}
				\end{threeparttable}
			\end{table}

			%%%%%%%%%%%%%%%%%%%%%%%%%%%%%%%%%%%%%%%%%%%%%%%%%%%%%%%%%%%%%
			\begin{table}[ht!]\centering
				\caption{Minimum wage effects on log total debts and log assets.\\
					Full regression table for Table \ref{tab:debts_assets}.}
				\label{tab:full_debts_assets}
				\begin{threeparttable}
					\begin{tabular}{l*{5}{c}}
						
						\hline\hline
						&\multicolumn{1}{p{2.5cm}}{\centering Log total\\debts}&\multicolumn{1}{p{2.5cm}}{\centering Log total\\assets}&\multicolumn{1}{p{2.5cm}}{\centering Log fixed\\assets costs}&\multicolumn{1}{p{2.5cm}}{\centering Log curret\\assets}&\multicolumn{1}{p{2.5cm}}{\centering Log \\cash}\\
						             & (1)&(2)&(3)&(4)\\

						\hline
						$Bite*Post$   &       -0.090\sym{**} &       0.067\sym{***}&       0.003         &       0.071\sym{***}&       0.284\sym{***}\\
						&     (0.024)         &     (0.015)         &     (0.028)         &     (0.018)         &     (0.054 )         \\
						
						$Bite*Year_{2014}$      &      -0.037         &      -0.009         &       0.010         &      -0.018         &       0.195\sym{***}\\
						&     (0.022)         &     (0.012)         &     (0.022)         &     (0.016)         &     (0.052 )         \\	[1ex] 
						$Bite*Pre$&         \multicolumn{5}{c}{Reference} \\[1ex] 
						$Constant$           &      15.452\sym{***}&      16.386\sym{***}&      14.776\sym{***}&      15.807\sym{***}&      13.168\sym{***}\\
						&     (0.001)         &     (0.001)         &     (0.001)         &     (0.001)         &     (0.003)         \\
						\hline
						Observations        &      184,702         &      184,702         &      184,702         &      184,702         &      184,702         \\
						
						\hline\hline
					\end{tabular}
					
					\begin{tablenotes}
						\item \textit{Notes:} Difference-in-differences regressions. The dependent variables are displayed above each column. A predetermined bite-specific trend is subtracted in all regressions. Firm fixed effects, county-year, and industry-year fixed effects are controlled. Firms are assigned to the county where their largest establishment is located. Industries are categorized with a two-digit industry code. Firm-level clustered standard errors are in parentheses. \sym{*}, \sym{**}, and \sym{***} denote statistical significance at 5\%, 1\% and 0.1\%, respectively.
						\\\textit{Data}: Linked data of BeH, BHP, and Amadeus, 2011-2018.
					\end{tablenotes}
				\end{threeparttable}
				
			\end{table}
			%%%%%%%%%%%%%%%%%%%%%%%%%%%%%%%%%%%%%%%%%%%%%%%%%%%%%%%%%%%%%

			\begin{table}[ht!]\centering
				
				\caption{Minimum wage effects on labor-related outcomes and log EBITDA.\\
					table for Table \ref{tab:bd_labor_share}.}
				\label{tab:full_value_added}
				\begin{threeparttable}
					\begin{tabular}{l*{6}{c}}
						\hline\hline
						&\multicolumn{1}{p{2cm}}{\centering Log \\employment} &\multicolumn{1}{p{2.2cm}}{\centering Log (fixed\\assets/empl.)} & \multicolumn{1}{p{2cm}}{\centering Log (labor\\costs/empl.)}
						&	\multicolumn{1}{p{2cm}}{\centering Log value\\added} &\multicolumn{1}{p{2cm}}{\centering Log\\EBITDA} &\multicolumn{1}{p{2cm}}{\centering Log total\\labor costs}\\
        & (1)&(2)&(3)&(4)&(5)&(6)\\
						\hline
						
						$Bite*Post$    &    -0.081\sym{***}&       0.063\sym{*} & 0.201\sym{***}  &       0.070\sym{***}&      -0.180\sym{***}&       0.120\sym{***}\\
						&     (0.018)         &     (0.023) & (0.013)			&     (0.017)         &     (0.029)         &     (0.023)         \\
						$Bite*Year_{2014}$       &    -0.014         &      0.013         &       0.057\sym{***}  & -0.006  &-0.036  & 0.043\sym{*} \\
						&  (0.014)          &     (0.022)         &   (0.013)      &(0.016) &(0.029)& (0.019)     \\[1ex] 
						$Bite*Pre$&       \multicolumn{6}{c}{Reference} \\[1ex] 
						$Constant$            &     4.391\sym{***}&    10.387\sym{***}&      10.381\sym{***} &15.346\sym{***} &14.187\sym{***}&14.77\sym{***}   \\
						&     (0.001)         &     (0.001)         &     (0.001)    & (0.001)  & (0.002)  & (0.001)      \\
						\hline
						Observations        &          184,702   &          184,702     &      184,702    &      189,045         &      169,645         &      184,702         \\
						
						\hline\hline
						
					\end{tabular}
					\begin{tablenotes}
						\item \textit{Notes:} Difference-in-differences regressions. The dependent variables are displayed above each column.  A predetermined bite-specific trend is subtracted in all regressions. Firm fixed effects, county-year, and industry-year fixed effects are controlled. Firms are assigned to the county where their largest establishment is located. Industries are categorized with a two-digit industry code. Firm-level clustered standard errors are in parentheses. \sym{*}, \sym{**}, and \sym{***} denote statistical significance at 5\%, 1\% and 0.1\%, respectively.
						\\\textit{Data}: Linked data of BeH, BHP, and Amadeus, 2011-2018.
					\end{tablenotes}
				\end{threeparttable}
			\end{table}

			%%%%%%%%%%%%%%%%%%%%%%%%%%%%%%%%%%%%%%%%%%%%%%%%%%%%%%%%%%%%%
			
			\begin{table}[ht!]\centering
				
				\caption{Minimum wage effects on log tangible assets and profitability.\\
					Full regression table for Table \ref{tab:other}.}
				\label{tab:full_other}
				\begin{threeparttable}
					\begin{tabular}{l*{4}{c}}
						\hline\hline
						&	\multicolumn{1}{p{3cm}}{\centering Log tangible\\assets} &	\multicolumn{1}{p{3cm}}{\centering EBITDA\\/ Total A.}  &	\multicolumn{1}{p{3cm}}{\centering EBIT\\/Total A.}	&\multicolumn{1}{p{3cm}}{\centering Net Income\\/Total A.} \\
        & (1)&(2)&(3)&(4)\\
						\hline
						
						$Bite*Post$                       &     -0.036        &      -0.036\sym{***}&      -0.034\sym{***}&      -0.027\sym{***}\\
					&     (0.030)         &     (0.004)         &     (0.004)         &     (0.003)         \\
						$Bite*Year_{2014}$      &      -0.006         &      -0.010\sym{*}  &      -0.011\sym{**} &      -0.008\sym{*}  \\
						&     (0.024)         &     (0.004)         &     (0.004)         &     (0.004)         \\[1ex] 
						$Bite*Pre$&       \multicolumn{4}{c}{Reference} \\[1ex] 
						$Constant$                     &           14.364\sym{***}&       0.128\sym{***}&       0.083\sym{***}&       0.052\sym{***}\\
						&     (0.002)         &     (0.000)         &     (0.000)         &     (0.000)         \\
						\hline
						Observations        &      184,702         &      184,702         &      184,702         &      173,208         \\
						
						\hline\hline
						
					\end{tabular}
					\begin{tablenotes}
						\item \textit{Notes:} Difference-in-differences regressions. The dependent variables are displayed above each column. A predetermined bite-specific trend is subtracted in all regressions. Firm fixed effects, county-year, and industry-year fixed effects are controlled. Firms are assigned to the county where their largest establishment is located. Industries are categorized with a two-digit industry code. Firm-level clustered standard errors are in parentheses. \sym{*}, \sym{**}, and \sym{***} denote statistical significance at 5\%, 1\% and 0.1\%, respectively.
						\\\textit{Data}: Linked data of BeH, BHP, and Amadeus, 2011-2018.
					\end{tablenotes}
				\end{threeparttable}
				
			\end{table}
			
			%%%%%%%%%%%%%%%%%%%%%%%%%%%%%%%%%%%%%%%%%%%%%%%%%%%%%%%%%%%%%
			\begin{table}[ht!]\centering
				
				\caption{Minimum wage effects on long/short term liabilities.\\
					Full regression table for Table \ref{tab:long}.}
				\label{tab:full_long}
				\begin{threeparttable}
					\begin{tabular}{l*{2}{c}}
						\hline\hline
						&	\multicolumn{1}{p{5.5cm}}{\centering Log long-term debts} &\multicolumn{1}{p{5.5cm}}{\centering Log short-term liabilities}\\
        & (1)&(2)\\
						\hline
						$Bite*Post$          &      -0.727\sym{**} &      0.113         \\
					&     (0.180)         &     (0.109)         \\
					
						$Bite*Year_{2014}$     &      -0.283         &       0.155         \\
						&     (0.170)         &     (0.127)         \\[1ex] 
						$Bite*Pre$&     \multicolumn{2}{c}{Reference} \\[1ex] 
						$Constant$            &     9.263\sym{***}&      14.533\sym{***}\\
						&     (0.010)         &     (0.006)         \\
					\hline
					Observations        &      184,702         &      184,702         \\

						\hline\hline
						
					\end{tabular}
					\begin{tablenotes}
						\item \textit{Notes:} Difference-in-differences regressions. The dependent variables are displayed above each column. A predetermined bite-specific trend is subtracted in all regressions. Firm fixed effects, county-year, and industry-year fixed effects are controlled. Firms are assigned to the county where their largest establishment is located. Industries are categorized with a two-digit industry code. Firm-level clustered standard errors are in parentheses. \sym{*}, \sym{**}, and \sym{***} denote statistical significance at 5\%, 1\% and 0.1\%, respectively.
						\\\textit{Data}: Linked data of BeH, BHP, and Amadeus, 2011-2018.
					\end{tablenotes}
				\end{threeparttable}
				
			\end{table}
			
			%%%%%%%%%%%%%%%%%%%%%%%%%%%%%%%%%%%%%%%%%%%%%%%%%%%%%%%%%%%%%
			\begin{table}[ht!]\centering
				
				\caption{Minimum wage effects on financial leverage and labor share.\\
					Full regression table for Table \ref{tab:os}.}
				\label{tab:full_os}
				\begin{threeparttable}
					\begin{tabular}{l*{4}{c}}
						\hline\hline
						
						&\multicolumn{2}{c}{Higher share of OS jobs}&\multicolumn{2}{c}{Lower share of OS jobs}\\
						\cmidrule(lr){2-3} \cmidrule(lr){4-5}
						
						&\multicolumn{1}{c}{Financial lvg}&\multicolumn{1}{c}{Labor share}&\multicolumn{1}{c}{Financial lvg}&\multicolumn{1}{c}{Labor share}\\
        & (1)&(2)&(3)&(4)\\
						
						\hline
						$Bite*Post$          &      -0.027\sym{***} &       0.124\sym{***}&      -0.133\sym{***}&       0.109\sym{***}\\
						&     (0.008)         &     (0.012)         &     (0.011)         &     (0.016)       \\
						$Bite*Year_{2014}$              &      -0.007         &       0.036\sym{*}  &      -0.035\sym{***}&       0.040         \\
						&     (0.007)         &     (0.016)         &     (0.009)         &     (0.021)         \\[1ex] 
						$Bite*Pre$&      \multicolumn{4}{c}{Reference} \\[1ex] 
						$Constant$           &       0.504\sym{***}&       0.660\sym{***}&       0.481\sym{***}&       0.665\sym{***}\\
						&     (0.000)         &     (0.001)         &     (0.000)         &     (0.001)         \\
						\hline
						Observations        &       91,586         &       91,586        &       81,608       &       81,608          \\
						\hline\hline

					\end{tabular}
					\begin{tablenotes}
						\item \textit{Notes:} Difference-in-differences regressions. The dependent variables are displayed above each column. The sample is split based on firms' share of outsourceable (OS) jobs. A predetermined bite-specific trend is subtracted in all regressions. Firm fixed effects, county-year, and industry-year fixed effects are controlled. Firms are assigned to the county where their largest establishment is located. Industries are categorized with a two-digit industry code. Firm-level clustered standard errors are in parentheses. \sym{*}, \sym{**}, and \sym{***} denote statistical significance at 5\%, 1\% and 0.1\%, respectively.
						\\\textit{Data}: Linked data of BeH, BHP, and Amadeus, 2011-2018.
					\end{tablenotes}
				\end{threeparttable}
				
			\end{table}

			%%%%%%%%%%%%%%%%%%%%%%%%%%%%%%%%%%%%%%%%%%%%%%%%%%%%%%%%%%%%%
			\begin{table}[ht!]\centering
				\caption{Minimum wage effect on financial leverage and labor share.\\
					Full regression table for Table \ref{tab:size_fi_lvg}.}
				\label{tab:full_size_fi_lvg}
				\begin{threeparttable}
					\begin{tabular}{l*{3}{c}}
						\hline\hline
						&\multicolumn{1}{c}{Firm size: $<$50}&\multicolumn{1}{c}{Firm size: 50-249}&\multicolumn{1}{c}{Firm size: $>=$250}\\
        & (1)&(2)&(3)\\
						\hline
						Financial leverage & &&\\
						\hline
					$Bite*Post$           &      -0.148\sym{***}&      -0.019\sym{*}         &       -0.017         \\
					&     (0.011)         &     (0.009)         &     (0.017)         \\
						$Bite*Year_{2014}$            &      -0.046\sym{***}&       -0.001         &      -0.026         \\
						&     (0.010)         &     (0.007)         &     (0.014)         \\[1ex] 
					$Bite*Pre$&       \multicolumn{3}{c}{Reference} \\[1ex] 
						$Constant$           &       0.519\sym{***}&       0.497\sym{***}&       0.426\sym{***}\\
						&     (0.001)         &     (0.000)         &     (0.001)         \\
					\hline
					Labor share & &&\\
					\hline
					
					\hline
					$Bite*Post$           &       0.184\sym{***}&       0.085\sym{***}&      -0.019         \\
					&     (0.019)         &     (0.012)         &     (0.013)         \\
						$Bite*Year_{2014}$          &       0.068\sym{**} &       0.015         &      -0.036\sym{*}         \\
						&     (0.026)         &     (0.013)         &     (0.014)         \\[1ex] 
						$Bite*Pre$&         \multicolumn{3}{c}{Reference} \\[1ex] 
						$Constant$           &       0.496\sym{***}&       0.711\sym{***}&       0.786\sym{***}\\
					&     (0.001)         &     (0.001)         &     (0.001)         \\
					\hline
					Observations        &       39,280         &       93,670         &       19,274         \\
					
					\hline
					Log EBITDA & &&\\
					\hline
					$Bite*Post$            &      -0.230\sym{***}&      -0.136\sym{**} &       0.118         \\
					&     (0.058)         &     (0.043)         &     (0.092)         \\
					
						$Bite*Year_{2014}$         &      -0.089         &      -0.014         &       0.158         \\
						&     (0.058)         &     (0.042)         &     (0.098)         \\[1ex]
						$Bite*Pre$&           \multicolumn{3}{c}{Reference} \\[1ex] 
						$Constant$            &      13.531\sym{***}&      14.195\sym{***}&      15.259\sym{***}\\
						&     (0.004)         &     (0.002)         &     (0.005)         \\
					\hline
					Observations        &       35,593        &       87,303         &       17,389        \\

						\hline\hline
					\end{tabular}
					\begin{tablenotes}
						\item \textit{Notes:} Difference-in-differences regressions. The dependent variables are displayed above each panel. The sample is split based on firm size categories. A predetermined bite-specific trend is subtracted in all regressions. Firm fixed effects, county-year, and industry-year fixed effects are controlled. Firms are assigned to the county where their largest establishment is located. Industries are categorized with a two-digit industry code. Firm-level clustered standard errors are in parentheses. \sym{*}, \sym{**}, and \sym{***} denote statistical significance at 5\%, 1\% and 0.1\%, respectively.
						\\\textit{Data}: Linked data of BeH, BHP, and Amadeus, 2011-2018.
					\end{tablenotes}
				\end{threeparttable}
			\end{table}

			\clearpage
			
			\section{Coefficients for $Bite_j*Year_{kt}$ for section \ref{sec:further} \label{app:event_graphs}}
			\renewcommand*{\thefigure}{\thesection\arabic{figure}}
			\renewcommand*{\thetable}{\thesection\arabic{table}}
			\setcounter{figure}{0}
			\setcounter{table}{0}		
			\begin{figure}[ht!]
				\captionabove{Coefficients for $Bite_j*Year_{kt}$ for Table \ref{tab:debts_assets}}
    \label{fig:event_bd_fl}
				%1-------------------------------------
				\begin{subfigure}{.5\textwidth}
					\centering
					
					% include the third image
					\includegraphics[height=4cm]{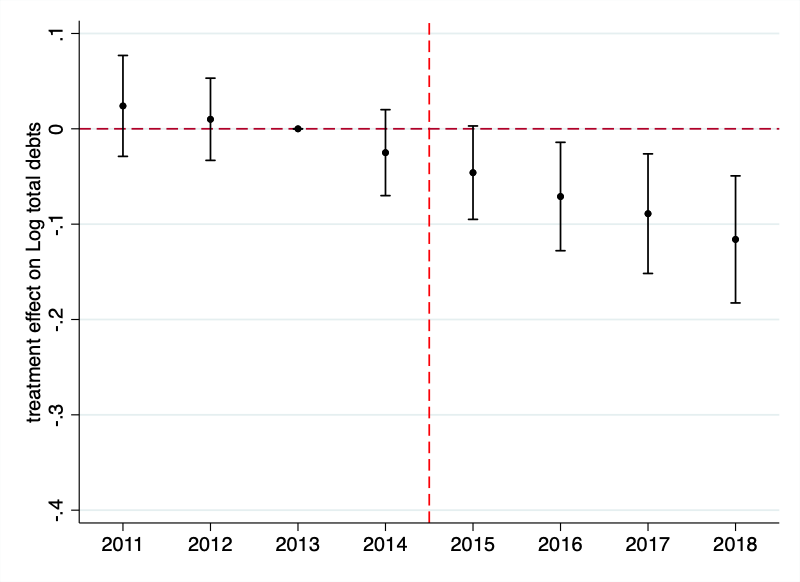}  
					\caption{Log total debts}
					%	\label{fig:pl_mech_labor_lvg_a}
				\end{subfigure}
				%2-------------------------------------		
				\begin{subfigure}{.5\textwidth}
					\centering
					% include the fourth image
					\includegraphics[height=4cm]{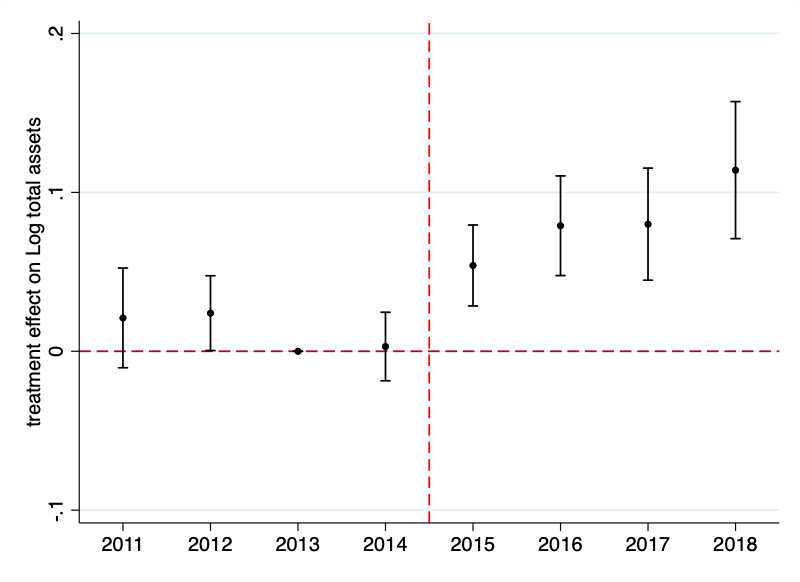}  
					\caption{Log total assets}
					%			\label{fig:pl_mech_labor_lvg_a}
				\end{subfigure}
				%3-------------------------------------
				\begin{subfigure}{.5\textwidth}
					\centering
					% include the fourth image
					\includegraphics[height=4cm]{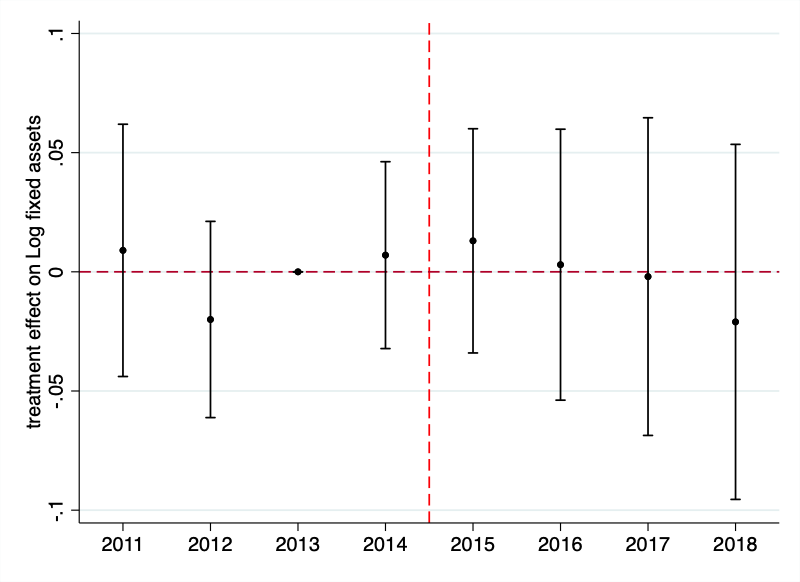}  
					\caption{Log fixed assets}
					%	\label{fig:pl_mech_labor_lvg_a}
				\end{subfigure}
				%4-------------------------------------
				\begin{subfigure}{.5\textwidth}
					\centering
					% include the fourth image
					\includegraphics[height=4cm]{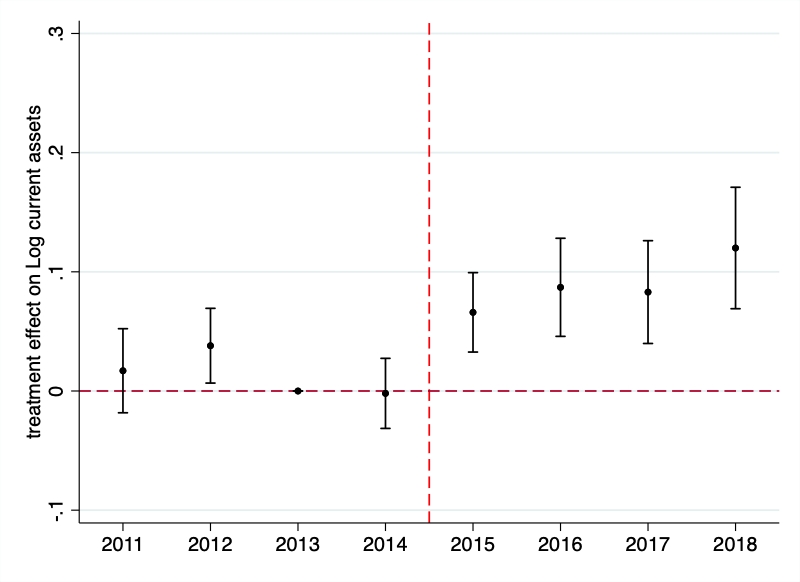}  
					\caption{Log current assets}
					%	\label{fig:pl_mech_labor_lvg_a}
				\end{subfigure}
				%5-------------------------------------			
				\begin{subfigure}{.5\textwidth}
					\centering
					% include the fourth image
					\includegraphics[height=4cm]{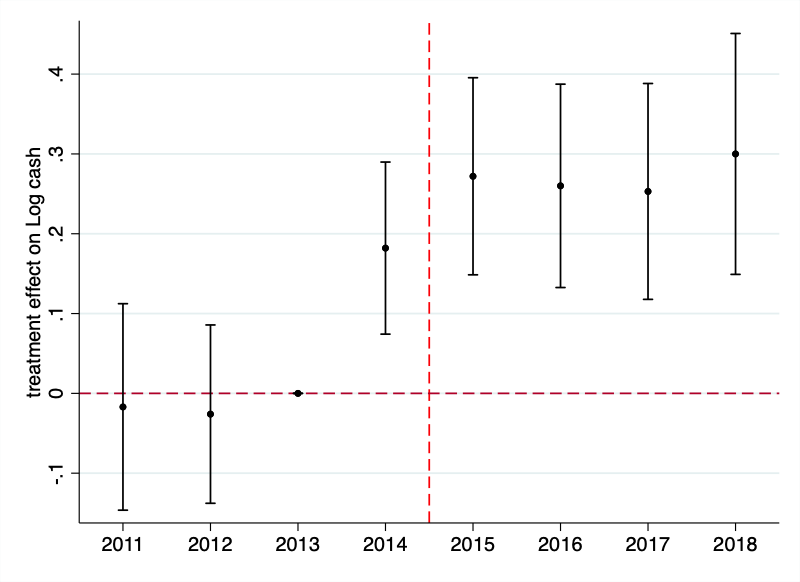}  
					\caption{Log cash}
					%	\label{fig:pl_mech_labor_lvg_a}
				\end{subfigure}
				\caption*{\textit{Notes}: The figure displays the detrended difference-in-differences (DiD) regression coefficients of $Bite_{j} * Year_{k,t}$ with 95\% confidence intervals. The dependent variables are displayed below each figure.
\\					\textit{Data}: Linked data of  BeH, BHP, and Amadeus, 2011-2018.}
			\end{figure}		
			
			\begin{figure}
				\captionabove{Coefficients for $Bite_j*Year_{kt}$ for Table \ref{tab:bd_labor_share}}
        \label{fig:event_bd_ls}

				%1-------------------------------------
				\begin{subfigure}{.5\textwidth}
					\centering
					\includegraphics[width=.8\linewidth]{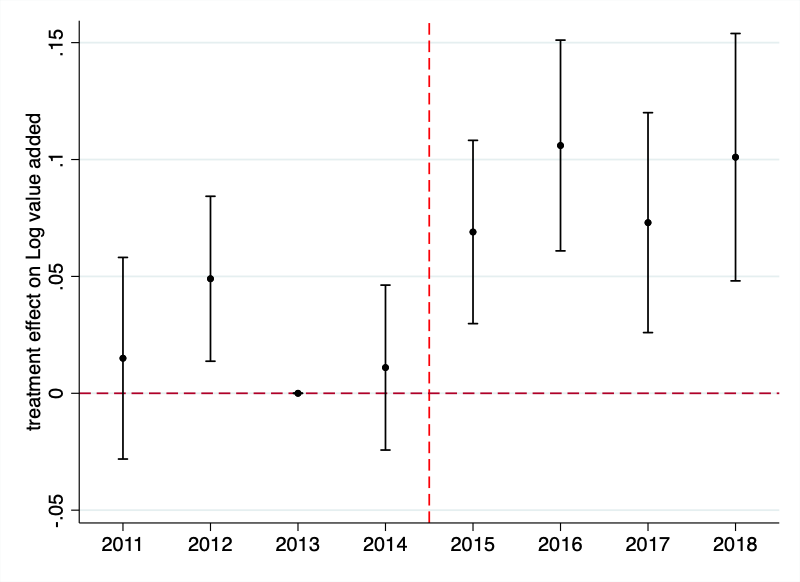}  
					\caption{Log value-added}
					%	\label{fig:pl_mech_labor_lvg_a}
				\end{subfigure}
				%2-------------------------------------		
				\begin{subfigure}{.5\textwidth}
					\centering
					% include fourth image
					\includegraphics[width=.8\linewidth]{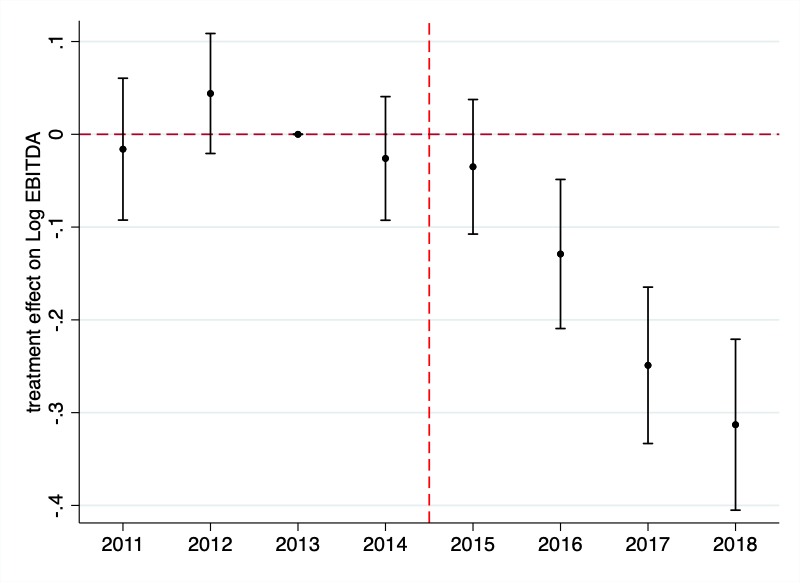}  
					\caption{Log EBITDA}
					%			\label{fig:pl_mech_labor_lvg_a}
				\end{subfigure}
				%3-------------------------------------
				\begin{subfigure}{.5\textwidth}
					\centering
					% include fourth image
					\includegraphics[width=.8\linewidth]{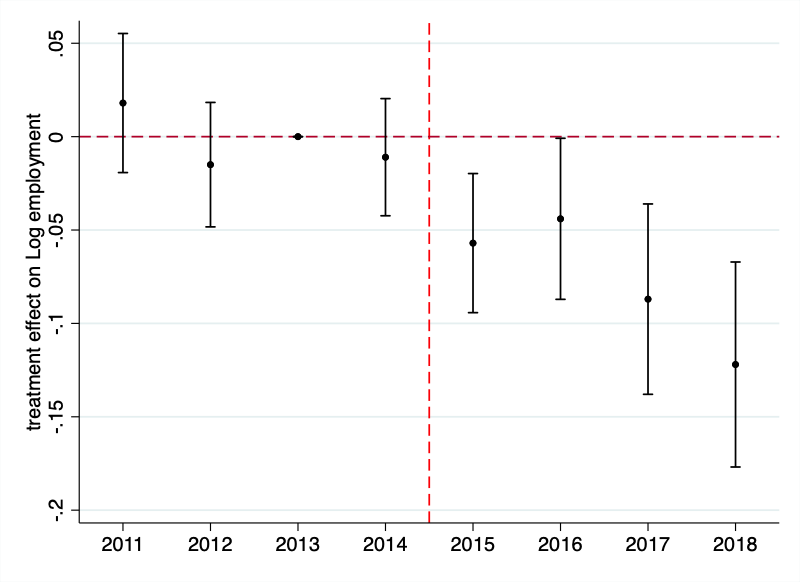}  
					\caption{Log employment (IAB)}
					%	\label{fig:pl_mech_labor_lvg_a}
				\end{subfigure}
				%4-------------------------------------
				\begin{subfigure}{.5\textwidth}
					\centering
					% include fourth image
					\includegraphics[width=.8\linewidth]{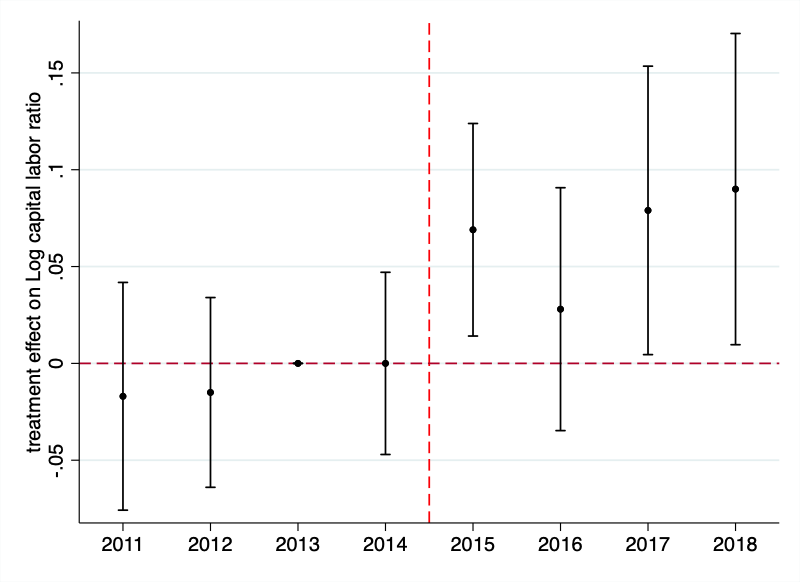}  
					\caption{Log (fixed assets/empl.)}
					%	\label{fig:pl_mech_labor_lvg_a}
				\end{subfigure}

				\caption*{\textit{Notes}: The figure displays the detrended difference-in-differences (DiD) regression coefficients of $Bite_{j} * Year_{k,t}$ with 95\% confidence intervals. The dependent variables are displayed below each figure.
			\\	\textit{Data}: Linked data of  BeH, BHP, and Amadeus, 2011-2018.}
			\end{figure}

			%%%%%%%%%%%%%%%%%%%%%%%%%%%%%%%%%%%%%%%%%%%%%
			%4_7_other channels
			%%%%%%%%%%%%%%%%%%%%%%%%%%%%%%%%%%%%%%%%%%%%%
			\begin{figure}
				\captionabove{Coefficients for $Bite_j*Year_{kt}$ for Table \ref{tab:other}}
        \label{fig:event_other}

				%1-------------------------------------
				\begin{subfigure}{.5\textwidth}
					\centering
					
					% include third image
					\includegraphics[width=.8\linewidth]{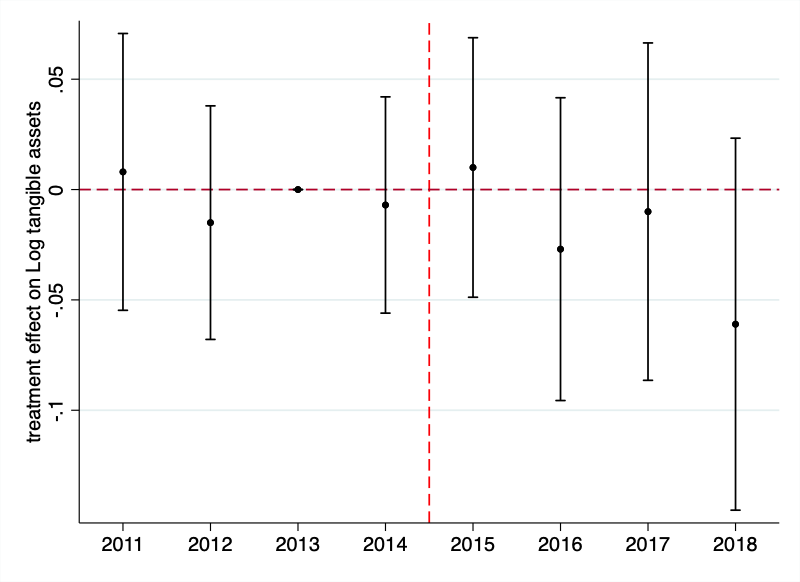}  
					\caption{Log tangible assets}
					%	\label{fig:pl_mech_labor_lvg_a}
				\end{subfigure}
				%2-------------------------------------		
				\begin{subfigure}{.5\textwidth}
					\centering
					% include fourth image
					\includegraphics[width=.8\linewidth]{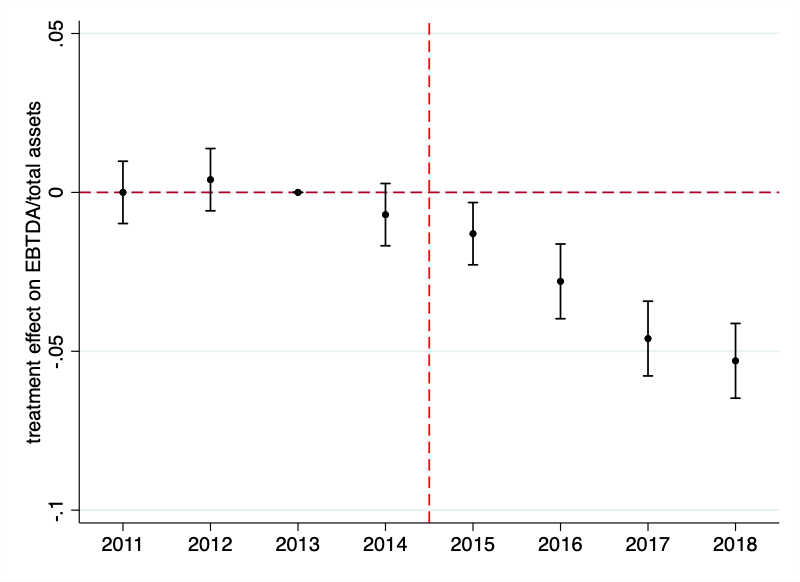}  
					\caption{EBITDA/Total assets}
					%			\label{fig:pl_mech_labor_lvg_a}
				\end{subfigure}
				%3-------------------------------------
				\begin{subfigure}{.5\textwidth}
					\centering
					% include fourth image
					\includegraphics[width=.8\linewidth]{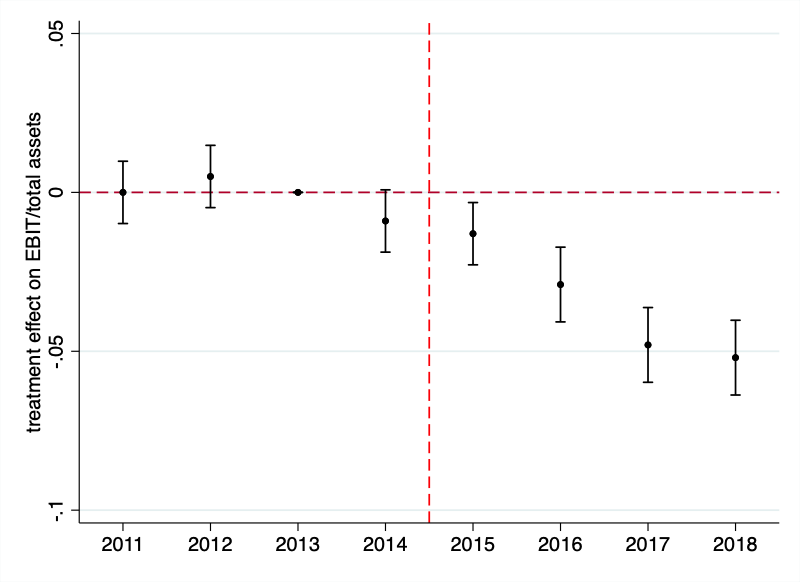}  
					\caption{EBIT/Total assets}
					%	\label{fig:pl_mech_labor_lvg_a}
				\end{subfigure}
				%4-------------------------------------
				\begin{subfigure}{.5\textwidth}
					\centering
					% include fourth image
					\includegraphics[width=.8\linewidth]{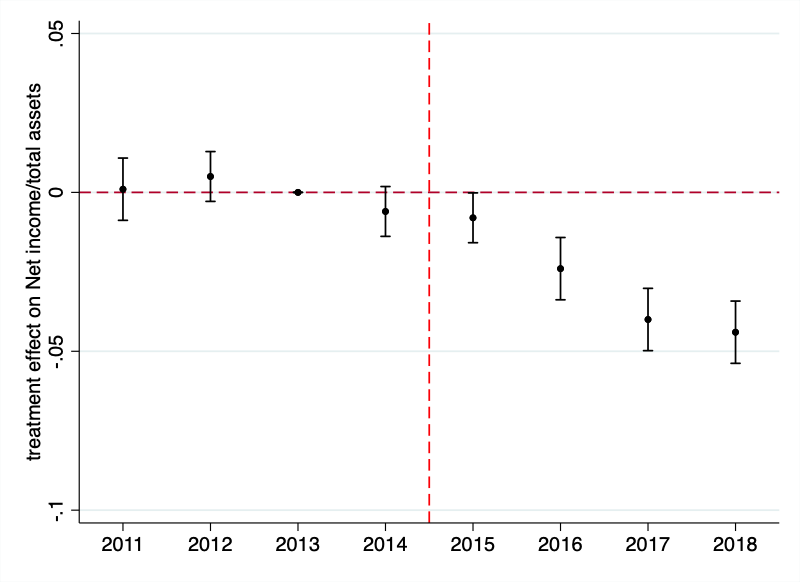}  
					\caption{Net income/Total assets}
					%	\label{fig:pl_mech_labor_lvg_a}
				\end{subfigure}

				\caption*{\textit{Notes}: The figure displays the detrended difference-in-differences (DiD) regression coefficients of $Bite_{j} * Year_{k,t}$ with 95\% confidence intervals. The dependent variables are displayed below each figure.
			\\	\textit{Data}: Linked data of  BeH, BHP, and Amadeus, 2011-2018.}
			\end{figure}

			%%%%%%%%%%%%%%%%%%%%%%%%%%%%%%%%%%%%%%%%%%%%%
			%4_3_long_term_debts
			%%%%%%%%%%%%%%%%%%%%%%%%%%%%%%%%%%%%%%%%%%%%%
			\begin{figure}
				\captionabove{Coefficients for $Bite_j*Year_{kt}$ for Table \ref{tab:long}}
        \label{fig:event_long}

				%1-------------------------------------
				\begin{subfigure}{.5\textwidth}
					\centering
					
					% include third image
					\includegraphics[width=.8\linewidth]{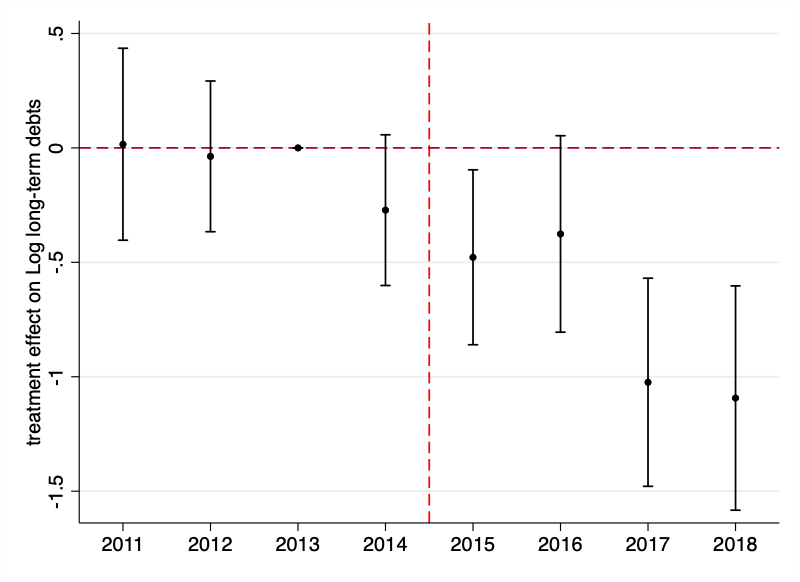}  
					\caption{Log long-term debts}
					%	\label{fig:pl_mech_labor_lvg_a}
				\end{subfigure}
				%2-------------------------------------		
				\begin{subfigure}{.5\textwidth}
					\centering
					% include fourth image
					\includegraphics[width=.8\linewidth]{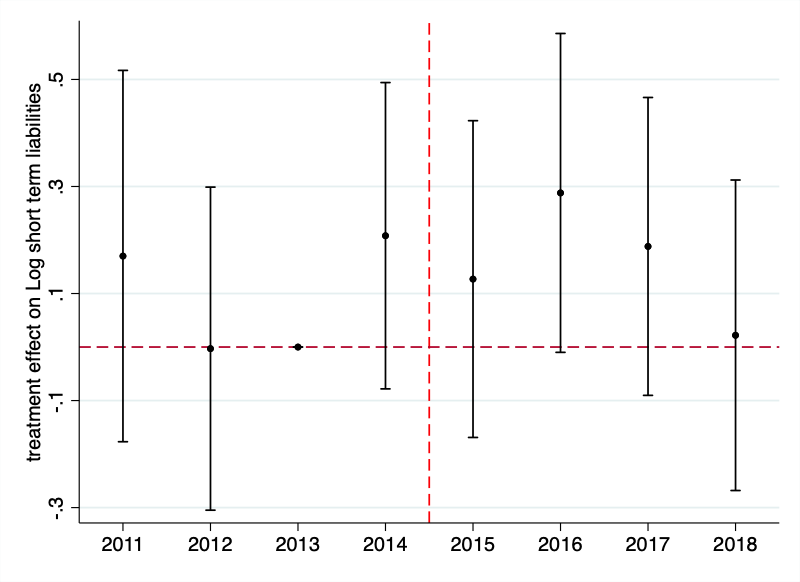}  
					\caption{Log short-term liabilities}
					%			\label{fig:pl_mech_labor_lvg_a}
				\end{subfigure}

				\caption*{\textit{Notes}: The figure displays the detrended difference-in-differences (DiD) regression coefficients of $Bite_{j} * Year_{k,t}$ with 95\% confidence intervals. The dependent variables are displayed below each figure.
									\\	\textit{Data}: Linked data of  BeH, BHP, and Amadeus, 2011-2018.}
			\end{figure}

			%%%%%%%%%%%%%%%%%%%%%%%%%%%%%%%%%%%%%%%%%%%%%
			%4_5_flex_labor
			%%%%%%%%%%%%%%%%%%%%%%%%%%%%%%%%%%%%%%%%%%%%%
			\begin{figure}
				\captionabove{Coefficients for $Bite_j*Year_{kt}$ for Table \ref{tab:os}}
        \label{fig:event_os}

				%1-------------------------------------
				\begin{subfigure}{.5\textwidth}
					\centering
					
					% include third image
					\includegraphics[width=.8\linewidth]{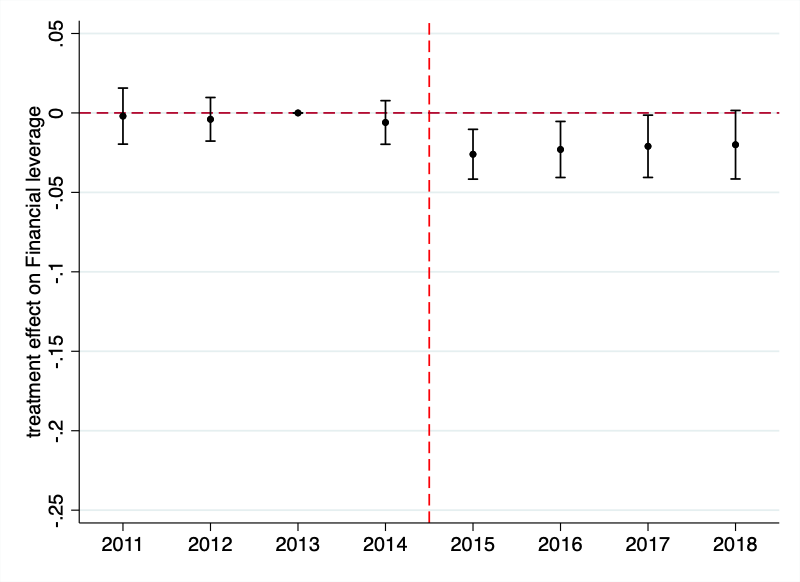}  
					\caption{Financial leverage, \\firms with high share of outsourceable jobs}
					%	\label{fig:pl_mech_labor_lvg_a}
				\end{subfigure}
				%2-------------------------------------		
				\begin{subfigure}{.5\textwidth}
					\centering
					% include fourth image
					\includegraphics[width=.8\linewidth]{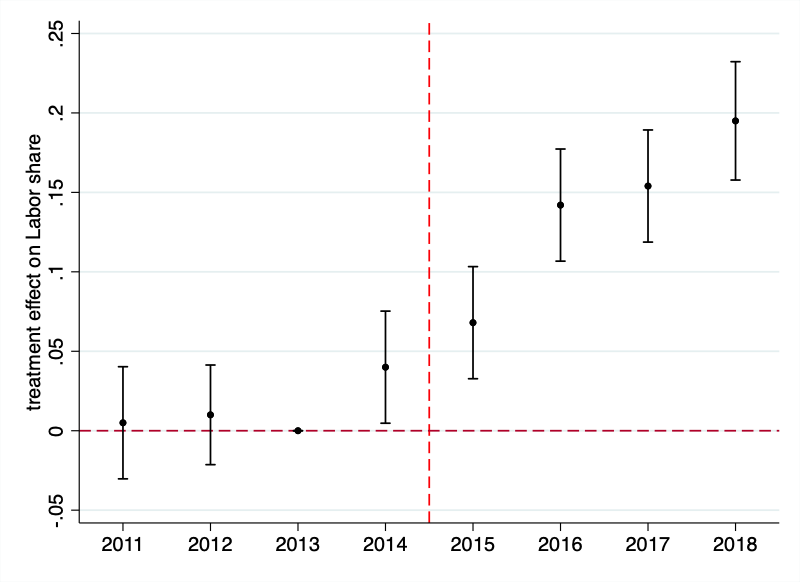}  
					\caption{Labor share, \\firms with high share of outsourceable jobs}
					%			\label{fig:pl_mech_labor_lvg_a}
				\end{subfigure}
				%3-------------------------------------
				\begin{subfigure}{.5\textwidth}
					\centering
					% include fourth image
					\includegraphics[width=.8\linewidth]{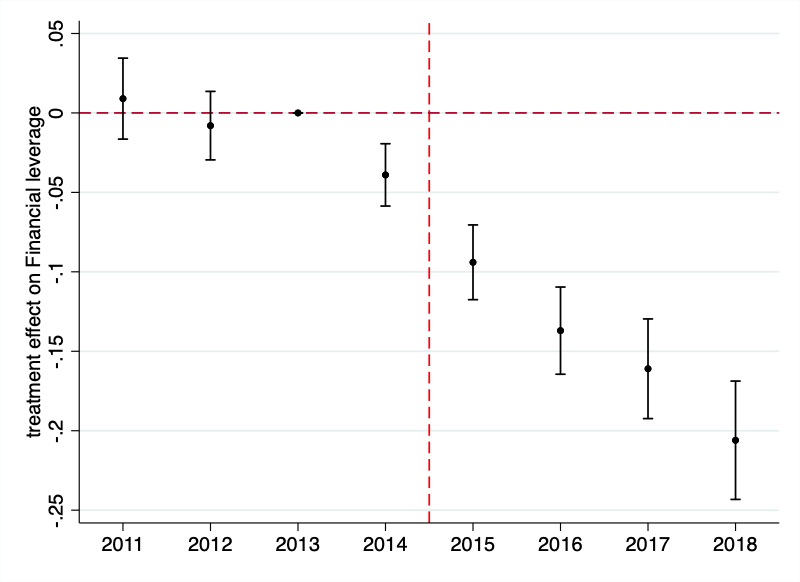}  
					\caption{Financial leverage, \\firms with low share of outsourceable jobs}
					%	\label{fig:pl_mech_labor_lvg_a}
				\end{subfigure}
				%4-------------------------------------
				\begin{subfigure}{.5\textwidth}
					\centering
					% include fourth image
					\includegraphics[width=.8\linewidth]{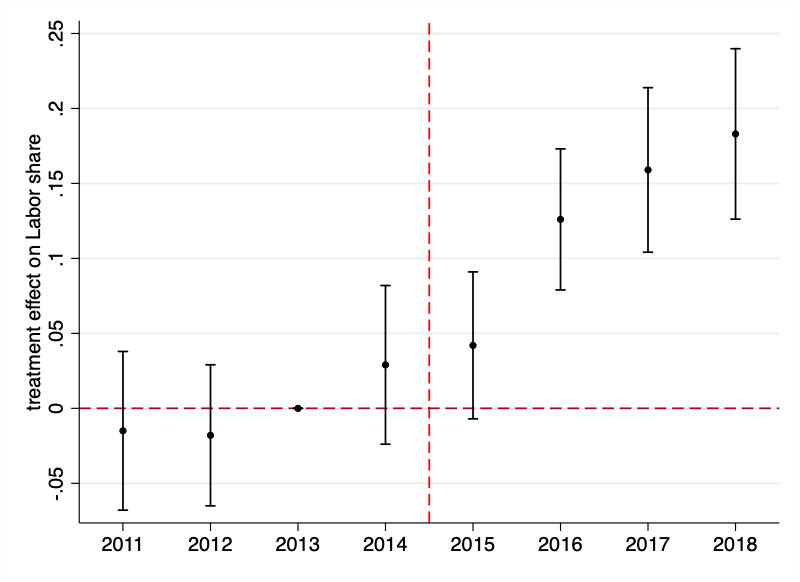}  
					\caption{Labor share, \\firms with low share of outsourceable jobs}
					%	\label{fig:pl_mech_labor_lvg_a}
				\end{subfigure}
				\caption*{\textit{Notes}: The figure displays the detrended difference-in-differences (DiD) regression coefficients of $Bite_{j} * Year_{k,t}$ with 95\% confidence intervals. The dependent variables are displayed below each figure.
							\\\textit{Data}: Linked data of  BeH, BHP, and Amadeus, 2011-2018.}
			\end{figure}

			%%%%%%%%%%%%%%%%%%%%%%%%%%%%%%%%%%%%%%%%%%%%%
			%4_6_sizes
			%%%%%%%%%%%%%%%%%%%%%%%%%%%%%%%%%%%%%%%%%%%%%
			\begin{figure}
				\captionabove{Coefficients for $Bite_j*Year_{kt}$ for Table \ref{tab:size_fi_lvg}}
        \label{fig:event_size}

				%4-------------------------------------
				\begin{subfigure}{.5\textwidth}
					\centering
					\includegraphics[width=.8\linewidth]{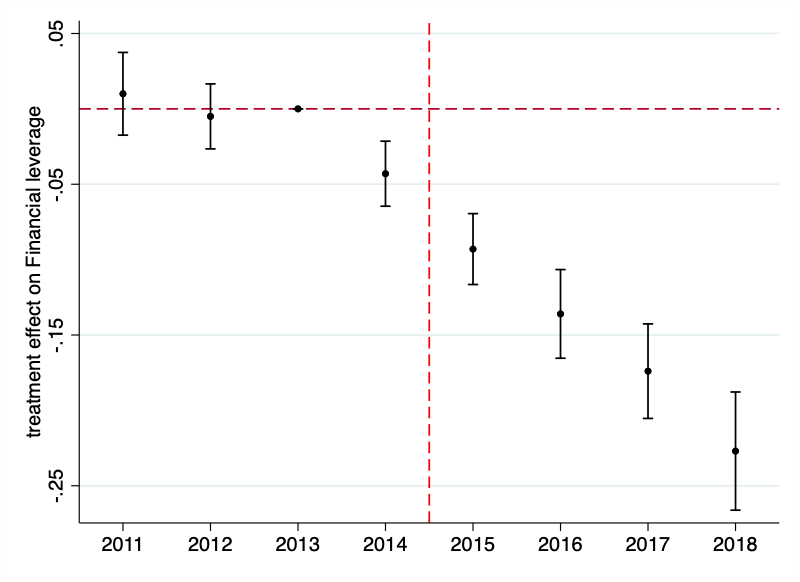}  
					\caption{Fin lvg, size$<$50}
					%	\label{fig:pl_mech_labor_lvg_a}
				\end{subfigure}
				\hfill
				%5-------------------------------------
				\begin{subfigure}{.5\textwidth}
					\centering
					% include fourth image
					\includegraphics[width=.8\linewidth]{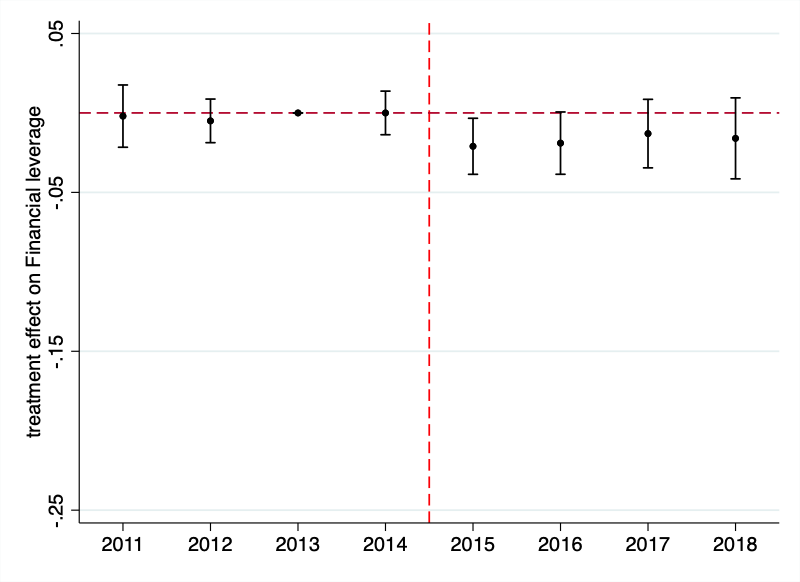}  
					\caption{Fin lvg, size:50-249}
					%	\label{fig:pl_mech_labor_lvg_a}
				\end{subfigure}
				\hfill
				%6-------------------------------------
				\begin{subfigure}{.5\textwidth}
					\centering
					% include fourth image
					\includegraphics[width=.8\linewidth]{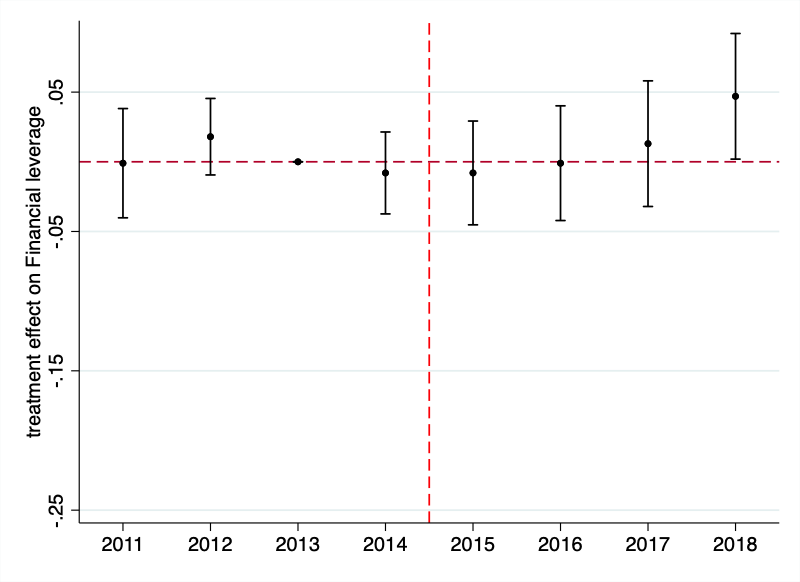}  
					\caption{Fin lvg: size$>=$250}
					%	\label{fig:pl_mech_labor_lvg_a}
				\end{subfigure}
				\hfill
				%7-------------------------------------
				\begin{subfigure}{.5\textwidth}
					\centering
					% include fourth image
					\includegraphics[width=.8\linewidth]{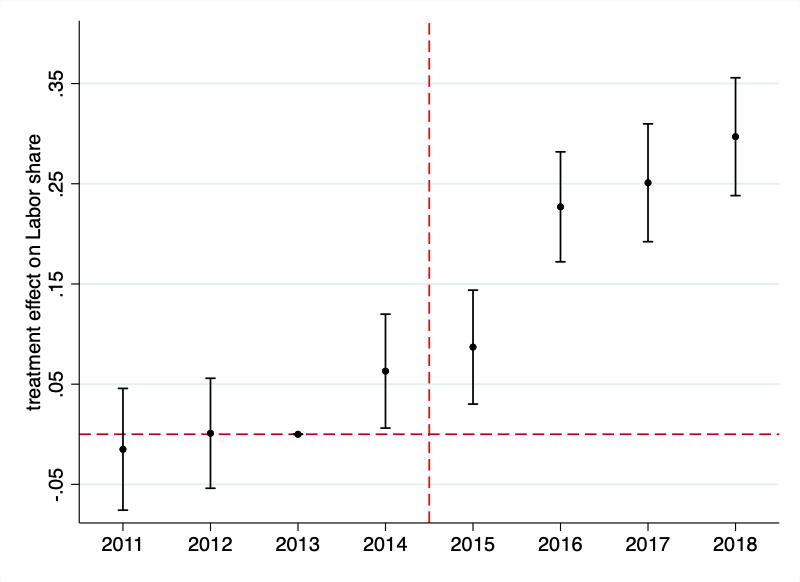}  
					\caption{Labor share, size$<$50}
					%	\label{fig:pl_mech_labor_lvg_a}
				\end{subfigure}
				\hfill
				%8-------------------------------------
				\begin{subfigure}{.5\textwidth}
					\centering
					% include fourth image
					\includegraphics[width=.8\linewidth]{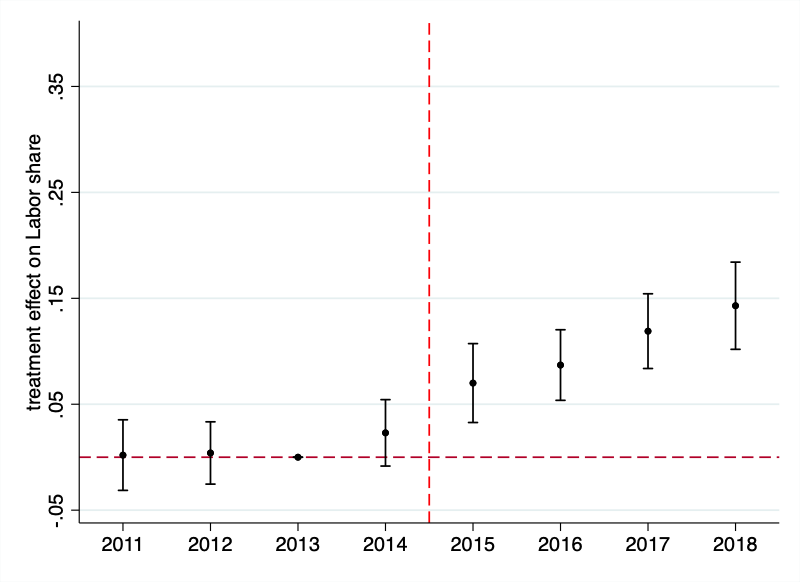}  
					\caption{Labor share, size:50-249}
					%	\label{fig:pl_mech_labor_lvg_a}
				\end{subfigure}
				\hfill
				%9-------------------------------------
				\begin{subfigure}{.5\textwidth}
					\centering
					% include fourth image
					\includegraphics[width=.8\linewidth]{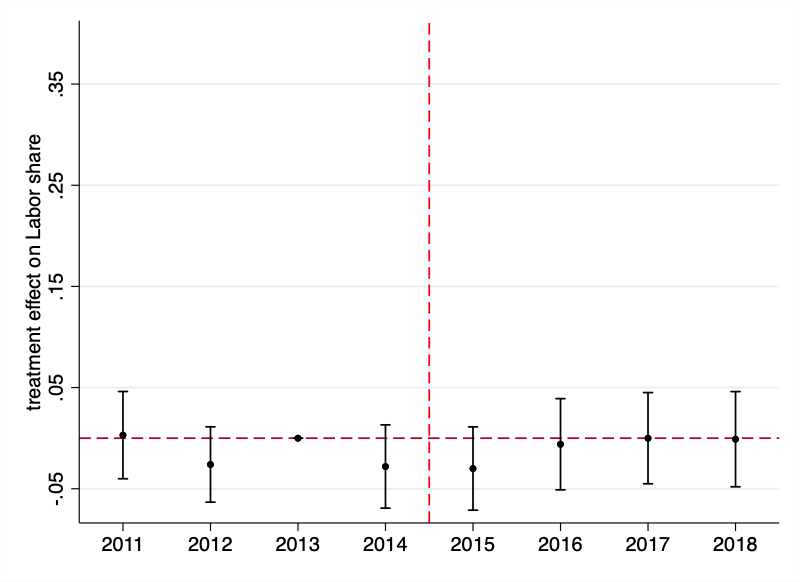}  
					\caption{Labor share, size$>=$250}
					%	\label{fig:pl_mech_labor_lvg_a}
				\end{subfigure}
				\caption*{\textit{Notes}: The figure displays the detrended difference-in-differences (DiD) regression coefficients of $Bite_{j} * Year_{k,t}$ with 95\% confidence intervals. The dependent variables are displayed below each figure.
				\\	\textit{Data}: Linked data of  BeH, BHP, and Amadeus, 2011-2018.}
			\end{figure}

			\begin{figure}
				\captionabove{Coefficients for $Bite_j*Year_{kt}$ for Table \ref{tab:size_fi_lvg}}

				%1-------------------------------------
				\begin{subfigure}{.5\textwidth}
					\centering
					
					% include third image
					\includegraphics[width=.8\linewidth]{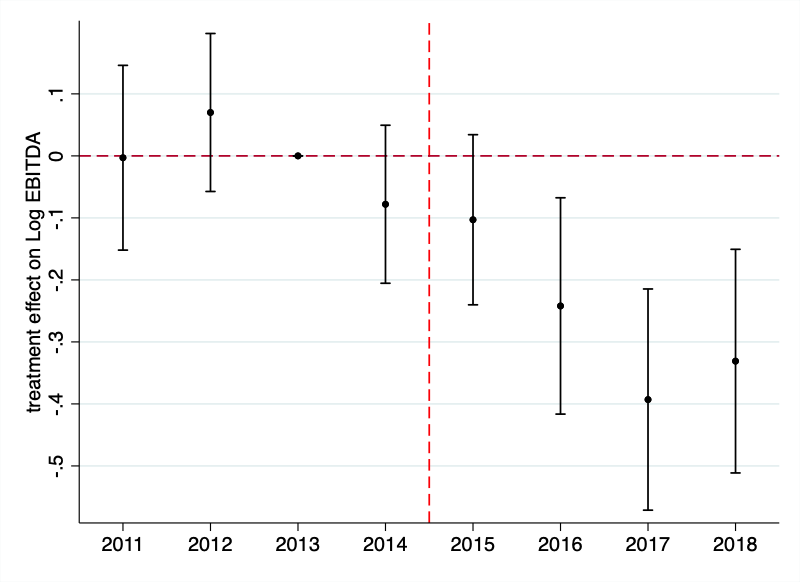}  
					\caption{Log EBITDA, size$<$50}
					%	\label{fig:pl_mech_labor_lvg_a}
				\end{subfigure}
				%2-------------------------------------		
				\begin{subfigure}{.5\textwidth}
					\centering
					% include fourth image
					\includegraphics[width=.8\linewidth]{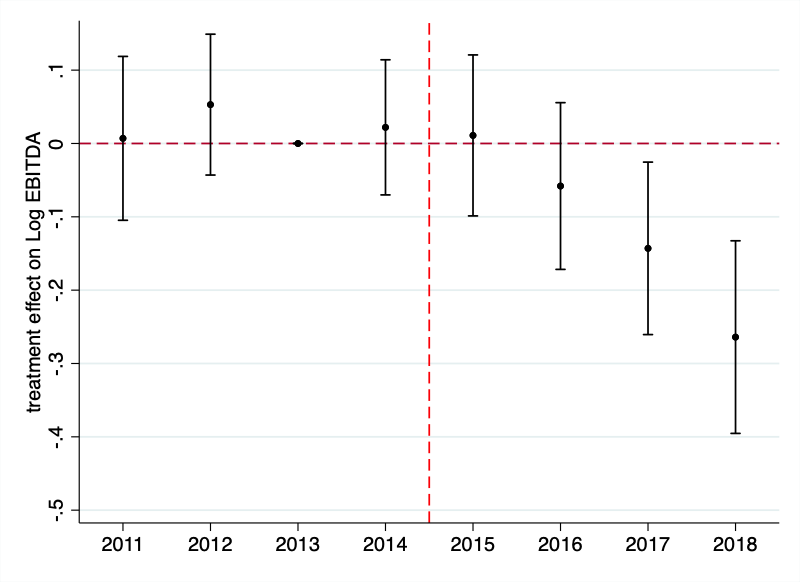}  
					\caption{Log EBITDA, size 50-249}
					%			\label{fig:pl_mech_labor_lvg_a}
				\end{subfigure}
				%3-------------------------------------
				\begin{subfigure}{.5\textwidth}
					\centering
					% include fourth image
					\includegraphics[width=.8\linewidth]{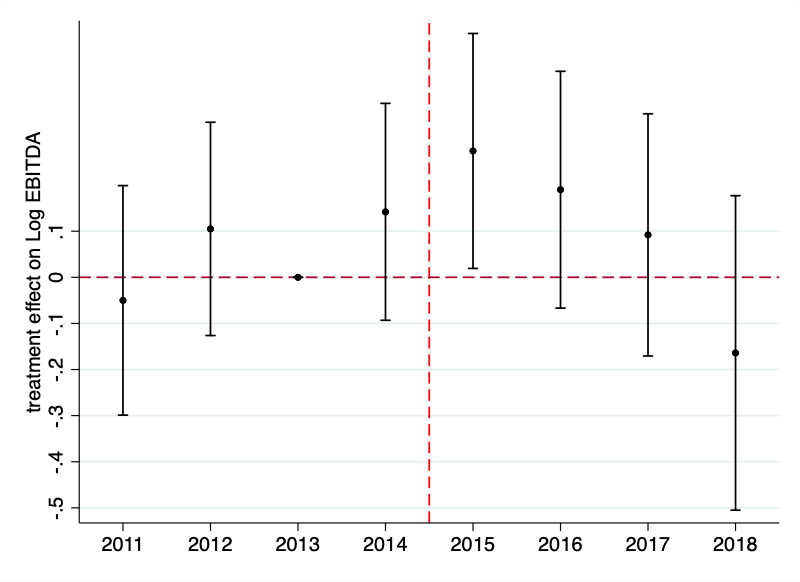}  
					\caption{Log EBITDA, size$>=$250}
					%	\label{fig:pl_mech_labor_lvg_a}
				\end{subfigure}	
				\caption*{\textit{Notes}: The figure displays the detrended difference-in-differences (DiD) regression coefficients of $Bite_{j} * Year_{k,t}$ with 95\% confidence intervals. The dependent variables are displayed below each figure.
					\\\	\textit{Data}: Linked data of  BeH, BHP, and Amadeus, 2011-2018.}
			\end{figure}		
		\end{document}